\documentclass[aps,pre,superscriptaddress,twocolumn,showkeys,floatfix]{revtex4-2}
\usepackage{enumitem}
\usepackage{graphicx,amssymb,amsmath,amssymb,natbib,bm}
\usepackage{epstopdf}
\usepackage[caption=false]{subfig}
\usepackage{float}
\usepackage{mathscinet}
\usepackage{color}
\usepackage{comment}

\newcommand{\sref}[1]{Sec.~\ref{#1}}

\begin{document}

\title{The effect of algorithmic bias and network structure on coexistence, consensus, and polarization of opinions}
%\title{Emergence of polarization in information spreading due to algorithmic bias}
%Effect of algorithmic bias in the dynamics of binary-state models on complex networks}
\author{Antonio F. Peralta}
\email{peraltaaf@ceu.edu}
\affiliation{Department of Network and Data Science, Central European University, A-1100 Vienna, Austria}
\author{Matteo Neri}
%\email{peraltaaf@ceu.edu}
\affiliation{Department of Network and Data Science, Central European University, A-1100 Vienna, Austria}
\author{J\'anos Kert\'esz}
%\email{kerteszj@ceu.edu}
\affiliation{Department of Network and Data Science, Central European University, A-1100 Vienna, Austria}
\affiliation{Complexity Science Hub, A-1080 Vienna, Austria}
\author{Gerardo I\~{n}iguez}
\email{iniguezg@ceu.edu}
\affiliation{Department of Network and Data Science, Central European University, A-1100 Vienna, Austria}
\affiliation{Department of Computer Science, Aalto University School of Science, FI-00076 Aalto, Finland}
\affiliation{Centro de Ciencias de la Complejidad, Universidad Nacional Auton\'{o}ma de M\'{e}xico, 04510 Ciudad de M\'{e}xico, Mexico}
\date{\today}

\begin{abstract}
Individuals of modern societies share ideas and participate in collective processes within a pervasive, variable, and mostly hidden ecosystem of content filtering technologies that determine what information we see online. Despite the impact of these algorithms on daily life and society, little is known about their effect on information transfer and opinion formation. It is thus unclear to what extent algorithmic bias has a harmful influence on collective decision-making, such as a tendency to polarize 
%online 
debate. Here we introduce a general theoretical framework to systematically link models of 
%information spreading, 
opinion dynamics, social network structure, and content filtering. We showcase the flexibility of our framework by exploring a family of binary-state opinion dynamics models where information exchange lies in a spectrum from pairwise to group interactions. All models show an opinion polarization regime driven by algorithmic bias and modular network structure. The role of content filtering is, however, surprisingly nuanced; for pairwise interactions it leads to polarization, while for group interactions it promotes coexistence of opinions. This allows us to pinpoint which social interactions are robust against algorithmic bias, and which ones are susceptible to bias-enhanced opinion polarization. Our framework gives theoretical ground for the development of heuristics to tackle harmful effects of online bias, such as information bottlenecks, echo chambers, and opinion radicalization.
\end{abstract}
\maketitle

\section{Introduction}\label{sec_intro}

Information spreading, opinion formation and other dynamical phenomena occurring on top of social networks have long been studied via agent-based modeling within the framework of statistical mechanics \cite{Castellano:2009,Porter:2016,cimini2019statistical}. The main goal of these stylized models is to discern how 
%simple 
local mechanisms governing the actions of individuals may lead to emergent collective behavior at the societal level, such as the rise of consensus of opinion within a group. The structure of society is typically represented by a network \cite{Newman:2003b,Boccaletti:2006,Starnini:2013,Latora:2017,lambiotte2021modularity}, %{\color{red}JK: Newer refs?}
where nodes correspond to individuals or groups thereof, and edges reflect the varied interactions between them (e.g., sharing information and opinions about political issues).
%or any other topic relevant to the context of study. 
Traditionally, information spreading over social networks has been 
%considered to be 
the result of face-to-face or phone conversations and consumption of mass media such as TV, radio or newspapers. In recent years, however, communication technologies have dramatically changed the way people interact, with a larger portion of information exchange taking place in online social media platforms like Google, Twitter, and Facebook \cite{lazer2009life,conte2012manifesto}.

Online social networks tune their services to maximize usage, rather than to serve accurate or balanced information. In order to achieve their business goals, they control the information users receive by means of filtering algorithms that attempt to deliver relevant and engaging content \cite{nikolov2018quantifying}. These algorithms collect personalized data on individual preferences and use it to selectively expose users to material that is either popular or similar to what they have consumed before \cite{Bozdag:2013,Moller:2018}. Such filtering leads to {\it algorithmic bias}, the tendency to receive information individuals already agree with \cite{Pariser:2011,Bakshy:2015}. The consequences of algorithmic bias at the societal scale are a matter of recent debate \cite{DelVicario:2016,bail2018exposure,ciampaglia2018algorithmic,blex2020positive}, but likely include emergence of the so-called ``filter bubbles'' or ``echo chambers'', groups with polarized views that reinforce their own opinions and rarely communicate with each other \cite{baumann2020modeling,cinelli2021echo}. Collective phenomena like fragmentation and polarization of opinion groups, increasingly visible features of the current socio-political landscape worldwide, are partly the outcome of the interplay between social behavior and algorithmic filtering happening online.

Previous efforts to explore the effect of algorithmic bias on information spreading \cite{Sirbu:2019} have considered bounded confidence mechanisms with continuous opinion variables \cite{Deffuant:2000}, where individuals interact only if their opinions are similar enough, and the degree of required similarity is related to the intensity of filtering. Under bounded confidence, algorithmic bias favors fragmentation and polarization, but slows down opinion formation. Binary-state models~\cite{Gleeson:2013}, widely used to study social interactions~\cite{Castellano:2009,Porter:2016,cimini2019statistical}, have also been explored in the context of algorithmic bias \cite{Perra:2019}.
%Such a nuanced the role of algorithms in online social behavior is also apparent in the case of binary opinion dynamics . 
When filtering promotes content similar to the opinion of a user, structural correlations lead to polarization, and network heterogeneity tends to decrease it.

While these results suggest that polarization arises from a mix of social behavioral patterns and the online algorithms constraining them, we still lack a general theoretical framework systematically linking models of information spreading, network structure, and algorithmic bias. This is particularly relevant given the diversity of mechanisms arguably driving the way people exchange information, including homophily \cite{mcpherson2001birds,asikainen2020cumulative} and social contagion \cite{watts2002simple,unicomb2021dynamics}, which may in turn lead to radically different patterns of polarization. Here we propose such a formalism by extending the theoretical description of binary-state dynamics, based on mean-field \cite{Sood:2008,Vespignani:2012}, pair \cite{Vazquez:2008a,Pugliese:2009} and higher-order \cite{Gleeson:2011, Gleeson:2013,Ruan:2015,Unicomb:2018} approximations, with a simple but flexible notion of algorithmic bias. Our formalism can be applied to a wide variety of models of information spreading, social networks with arbitrary degree distributions and modular structure, and implementations of online content filtering.

We showcase the potential and flexibility of our framework by focusing on a wide family of binary-state models of opinion formation, where the nature of information exchange lies in a spectrum from pairwise to group interactions in the presence of noise. In the extreme of pairwise interactions, represented by the {\it noisy voter model} \cite{Kirman:1993, Granovsky:1995}, opinion-switching depends on a herding or imitation mechanism where individuals copy the opinions of their neighbors. In the extreme of group interactions, implemented by the {\it majority-vote model} \cite{deOliveira:1992}, individuals change opinion if most of their neighbors have opposing views. We finally consider the {\it language model} \cite{Abrams:2003,Castellano:2009b,Schweitzer:2009,Nyczka:2012,Jedrzejewski:2017,Peralta:2018}, a non-linear extension of noisy voter dynamics, which has a behavior interpolating between the voter and the majority-vote model as a function of a model parameter. When studied over networks, these models exhibit a rich 
%and well-known 
phenomenology ranging from continuous/discontinuous coexistence-consensus-polarization transitions to non-trivial scaling behavior \cite{Peralta:2018} making them ideal ground to explore the generic role of algorithmic bias on the dynamics of information spreading.

The paper is organized as follows. In \sref{sec_model} we present the studied binary-state models and summarize their known properties. We also introduce the notion of algorithmic bias and its effect on the transition rates of the dynamics. In \sref{sec_methods} we derive mean-field rate equations of global opinion variables for both homogeneous and modular networks. In \sref{sec_theo_results} we analyze the stationary solutions of the mean-field equations and their linear stability, focusing on a transition to polarization driven by the joint effects of content filtering and modular structure. In \sref{sec_com_num} we gauge the accuracy of our theoretical results with numerical simulations on both synthetic and real-world networks. Overall, we show that algorithmic bias has opposite effects depending on the mechanism governing information spreading: for pairwise interactions it leads to polarization, while for group interactions it promotes coexistence.

\section{Model}\label{sec_model}

\subsection{Binary-state dynamics}\label{sec_binary_state}

\begin{figure}[t]
\begin{center}
\includegraphics[width=0.44\textwidth]{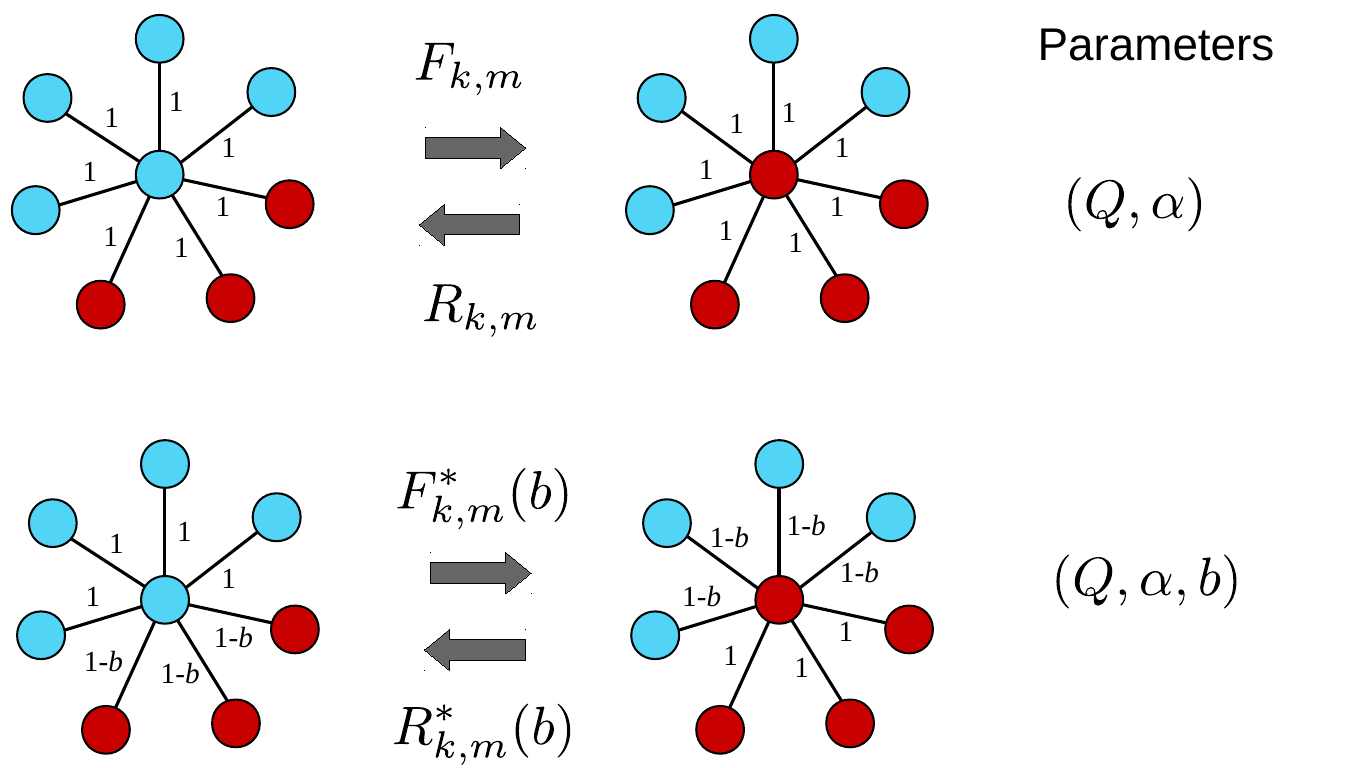}
\caption{Schematic representation of dynamics of information spreading with a minimal notion of algorithmic bias. With no bias ($b=0$; upper plot), the central node (with degree $k$ and $m$ infected neighbors) interacts with any of its neighbors with probability $1$, regardless of state (denoted by color and shading) and according to transitions rates $F_{k,m}$ and $R_{k,m}$. With bias ($b>0$; lower plot), the node interacts with any of its neighbors in the opposite state with probability $1-b$, resulting in bias-dependent effective rates [$F_{k,m}^{*}(b)$ and $R_{k,m}^{*}(b)$]. The dynamics is determined by noise $Q$ and a parameter $\alpha$ regulating pairwise and group interactions.}
\label{fig_rates_scheme}
\end{center}
\end{figure}

In order to characterize the dynamics of information spreading in a networked population of $N$ individuals, we take a binary-state approach where each individual $i=1, \dots, N$ holds a variable $s_{i}(t)=0, 1$ at time $t$. The interpretation of this state is varied and depends on the context and model chosen \cite{Castellano:2009,Gleeson:2011,Gleeson:2013}, with $s=0$ typically denoting a state of susceptibility or inactivity, and $s=1$ a state of infection or activity. In the case of opinion dynamics, states encode the tendency to agree with some binary opinion ($s=0$) or its opposite ($s=1$). Individuals influence each other and may eventually be convinced to change opinion. We consider a social network with adjacency matrix $A_{ij}$, equal to $1$ if $i$ and $j$ are connected and $0$ otherwise. The rate (probability per unit time) at which node $i$ changes state is a function of network degree, $k_{i}=\sum_{j=1}^{N}A_{ij}$, and the number of infected neighbors (in state $1$), $m_{i}=\sum_{j=1}^{N} A_{ij} s_{j}$ (Fig. \ref{fig_rates_scheme}). We define the rates of infection ($F_{k,m}$) and recovery ($R_{k,m}$) as the rates of state switching from $s=0 \rightarrow 1$ and from $s=1 \rightarrow 0$, respectively. The functional form of these transition rates determines how individuals behave collectively as a result of interactions with their neighbors. The macroscopic, dynamical behavior of the social system is encoded in the global opinion variable
$\rho = N^{-1} \sum_{i=1}^{N} s_{i} \in [0, 1]$, i.e. the fraction of nodes in state 1. 

%\begin{itemize}
%    \item Noisy voter
%    \begin{equation}
%    \label{voter_rate}
%    F_{k,m} = Q + (1-2 Q) \frac{m}{k}.    
%    \end{equation}
%    \item Majority-vote
%    \begin{equation}
%    \label{majority_rate}  
%    F_{k,m} = \begin{cases}
%Q \hspace{0.2cm} &\text{if} \hspace{0.5cm}  m < k/2, \\
%1/2 \hspace{0.2cm} &\text{if} \hspace{0.5cm} m=k/2, \\
%1-Q \hspace{0.2cm} &\text{if} \hspace{0.5cm}  m > k/2.
%\end{cases}
%    \end{equation}
%    \item Language
%    \begin{equation}
%    \label{language_rate}
%    F_{k,m} = Q + (1-2 Q) \left( \frac{m}{k} \right)^{\alpha}. 
%    \end{equation}
%\end{itemize}

\begin{table*}[t]
\setlength{\tabcolsep}{8pt}
\begin{center}
\begin{tabular}{|c|c|c|c|}
 \hline
 Model & $F_{k,m}$ & $R_{k,m}$ & Phenomenology\\ 
 \hline
 \hline
 Noisy voter \citep{Kirman:1993,Granovsky:1995}   & $Q + (1-2 Q) \dfrac{m}{k}$    & $Q + (1-2 Q) \dfrac{k-m}{k}$ & $\begin{array}{l}
 \text{ } \\
\text{No transition } (Q>0)\\
\text{ }
\end{array}$\\
 %\hline
 %\hline
 Language \cite{Abrams:2003,Castellano:2009b,Schweitzer:2009,Nyczka:2012,Jedrzejewski:2017,Peralta:2018} & $Q + (1-2 Q) \left(\dfrac{m}{k}\right)^{\alpha}$ & $Q + (1-2 Q) \left(\dfrac{k-m}{k}\right)^{\alpha}$ & $\begin{array}{l}
\text{No transition } (\alpha \le 1)\\
\text{Continuous transition } (1 < \alpha \le 5) \\
\text{Discontinuous transition } (\alpha >5) \\
\end{array}$ \\
 Majority-vote \citep{deOliveira:1992} &  $\begin{cases}
Q \hspace{0.2cm} &\text{if} \hspace{0.5cm}  m < k/2 \\
1/2 \hspace{0.2cm} &\text{if} \hspace{0.5cm} m=k/2 \\
1-Q \hspace{0.2cm} &\text{if} \hspace{0.5cm}  m > k/2
\end{cases}$  & $\begin{cases}
1-Q \hspace{0.2cm} &\text{if} \hspace{0.5cm}  m < k/2 \\
1/2 \hspace{0.2cm} &\text{if} \hspace{0.5cm} m=k/2 \\
Q \hspace{0.2cm} &\text{if} \hspace{0.5cm}  m > k/2
\end{cases}$ & $\begin{array}{l}
\vspace{0.2cm}
\\
\text{Continuous transition}
\\
\vspace{0.2cm}
\end{array}$\\
 \hline
\end{tabular}
\end{center}
\caption{Models considered in this work, along with references, transition rates ($F_{k,m}$ and $R_{k,m}$) and basic phenomenology of the coexistence-consensus transition in the mean-field limit (complete graph).}
\label{tab_models}
\end{table*}

We focus our attention on binary-state dynamics with up-down symmetry,
\begin{equation}
\label{up-down}
R_{k,m}=F_{k,k-m},
\end{equation}
meaning the probability to change state is a function of the number of neighbors in the opposite state, regardless of state. From this family we consider three prototypical opinion dynamics: the noisy voter, language, and majority-vote models (Table \ref{tab_models}).
All models include the parameter $Q \in [0, 1/2]$ (with $Q=F_{k,0}=R_{k,k}$), commonly interpreted as noise, `social temperature', or `independence' \citep{Nail:2000, Nyczka:2013, Jedrzejewski:2017}, and equal to the probability of changing state when all neighbors have the same (opposite) opinion. The three models differ in their infection rate $F_{k,m}$: (i) the noisy voter model has a linear dependence on the fraction of infected neighbors $m/k$, corresponding to a pairwise copying mechanism or blind imitation; (ii) the majority-vote model considers that individuals copy the majority state in their neighborhood; and (iii) the language model introduces a non-linear dependence $(m/k)^\alpha$ regulated by a tuning parameter $\alpha \in (0,\infty)$. For integer $\alpha$, the language model is driven by group interactions between an individual and $\alpha$ of its neighbors, and opinion unanimity in the group is required to change state. This particular case is also known as the $q-$voter model \cite{Castellano:2009b,Nyczka:2012,Nyczka:2013,Vieira:2020} (with $q = \alpha$). For $\alpha=1$, the language model recovers the noisy voter model and its pairwise interactions.

In many cases the models show a symmetry-breaking phase transition as a function of the noise parameter $Q$, usually between stationary states of opinion consensus [$\rho(t) \neq 1/2$] and coexistence [$\rho(t) = 1/2$] for $t\rightarrow \infty$ (see Table~\ref{tab_models}). The phenomenology of the transition depends on the model. We differentiate between two general behaviors, voter like and majority-vote like \cite{Castello:2006}, with the language model interpolating between the two. For example, in the mean-field limit we can tune $\alpha$ and move from voter like behavior (low $\alpha  \lesssim 1$; pairwise interactions), to majority-vote like (high $\alpha \leq 5$; group interactions). In the regime $\alpha >5$ the transition is always discontinuous. This classification, although qualitative, will help us further understand the phenomenology of information spreading for varying $\alpha$ in the presence of algorithmic bias.

\begin{figure*}[t]
\begin{center}
\includegraphics[width=0.99\textwidth]{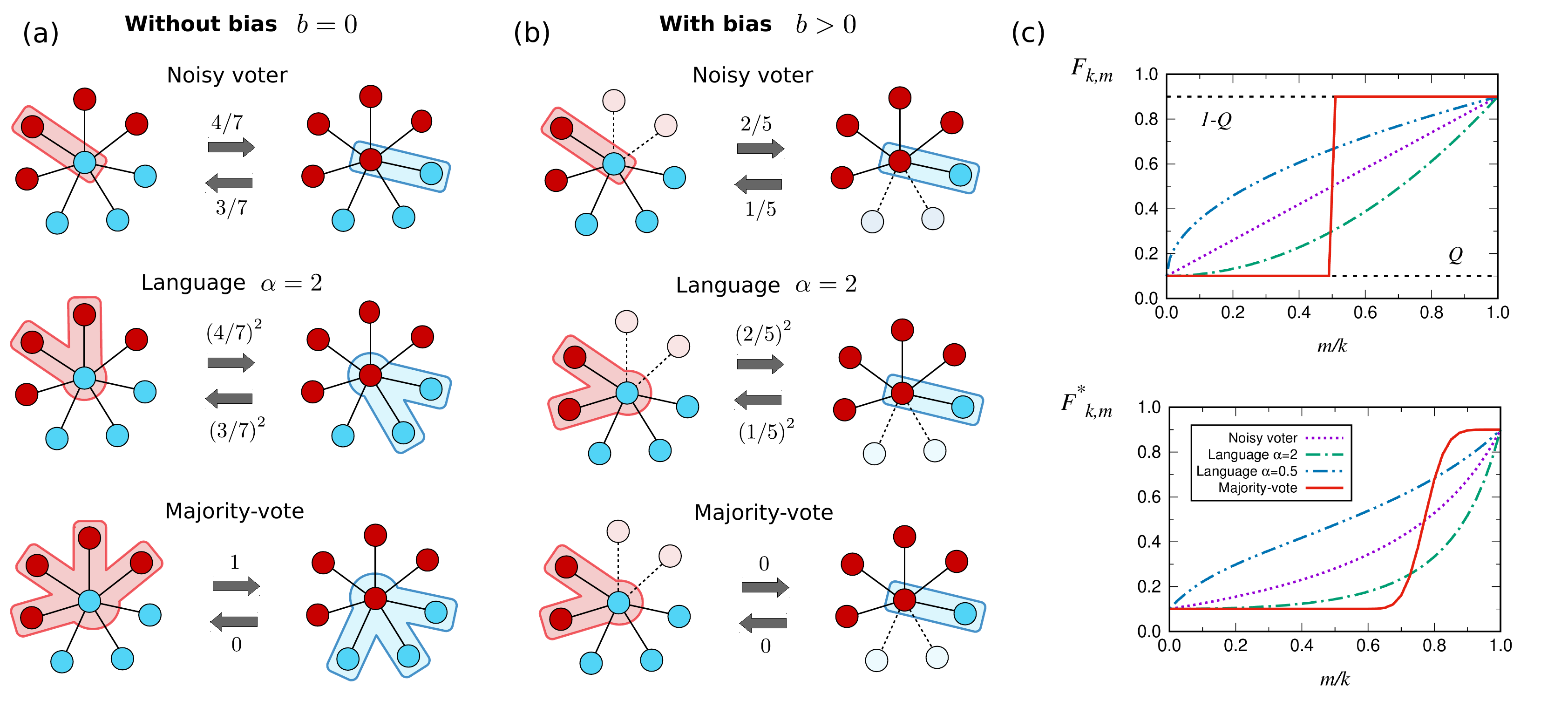}
\caption{Schematic representation of the considered binary-state dynamics of opinion formation, both in the absence ($b=0$; a) and presence ($b>0$; b) of algorithmic bias. The central node interacts with a set of neighbors (shaded area), ranging from pairwise to group interactions (top to bottom), leading to original/effective transition rates between states (denoted by numbers). In the presence of bias, some neighbors in the opposite state are not considered (dashed links). In (c) we show the functional form of the original ($F_{k,m}$; top) and effective ($F_{k,m}^{*}(b)$; bottom) infection rates as a function of the fraction of infected neighbors $m/k$ for all models, with bias intensity $b=0$ and $b=0.7$, respectively, degree $k=40$ and $Q=0.1$.}
\label{fig_paper_scheme}
\end{center}
\end{figure*}

\subsection{Algorithmic bias}\label{sec_algorithmic}

Online platforms use content filtering algorithms, particularly semantic and collaborative filtering \cite{Bozdag:2013,Moller:2018}, to preferably display content similar to what an individual (or alike users) have consumed before, leading to algorithmic bias \cite{Pariser:2011,Bakshy:2015}. In the context of binary-state dynamics of information spreading over social networks, we can minimally implement this bias as a tunable preference to filter the interactions between an individual and its neighbors, depending on their state. We introduce the {\it bias intensity} $b$, defined as the probability that a node does not interact with a neighbor in the opposite state (due to content filtering by the platform). If the dynamics is originally driven by rates $F_{k,m}$ and $R_{k,m}$, bias leads to the effective transition rates
\begin{align}
\label{efective_rate_def_infection}
&F_{k,m}^{*}(b) = \sum_{i=0}^{m} B_{m,i}(1-b) F_{k-m+i,i},\\
\label{efective_rate_def_recovery}
&R_{k,m}^{*}(b) = \sum_{s=0}^{k-m} B_{k-m,s}(1-b) R_{m+s,m},
\end{align}
where $B_{m,i}(1-b)=\binom{m}{i} (1-b)^{i} b^{m-i}$ is the binomial distribution. Eqs.~(\ref{efective_rate_def_infection})-(\ref{efective_rate_def_recovery}) are simply the average transition rates after removing a number of randomly selected neighbors in the opposite state with probability $b$. Within our framework, the effects of algorithmic bias amount to a binary-state dynamics driven by the effective rates $F_{k,m}^{*}(b)$ and $R_{k,m}^{*}(b)$. Note that this definition of bias respects up-down symmetry [Eq.~(\ref{up-down})], since $R_{k,m}^{*}(b) = F^{*}_{k,k-m}(b)$. In general, it is not possible to obtain a simple closed expression for the effective rates of the models listed in Table \ref{tab_models}. In Fig. \ref{fig_paper_scheme} we show a schematic representation of the effect of algorithmic bias on the considered models, together with the shape of the original and effective transition rates as a function of the fraction of infected neighbors $m/k$. In the presence of bias ($b>0$), the language model has effective rates similar to the original rates for a higher value of $\alpha$. Since algorithmic bias `hides' neighbors in the opposite state, more such neighbors are needed for an individual to change state (i.e. a higher $\alpha$ value in the original dynamics). Overall we have $F_{k,m}^{*}(b) \le F_{k,m}$, meaning bias impedes interactions between individuals that result in a change of opinion. This property is valid for monotonically increasing rates, as $i/(k-m+i) \leq m/k$ for $i=0, \dots, m$ in Eq.~(\ref{efective_rate_def_infection}).
%(if $\alpha \ge 1$) by means of Jensen's inequality in Eq.~(\ref{efective_rate_def_infection}).

\section{Methods}\label{sec_methods}

\subsection{Numerical simulations}\label{sec_simulation}

The most direct implementation of our framework is by numerical simulation of the stochastic rules described in Section~\ref{sec_model}. At time $t$ of the simulation we perform the following steps: (i) Select an individual $i$ uniformly at random from all $N$ nodes, which has degree $k$ and number of infected neighbors $m$ at time $t$. (ii) If $s_{i}=0$ the individual switches state with probability $F_{k,m}^{*}$, and if $s_{i}=1$ with probability $R_{k,m}^{*}$ \footnote{Note that this requires that the $\text{max} \lbrace F^{*}_{k,m}, R^{*}_{k,m} \rbrace \leq 1$ for all nodes (all possible values of $k$ and $m$). If that is not the case, we must re-scale the rates and time dividing by the maximum value $\text{max} \lbrace F^{*}_{k,m}, R^{*}_{k,m} \rbrace$.}. (iii) Time increases by $\Delta t = 1/N$ (i.e. the time unit is one Monte Carlo step per node). This numerical method allows us to obtain trajectories of the state variables $\lbrace s_{i}(t) \rbrace_{i=1,\dots,N}$ and consequently the global state $\rho(t)$. We may then compute the average $\langle \rho(t) \rangle$ over stochastic realizations of the same initial conditions, as typically done in non-equilibrium ensembles (in what follows we drop the average brackets for simplicity of notation, unless otherwise stated).
%in order to have a simplified description of the dynamics.

\subsection{Mean-field description}\label{sec_mf}

\subsubsection{Homogeneous network structure}

In the simplest analytical treatment of binary-state dynamics, we assume that one dynamical variable is sufficient to describe the state of the system: the global opinion (or fraction of infected nodes) $\rho(t)$. Following the heterogeneous mean-field approximation \cite{Vespignani:2012,Gleeson:2013}, we obtain a closed differential equation for the dynamics by defining the average rate $f$ of switching state from $0$ to $1$. In the absence of algorithmic bias ($b=0$),
\begin{equation}
\label{mean_field_rate}
f[x] \equiv \sum_{k} \frac{P_{k} k}{z} \sum_{m=0}^{k} F_{k,m} B_{k,m}(x),
\end{equation}
where $P_{k}$ is the degree distribution of the network, $z = \sum_{k} P_{k} k$ is the average degree, and $x$ is the probability of finding a neighbor in state $1$.  If we consider a homogeneous, highly connected network with $z \gg 1$ (i.e. $P_{k}$ peaks around a high degree), then the binomial function $B_{k,m}(x)$ is also highly peaked around a large $m = z x$. Since the transition rates of the models in Table \ref{tab_models} only depend on the fraction of infected nodes $m/k$, we have $f[m/k] \approx F_{k,m}$. As we approach the mean-field limit, we replace the local (node) probability $x$ of finding a neighbor in state $1$ by the fraction of infected nodes in the network, $\rho$.
%For the model without bias $b=0$, the effective rate $f[m/k]$ in the highly connected case is equivalent to rate definition $F_{k,m}$. 
In the presence of algorithmic bias ($b>0$), Eq.~(\ref{efective_rate_def_infection}) and the assumption of a highly connected network lead to $F^{*}_{k,m} \approx F_{k-b m, (1-b) m}$. The biased version of Eq.~(\ref{mean_field_rate}) is then $f^{*}[x] = f \left[(1-b)x/(1-bx) \right]$, which reduces to $f^{*}[x]=f[x]$ for $b=0$.

Taking into account these approximations, we write a differential equation for the average over realizations of the global opinion $\rho(t)$,
%, that is:
\begin{equation}
\label{dyn_rho}
\frac{d \rho}{d t}= (1-\rho) f \left[ \frac{(1-b) \rho}{1-b \rho}  \right] - \rho f \left[ \frac{(1-b) (1-\rho)}{1-b (1-\rho)}  \right],
\end{equation}
where we assume that the probability $x$ of finding a neighbor in state $1$ is just the fraction of infected nodes in the network. Eq.~(\ref{dyn_rho}) can be thought of as a detailed balance condition: the change in time of the fraction of infected nodes $\rho$ is equal to the probability of selecting a susceptible node ($s=0$) times the rate of switching from $0$ to $1$, minus the probability of selecting an infected node ($s=1$) times the rate of switching from $1$ to $0$. A more detailed derivation of Eq.~(\ref{dyn_rho}), and an analysis of the accuracy of the highly connected approximation, can be found in the Supplemental Material (SM) \cite{SM_ref}, Sections S1.2.1 and S1.1. 
%{\color{blue} See Supplemental Material at \cite{S1.2.1, S1.1} for a more detailed derivation of Eq.~(\ref{dyn_rho}) and an analysis of the accuracy of the highly connected approximation.}

Note that the accuracy of Eq.~(\ref{dyn_rho}) depends on two assumptions: (i) a highly connected, homogeneous network with $P_{k}$ peaked around a large average degree $z \gg 1$ (exactly valid only in the case of a complete graph with $z=N-1$); and (ii) a negligible role of stochastic finite-size effects, with $\langle \rho(t)^{n} \rangle \approx \langle \rho(t) \rangle^{n}$ for any $n \geq 1$ (valid in the thermodynamic limit $N \rightarrow \infty$).

%Note that we made two approximations to derive Eq.~(\ref{dyn_rho}): (i) neglect stochastic finite-size effects by assuming $\langle \rho(t)^{n} \rangle \approx \langle \rho(t) \rangle^{n}$ for any $n \geq 1$, which is of general validity in the thermodynamic limit $N \rightarrow \infty$ {\color{red}JK: Isn't it that this $N\to \infty$ limit works only for the highly connected network? For a lattice it would not work. If so, I suggest exchanging i) and ii).}, and (ii) assume a highly connected network, with $P_{k}$ peaked around a high degree value $k \gg 1$, which is exact only in the case of a \emph{complete graph} with $k=N-1$.

\subsubsection{Modular network structure}

\begin{figure}[t!]
\vspace{20pt}
\begin{center}
\includegraphics[width=0.44\textwidth]{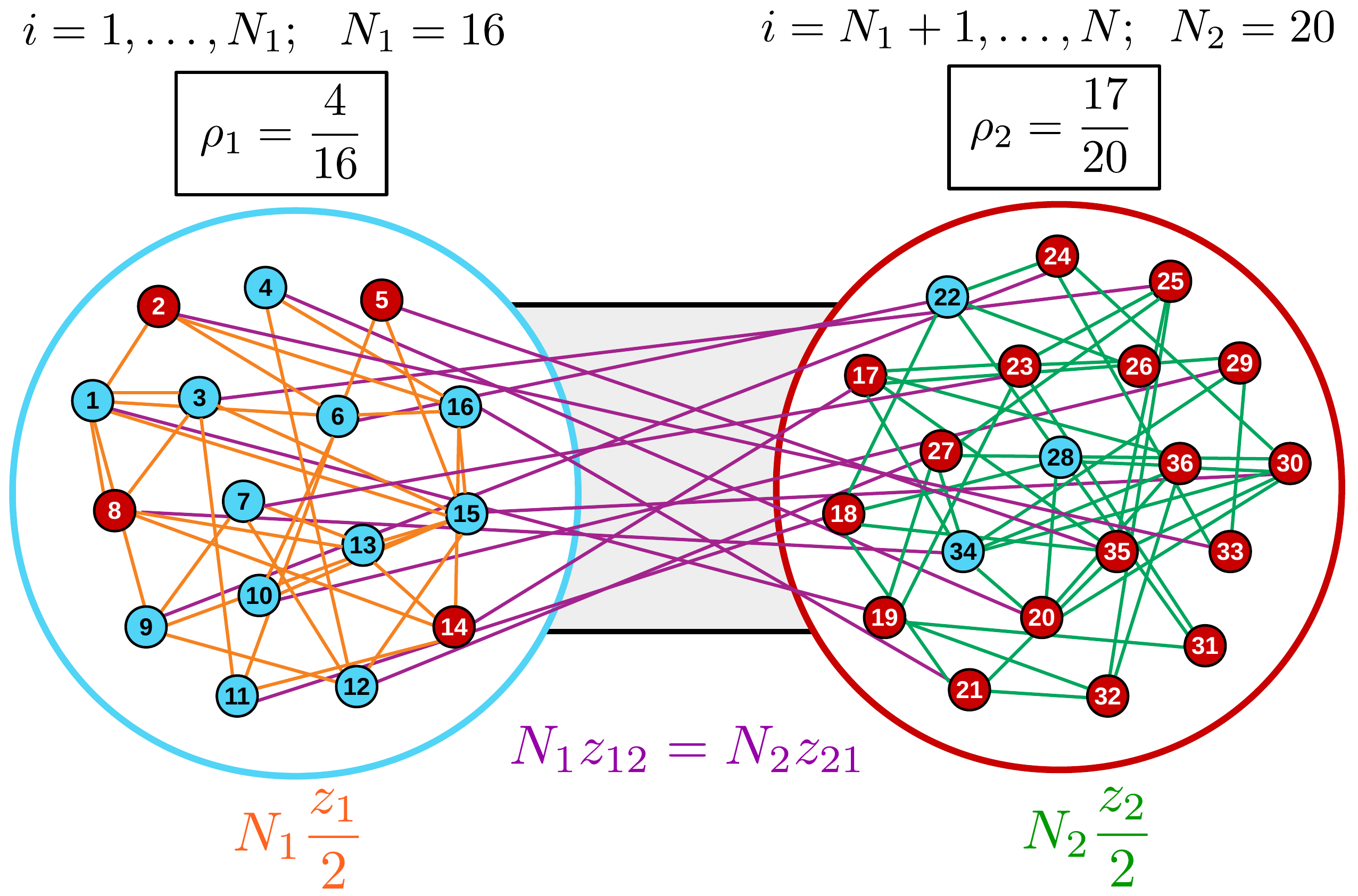}
\caption{Schematic representation of a simple network with modular (community) structure. Groups 1 and 2 have sizes $N_{1}$ and $N_{2}$, internal average degrees $z_{1}$ and $z_{2}$, are connected between them with average degrees $z_{12}$ and $z_{21}$, and have fractions of infected nodes $\rho_{1}$ and $\rho_{2}$, respectively.}
\label{fig_community_scheme}
\end{center}
\end{figure}

%Among all the possible structural properties of the social network that one might consider to be relevant for the dynamics, for the purpose of our study we will focus only on the degree distribution $P_{k}$ and the modular (community) structure of the adjacency matrix $A_{ij}$. 
%In the 
%description

Beyond the degree distribution $P_{k}$, we expect the modular structure of the network to have a major impact on opinion dynamics under the effect of algorithmic bias. In the simplest modular setting, we consider two communities (denoted $1$ and $2$) of sizes $N_{1}$ and $N_{2}$, respectively, with $N=N_{1}+N_{2}$, such that nodes $i=1, \dots, N_{1}$ are in group $1$, and nodes $i=N_{1}+1, \dots, N$ belong to group $2$. We define the internal average degree as the average number of links of a node with others in the same community, $z_{1} = N_{1}^{-1} \sum_{i=1}^{N_{1}} \sum_{j=1}^{N_{1}} A_{ij}$ and $z_{2} = N_{2}^{-1} \sum_{i=N_{1}+1}^{N} \sum_{j=N_{1}+1}^{N} A_{ij}$. The average degree of a node in group 1 with nodes in group 2 is $z_{12} = N_{1}^{-1} \sum_{i=1}^{N_{1}} \sum_{j=N_{1}+1}^{N} A_{ij}$, while the average degree of a node in group 2 with nodes in group 1 is $z_{21} = N_{2}^{-1} \sum_{i=N_{1}+1}^{N} \sum_{j=1}^{N_{1}} A_{ij}$. Since the network is undirected and the adjacency matrix is symmetric ($A_{ij}=A_{ji}$), we have the constraint $N_{1} z_{12} = N_{2} z_{21}$. Just as before, we characterize the state of the system by the fraction of infected nodes in each community, $\rho_{1}(t) = N_{1}^{-1} \sum_{i=1}^{N_{1}} s_{i}(t)$ and $\rho_{2}(t) = N_{2}^{-1} \sum_{i=N_{1}+1}^{N} s_{i}(t)$, with $\rho(t)=\dfrac{N_{1}}{N} \rho_{1}(t)+\dfrac{N_{2}}{N} \rho_{2}(t)$ (see Fig. \ref{fig_community_scheme} for a schematic of these definitions and a simple example).

%In order to obtain a closed set of differential equations for the dynamics of $\rho_{1}(t)$ and $\rho_{2}(t)$, we define the average rate of switching from $0$ to $1$:
%
%\begin{equation}
%\label{mean_field_rate}
%f[x] \equiv \sum_{k} P_{k} \sum_{m=0}^{k} F_{k,m} B_{k,m}(x),
%\end{equation}
%where $x$ is the probability to find a neighbor in state $1$. If we assume a highly connected network with $k \gg 1$, i.e. $P_{k}$ is peaked around a high degree value, then the binomial function $B_{k,m}(x)$ is also highly peaked around the value $m = k x$. As the rates Eqs.~(\ref{voter_rate}-\ref{language_rate}) only depend on the fraction $m/k$, this is simply $f[m/k] \approx F_{k,m}$. In a \emph{mean-field} type of approximation, we replace the local (node) probability $x$ of finding a neighbor in state $1$ by the global fraction of nodes in state $1$ in the network $\rho$. We will go a step further in this direction as we want to consider the community structure and also algorithmic bias. If we apply the highly connected assumption to the effective rate with bias Eq.~(\ref{efective_rate_def_infection}) we obtain $F^{*}_{k,m} \approx F_{k-b m, (1-b) m}$ and thus, the equivalent of Eq.~(\ref{mean_field_rate}) after applying algorithmic bias is $f^{*}[x] = f \left[(1-b)x/(1-bx) \right]$.

If, at some instant of time, the system is in a non-homogeneous state ($\rho_{1} \neq \rho_{2}$), the probabilities $x_1$ and $x_2$ of a node in group 1 or 2 finding an infected neighbor ($s=1$) are in principle different and calculated as
\begin{equation}
\label{prob_community}  
x_{1}=\frac{N_{1} z_{1} \rho_{1} + N_{2} z_{21} \rho_{2}}{N_{1} z_{1} + N_{2} z_{21}} = \frac{ \rho_{1} + p_{1} \rho_{2}}{1 + p_{1}},
\end{equation}
with $p_{1}=N_{2} z_{21}/N_{1} z_{1} = z_{12}/z_{1}$, and equivalently for $x_{2}$ by exchanging the index $1$ with $2$, i.e., $p_{2}=N_{1} z_{12}/N_{2} z_{2} = z_{21}/z_{2}$. Eq.~(\ref{prob_community}) is the ratio of the number of links coming out of infected nodes that end in community $1$ to the total number of links ending in community $1$. In the homogeneous case where $\rho=\rho_{1}=\rho_{2}$ (i.e. $p_{1}=p_{2}=1$), we recover $x=x_{1}=x_{2}=\rho$ as expected. We write the mean-field rate equations analogous to Eq.~(\ref{dyn_rho}) by considering the variables $\rho_{1}$ and $\rho_{2}$ and probabilities $x_{1}$ and $x_{2}$,
\begin{widetext}
\begin{align}
\label{dyn_rho1}
\frac{d\rho_{1}}{dt}&=(1-\rho_{1}) f \left[ \dfrac{(1-b)(\rho_1 + p_{1} \rho_2)}{1+p_{1}-b(\rho_1 + p_{1} \rho_2)} \right] - \rho_{1} f \left[ \dfrac{(1-b)(1-\rho_1 + p_{1} (1-\rho_2))}{1+p_{1}-b(1-\rho_1 + p_{1} (1-\rho_2))} \right],\\
\label{dyn_rho2}
\frac{d\rho_{2}}{dt}&=(1-\rho_{2}) f \left[ \dfrac{(1-b)(\rho_2 + p_{2} \rho_1)}{1+p_{2}-b(\rho_2 + p_{2} \rho_1)} \right] - \rho_{2} f \left[ \dfrac{(1-b)(1-\rho_2 + p_{2} (1-\rho_1))}{1+p_{2}-b(1-\rho_2 + p_{2} (1-\rho_1))} \right],
\end{align}
\end{widetext}
which reduce to Eq.~(\ref{dyn_rho}) in the homogeneous case ($\rho_{1}=\rho_{2}=\rho$). A more detailed derivation of Eqs.~(\ref{dyn_rho1})-(\ref{dyn_rho2}) can be found in the SM \cite{SM_ref}, Section S1.2.2. 

%{\color{blue} See Supplemental Material at \cite{S1.2.2} for a more detailed derivation of Eqs.~(\ref{dyn_rho1})-(\ref{dyn_rho2}).}

The solutions $\rho_{1}(t)$ and $\rho_{2}(t)$ of the coupled system in Eqs.~(\ref{dyn_rho1})-(\ref{dyn_rho2}) approximate the averages over realizations $\langle \rho_{1}(t) \rangle$ and $\langle \rho_{2}(t) \rangle$ obtained from numerical simulations (see Section~\ref{sec_simulation}). We expect this approximation to be accurate for highly connected networks and large system size $N$. As before, a large average degree is required by the mean-field assumption, and the thermodynamic limit to avoid finite-size effects. The study of Eqs.~(\ref{dyn_rho1})-(\ref{dyn_rho2}) will shed light on the dynamics of collective information spreading in the presence of algorithmic bias and the phenomenology of the models considered here.

The dynamics of $\rho_{1}(t)$ and $\rho_{2}(t)$ is completely determined by the parameters $(Q, \alpha, b)$ and initial conditions $\rho_{1}(0)$ and $\rho_{2}(0)$. In order to understand the macroscopic behavior of the models for all parameter values, we build a phase diagram, i.e. we divide the parameter space $(Q, \alpha, b)$ in regions associated with different stable fixed points $\rho_{1}(t)=\rho_{1}^{\text{st}}$ and $\rho_{2}(t)=\rho_{2}^{\text{st}}$. The role of initial conditions can be determined by a vector field or phase portrait, i.e. the right hand side of Eqs.~(\ref{dyn_rho1})-(\ref{dyn_rho2}) as a function of $\rho_{1}$ and $\rho_{2}$. The basin of attraction of a stable fixed point delimits the region of initial conditions that will be attracted to that point by the dynamics after long times. Before analyzing the phase diagrams and basins of attraction of all stable fixed points in Section~\ref{sec_theo_results}, we briefly discuss analytical approximations more accurate than the mean-field limit.

\subsection{Higher-order approximations}

Higher-order descriptions of binary-state dynamics are possible, at the expense of simplicity and tractability, by considering a set of deterministic evolution equations larger than Eq.~(\ref{dyn_rho}) or Eqs.~(\ref{dyn_rho1})-(\ref{dyn_rho2}), see Section S2 of the SM \cite{SM_ref}. %{\color{blue} see Supplemental Material at \cite{s2}.} 
In the highly-connected, infinite network size limit the results of all approximations coincide. Higher-order approximations involve evolution equations for random (uncorrelated) networks with an arbitrary degree distribution $P_{k}$, with degrees in the range $k_{\text{min}}\leq k \leq k_{\text{max}}$, and include:
%, evolving with the individual effective rates $F^{*}_{k,m}$ and $R^{*}_{k,m}$, Eqs.~(\ref{efective_rate_def_infection} \ref{efective_rate_def_recovery}).
(i) the pair approximation \cite{Vazquez:2008a,Peralta_pair:2018,Vieira:2020}, with $k_{\text{max}}-k_{\text{min}}+2$ variables, and (ii) approximate master equations \cite{Gleeson:2011,Gleeson:2013,Ruan:2015,Porter:2016,Unicomb:2018,unicomb2021dynamics,Peralta:2020}, with $(1+k_{\text{max}}-k_{\text{min}})(2+k_{\text{max}}+k_{\text{min}})$ variables. In the case of a homogeneous (single community) network structure, the methods developed in the references above can be directly applied to any binary-state model with effective transition rates $F^{*}_{k,m}$ and $R^{*}_{k,m}$ [see Eqs.~(\ref{efective_rate_def_infection})-(\ref{efective_rate_def_recovery})]. In the case of modular networks some modifications are needed, as described in the SM \cite{SM_ref}, Section~S2.2, where we develop a pair approximation scheme for modular $z-$regular networks. 

%{\color{blue} see Supplemental Material at \cite{s2.2} where we develop a pair approximation scheme for modular $z-$regular networks.}

%JK: Don't we want to say something? Pair appr. AME.}
%\subsection{Approximate master equation}

%highly connected

\section{Theoretical results}\label{sec_theo_results}

We now proceed with a detailed analysis of the fixed points (steady states) of the dynamical mean-field Eqs.~(\ref{dyn_rho1})-(\ref{dyn_rho2}) and their stability, in both the absence and presence of algorithmic bias. We first discuss the homogeneous solution $\rho = \rho_{1}=\rho_{2}$ (Section~\ref{sec_hom_sol} and Section S3.1 in the SM \cite{SM_ref}), 
%{\color{blue} (see Section~\ref{sec_hom_sol} and the Supplemental Material at \cite{s3.1})}
where we focus on the coexistence-consensus transition with order parameter $\vert \rho-1/2 \vert$, and then turn our attention to the polarized solutions $\rho_{1} \neq \rho_{2}$ (Section~\ref{sec_pol_sol} and Section S3.2 in the SM \cite{SM_ref})%{\color{blue} (see Section~\ref{sec_pol_sol} and the Supplemental Material at \cite{s3.2})}
, where we concentrate on a polarization transition with order parameter $P = \vert \rho_{1}-\rho_{2} \vert$.

%{\color{red}JK: Define the order parameter}

\subsection{Homogeneous solutions}\label{sec_hom_sol}

The homogeneous condition $\rho_{1}(t) = \rho_{2}(t) = \rho(t)$ can be satisfied by Eqs.~(\ref{dyn_rho1})-(\ref{dyn_rho2}). Among all possible homogeneous solutions of Eq.~(\ref{dyn_rho}), we highlight the state of opinion coexistence ($\rho=1/2$), which is always present independently of parameter values as a consequence of the up-down symmetry in the transition rates [see Eq.~(\ref{up-down})]. In order to understand the effect algorithmic bias has on these homogeneous solutions, we divide the following results in the cases $b=0$ and $b > 0$.

\subsubsection{No algorithmic bias ($b = 0$)}

%The results that we present in this section were obtained in previous works, see e.g. \citep{Peralta:2018,Peralta_pair:2018}, we will make a quick summary to understand the phenomenology, already mentioned in Table \ref{tab_models}, of the models without bias.

In the noisy voter model with $Q>0$, the only stable solution is coexistence ($\rho_{\text{st}}=1/2$) \cite{Peralta:2018,Peralta_pair:2018}. %For $Q=0$ any solution {\color{red}JK: ???} $0 \le \rho_{st} \le 1$ is possible with marginal stability 
%(the noiseless voter model 
%\citep{Clifford:1973,Holley:1975}. The stochastic formalism  \citep{Kirman:1993,Peralta_pair:2018} reveals a finite-size discontinuous transition between coexistence and consensus $\rho_{\text{st}}=\pm 1$ of opinions at $Q_{c}(N) = (2+N)^{-1}$. {\color{red}JK: Is this relevant?}
%In the thermodynamic limit $N \rightarrow \infty$ the transition disappears ($Q_{c}=0$) and only the coexistence state remains stable. 
In the majority-vote model, however, there is a well-defined continuous coexistence-consensus transition (supercritical pitchfork bifurcation) for a finite critical value of the noise $Q_{c}>0$: For $Q>Q_{c}$ the coexistence state $\rho_{\text{st}}=1/2$ is the only stable solution, while for $Q<Q_{c}$ the coexistence state loses its stability and two symmetry-breaking, imperfect consensus states appear as stable solutions, with $|\rho_{\text{st}}-1/2| \propto (Q_{c}-Q)^{1/2}$.
    
\begin{figure}[ht!]
\begin{center}
\includegraphics[width=0.4\textwidth]{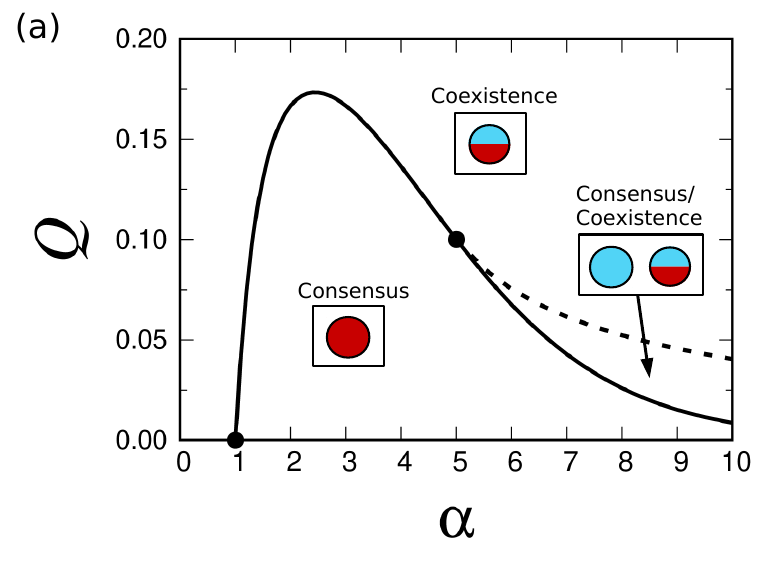}
\includegraphics[width=0.4\textwidth]{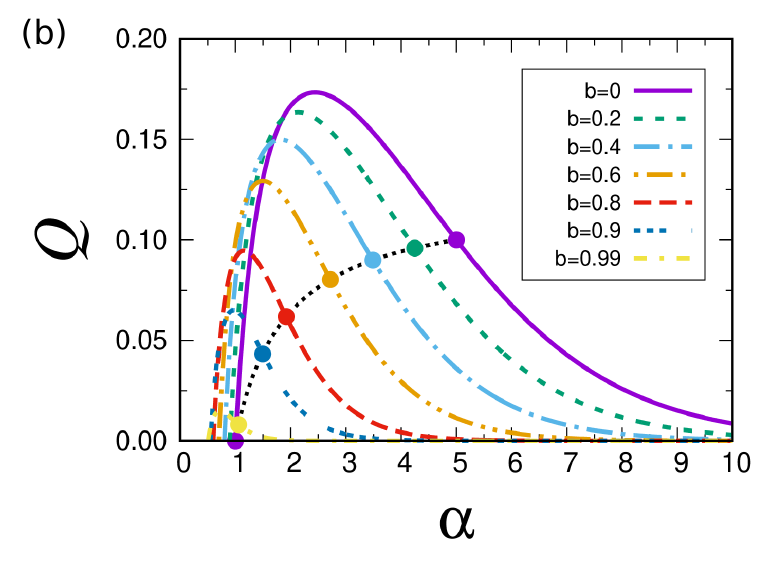}
\caption{Phase diagram of homogeneous solutions for the language model in parameter space $(Q, \alpha)$, both without bias  ($b=0$; a) and for increasing bias in the range $b \in [0, 1)$ (b). The solid line in (a) and the different dashed lines in (b) denote the transition values $Q_{c}(\alpha,b)$, while the dashed line in (a) corresponds to $Q_{t}(\alpha,b)$. The tricritical points $\alpha_{t}(0)=5$ and $\alpha_{t}(b)$ [dot in (a) and dotted line in (b), respectively] separate the continuous transition [$1 < \alpha < \alpha_{t}(b)$] from the discontinuous [$\alpha > \alpha_{t}(b)$].  Symbols inside squares indicate the global stable states that can be found inside the parameter region delimited by the transition lines. Circle colors (shading in grayscale) correspond to opinion states ($s = 0,1$), while a single circle represents a homogeneous system. There are two possible global states: (i) coexistence (mixed-color circles) and (ii) consensus (single-color circles). When more than one stable state is present in a parameter region, the initial condition determines the final state. Note that symmetric states (obtained by exchanging circle colors) are always possible in the same parameter region. The method to obtain the transition parameter values $Q_{c}, Q_{t}$ is discussed in the SM \cite{SM_ref}, Section S3.1. 
%{\color{blue} See Supplemental Material at \cite{s3.1} for a detailed discussion of the method to obtain the transition parameter values $Q_{c}, Q_{t}$.}
}
\label{fig_homogeneous_nobias}
\end{center}
\end{figure}

In the language model, for $\alpha \le 1$ there is no transition and the only stable state is coexistence (like in the noisy voter model). For $1<\alpha<5$ there is a continuous coexistence-consensus transition (supercritical pitchfork bifurcation, majority-vote like), while for $\alpha > 5$ the transition is discontinuous (subcritical pitchfork bifurcation). These regimes are separated by a tricritical point at $\alpha=5$. For the discontinuous case there are two transition lines at $Q = Q_{c}, Q_{t}$. For $Q>Q_{t}$ the coexistence state is the only stable solution. For $Q_{c}<Q<Q_{t}$ both coexistence and consensus are stable solutions (with a stationary state depending on initial conditions), while for $Q<Q_{c}$ only consensus is possible. In Fig. \ref{fig_homogeneous_nobias} we show the transition lines $Q = Q_{c}, Q_{t}$ as a function of $\alpha$, as well as the stable solutions in the parameter regions $(Q, \alpha)$ delimited by these transition lines (see Refs.~\cite{Peralta:2018,Peralta_pair:2018} for more details on the case without bias, and Table \ref{tab_models} for a summary of the phenomenology).

%{\color{red}{JK: I suggest introducing lettering some points in the upper figure (A, T, B, C) so that it becomes easy to explain where us the tricritical points, and what are $Q_{c}(\alpha,b)$, $Q_{t}(\alpha,b)$}.}

\subsubsection{Algorithmic bias ($b > 0$)}

\begin{figure}[ht!]
\begin{center}

\includegraphics[width=0.4\textwidth]{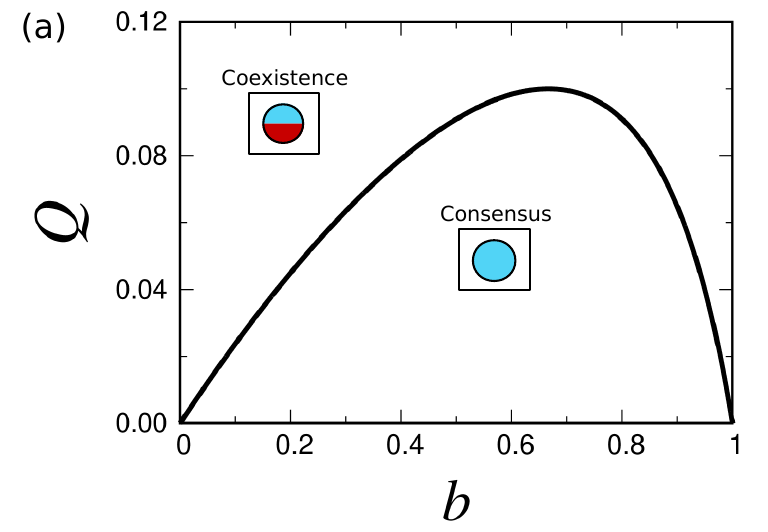}
\includegraphics[width=0.4\textwidth]{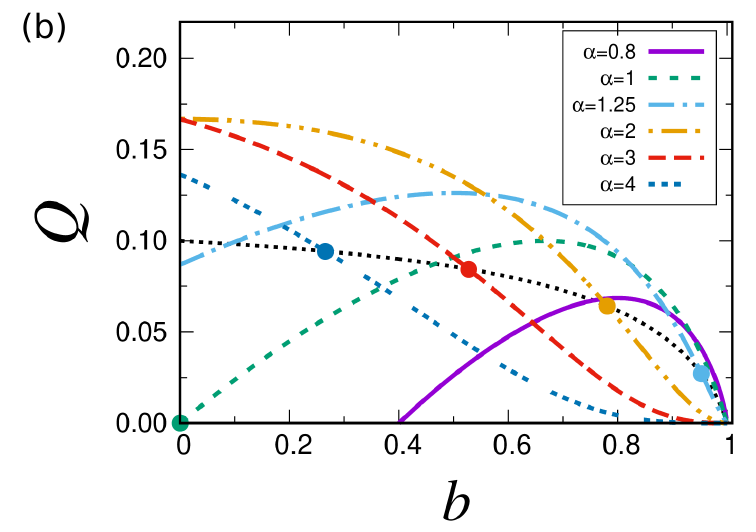}
\includegraphics[width=0.4\textwidth]{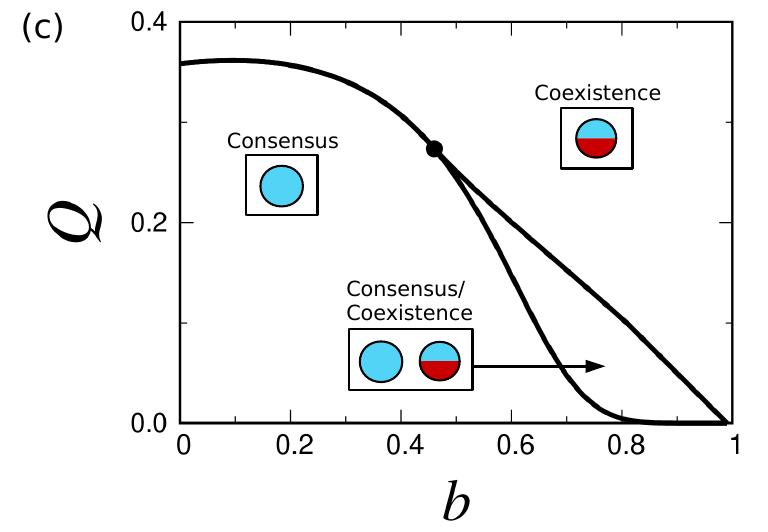}
\caption{Phase diagrams in parameter space $(Q, b)$ of the homogeneous solutions for the noisy voter (a), language (b), and majority-vote (c) models under algorithmic bias ($b > 0$). For the noisy voter and language models, the average rates used to obtain the transition lines are calculated via Eq.~(\ref{mean_field_rate}) in the highly connected limit ($z \rightarrow \infty$), while for the majority-vote model we use a $z-$regular network ($P_{k}=\delta_{k,z}$ with average degree $z=20$). Terminology and symbols are the same as in Fig. \ref{fig_homogeneous_nobias}.}
\label{fig_homogeneous_bias}
\end{center}
\end{figure}

The presence of algorithmic bias (coded by a nonzero intensity $b$) shifts the transition lines and changes their nature. In Fig.~\ref{fig_homogeneous_bias} we plot the phase diagrams of the three models in Table \ref{tab_models} as a function of $b$.
%The effect of bias depends on the model and it can be summarized as follows.
For the noisy voter model, there is always a 
%the finite-size transition of the model without bias turns into 
well defined continuous coexistence-consensus transition for $b>0$. 
The critical value $Q_{c}$ increases as a function of $b$ up to a maximum at $b=2/3$, after which it decreases to $Q_{c}=0$ for $b=1$. This means that for $b<2/3$, algorithmic bias promotes consensus of opinions in the noisy voter model.
    
In the majority-vote model, surprisingly, the opposite effect takes place, i.e. the critical point $Q_{c}$ decreases as a function of $b$. In other words, algorithmic bias promotes coexistence of opinions. For high values of $b$ the transition becomes discontinuous with the appearance of a tricritical point, similarly to the phenomenology of the language model without bias for $\alpha>5$ (see Fig. \ref{fig_homogeneous_bias}). The majority-vote model with high bias thus behaves similarly to the language model with large $\alpha > 5$.
    
The language model, depending on whether $\alpha$ is small or large, interpolates between the two behaviors described above, i.e. algorithmic bias promotes consensus (low $\alpha$) or opinion coexistence (high $\alpha$). With respect to the effect of bias, we can distinguish between {\it voter like} (low $\alpha$) and {\it majority-vote like} (high $\alpha \leq 5$) behavior. For $\alpha > 5$ the transition is always discontinuous with and without bias (see SM \cite{SM_ref}, Section~S3.3 and Fig. S2).
%{\color{blue} (see Supplemental Material at \cite{S3.3} and the corresponding Fig. S2)}. 
We also observe that for high enough bias the transition becomes discontinuous for $1 < \alpha < 5$, so bias also favors the discontinuity of the transition. 

%As shown in this section, algorithmic bias has a non-trivial effect on the consensus-coexistence transition of opinion models. Depending on the specific rules of the copying mechanism, and particularly if it is pairwise or group based, bias will favor consensus or coexistence of opinion respectively. {\color{red}JK: This paragraph could go into the summary.}

\subsection{Polarized solutions}\label{sec_pol_sol}

\begin{figure*}[ht!]
\begin{center}

\includegraphics[width=0.4\textwidth]{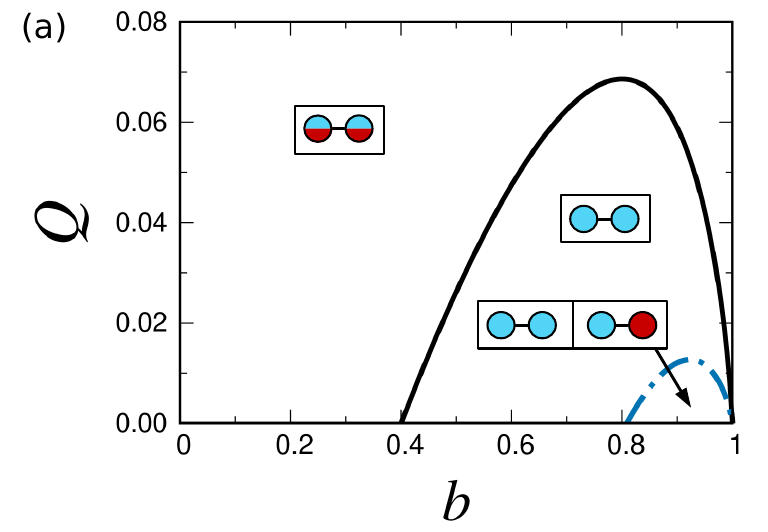}
\includegraphics[width=0.4\textwidth]{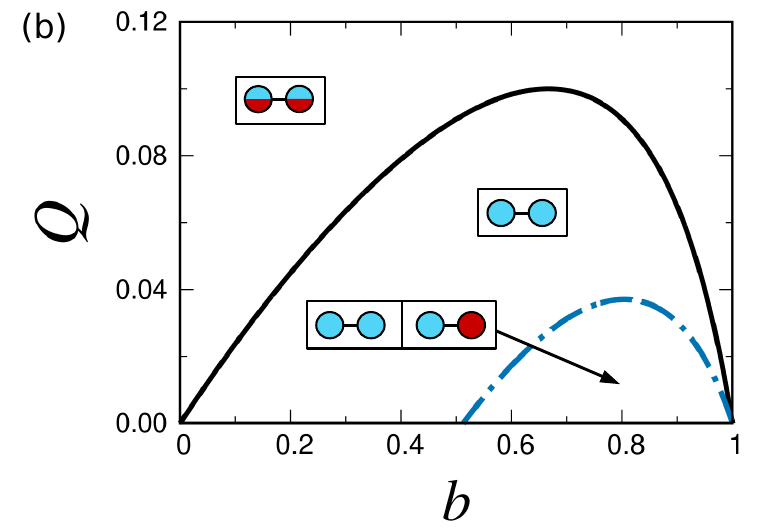}

\includegraphics[width=0.4\textwidth]{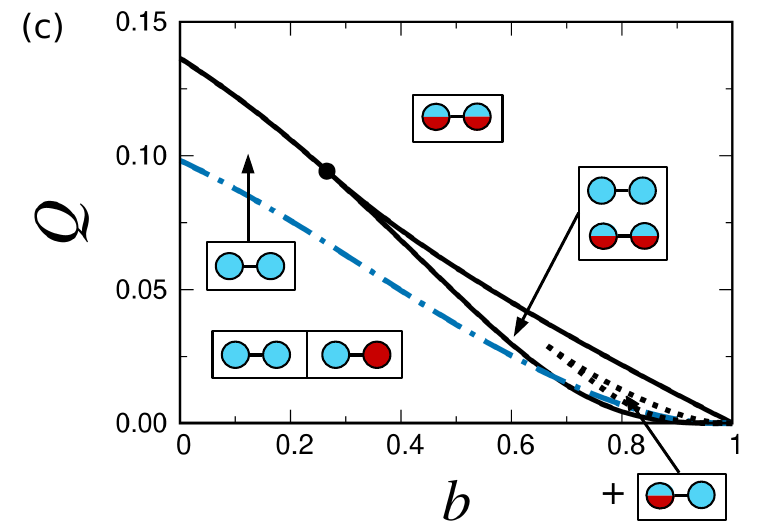}
\includegraphics[width=0.4\textwidth]{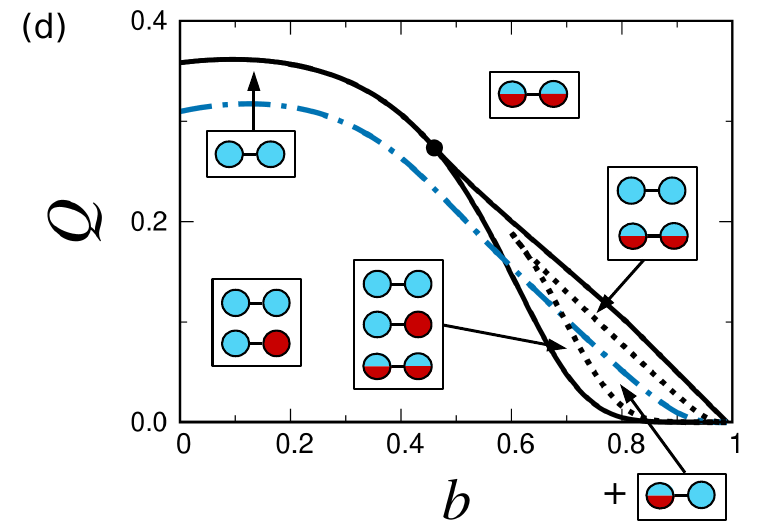}

\caption{Phase diagrams, in noise-bias parameter space $(Q, b)$, of homogeneous and polarized solutions for the language model with low $\alpha=0.8$ (a), noisy voter model (b), language model with high $\alpha=4$ (c), and majority-vote model (d) in the presence of algorithmic bias and modular network structure (with fixed $p=0.1$), where the order implies a gradual move from pairwise to group interactions. We observe two polarized states: (i) standard polarization (circles of different color; delimited by a dash-dotted line), and (ii) partial polarization (circles with mixed and full colors, delimited by a black dotted line). Symmetric states (obtained by exchanging colors) are always possible in the same parameter region. The stationary state ($t \rightarrow \infty$) in a parameter region with several possible stable states is determined by the initial condition (see Fig. \ref{fig_pola_scheme}). For the noisy voter and language models, average rates used to obtain the transition lines are calculated via Eq.~(\ref{mean_field_rate}) in the highly connected limit ($z \rightarrow \infty$), while for the majority-vote model we use a $z-$regular network ($P_{k}=\delta_{k,z}$ with average degree $z=20$). The method to obtain the transition parameter values $Q^{*}_{p}$ and $Q_{p}$ is described in the SM \cite{SM_ref}, Section S3.2. 
%{\color{blue} See Supplemental Material at \cite{s3.2} for a detailed discussion of the method to obtain the transition parameter values $Q^{*}_{p}$ and $Q_{p}$.}
} 
\label{fig_heterogeneous_bias}
\end{center}
\end{figure*}

\begin{figure*}[ht!]
\begin{center}
\includegraphics[width=0.9\textwidth]{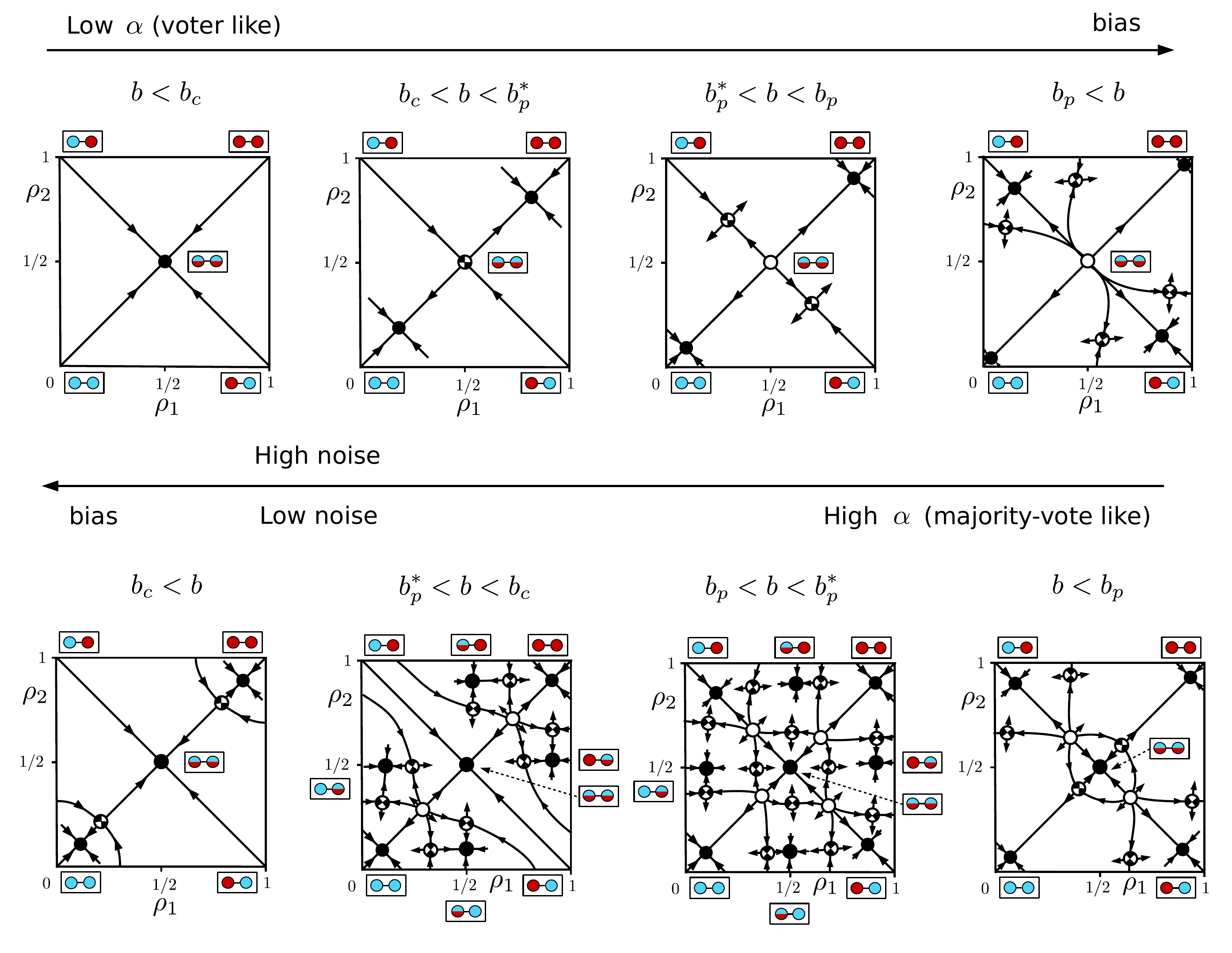}
\caption{Schematic representation of the various fixed points, linear stability and shape of associated vector fields for Eqs.~(\ref{dyn_rho1})-(\ref{dyn_rho2}), a mean-field approximation of binary-state dynamics of information spreading under algorithmic bias. We show all possible phases included in Fig. \ref{fig_heterogeneous_bias}, together with phase transitions that occur for fixed noise $Q$ and varying bias intensity $b$ (examples of the associated vectors fields for given values of the parameters can be found in SM \cite{SM_ref}, Section S3.4)%{\color{blue} (see Supplemental Material at \cite{s3.4} for examples of the associated vectors fields for given values of the parameters)}
.}
\label{fig_pola_scheme}
\end{center}
\end{figure*}

In order to understand the interplay between algorithmic bias and modular network structure in binary dynamics of information spreading, we relax the homogeneous condition and allow $\rho_{1}$ and $\rho_{2}$ to vary freely. Solutions that do not fulfill the condition $\rho_{1} = \rho_{2}$ can be considered as {\it polarized}, since groups have different average opinions. Extreme polarization happens for $\rho_{1}=1$ and $\rho_{2}=0$ (or the other way around), i.e. full consensus in community $1$ and full consensus of the opposite opinion in community $2$. We measure the degree of polarization in the social system with the order parameter $P = \vert\rho_{1}-\rho_{2} \vert \in [0, 1]$, such that the homogeneous case ($\rho_{1}=\rho_{2}$) corresponds to $P=0$, and extreme polarization to $P=1$. Any other value ($0 < P < 1$) represents polarization to a certain degree.

Assuming communities of equal size ($N_{1}=N_{2}=N/2$) and connectivity ($z_{1} = z_{2}$, $p=p_{1}=p_{2}=z_{12}/z_{1}=z_{21}/z_{2}$), it is straightforward to show that a possible solution of Eqs.~(\ref{dyn_rho1})-(\ref{dyn_rho2}) fulfills the condition $\rho_{1}(t)+\rho_{2}(t)=1$, which we call the \emph{polarization line}. We are mainly interested in stationary solutions across this line, but there are polarized solutions outside of it. Note that even if we find a fixed point in the polarization line, stability analysis needs to be performed in all directions, not only across the line.

In all considered models, for particular values of the noise $Q$, bias intensity $b$, and connectivity parameter $p$, we find a transition to polarization. If we vary the noise $Q$ and keep other parameters fixed, there are two critical values: $Q_{p}$ and $Q^{*}_{p}$ with $Q_{p} < Q^{*}_{p}$. For $Q>Q^{*}_{p}$, there are no fixed points along the polarization line besides the trivial coexistence state $\rho_{1}^{\text{st}}=\rho_{2}^{\text{st}}=1/2$. For $Q_{p}<Q<Q^{*}_{p}$ two polarized fixed points appear, stable along the polarization line and unstable in the perpendicular direction (the {\it homogeneous line}), meaning they are saddle points in $(\rho_{1},\rho_{2})$ space. For $Q<Q_{p}$ the same fixed points become stable in both directions, representing a polarized state of opinion. 

The polarization transition line $Q_{p}(\alpha,b,p)$, or equivalently $b_{p}(Q,\alpha,p)$, is displayed in Fig. \ref{fig_heterogeneous_bias} for all models. In Fig. \ref{fig_pola_scheme} we show a schematic representation of the fixed points and their stability analysis, i.e. the phase portrait, of all phases in Fig. \ref{fig_heterogeneous_bias}. We summarize the behavior of the polarization transition in each model as follows.

In the noisy voter model, the polarization transition appears for a fixed value of the bias intensity $b>b_{p}(0,1,p)$, while for $b<b_{p}(0,1,p)$ the polarized state disappears. This means that polarization emerges as a consequence of algorithmic bias. The noise value at the transition $Q_{p}(1,b,p)$ increases as a function of $b$ and has a maximum, similarly to the homogeneous coexistence-consensus transition line. Overall, polarization is induced and promoted by algorithmic bias in the presence of pairwise interactions.
    
In the majority-vote model, polarization already exists without bias ($b=0$). As bias increases, the transition value $Q_{p}$ changes slightly (the transition line is very flat as a function of $b$) and may have a small maximum depending on $p$. For high enough bias $Q_{p}(b,p)$ decreases, meaning that bias may inhibit polarization, as opposed to the noisy voter model. For large $b$ a new stable polarized state appears, where one of the groups has consensus and the other one coexistence. This new state is possible because consensus and coexistence are both stable in the homogeneous phase. In Fig. \ref{fig_heterogeneous_bias}, inside the consensus/coexistence region of the homogeneous phase, we observe a smaller region where the new state is stable. In Fig. \ref{fig_pola_scheme} we show how these new states appear, together with the stability analysis of all additional fixed points outside of the polarization line.
    
The behavior of the language model depends on the value of $\alpha$, as before. For low values, we find similar behavior as the noisy voter model with an emergence of polarization for increasing bias. High values of $\alpha$ display the opposite behavior, with polarization disappearing for large enough $b$, similarly to the majority-vote model. Note that the previous statements and the results of Fig. \ref{fig_heterogeneous_bias} correspond to a fixed value of the connectivity parameter $p$. The polarization transition lines strongly depend on the value of $p$. For example, the value of $\alpha$ that separates (smoothly) voter like and majority-voter like behaviors increases with $p$. If the value of $p$ is high enough, but a polarized solution still exists despite the high degree of mixing between groups, the polarization transition has a maximum as function of $b$, even for large $\alpha$ (for more details on the separation between $\alpha$ regimes see SM \cite{SM_ref}, Section S3.3 and Fig. S2).
%{\color{blue} (see Supplemental Material at \cite{S3.3} for more details on the separation between $\alpha$ regimes)}

\section{Comparison with numerical simulations}\label{sec_com_num}

In Section~\ref{sec_theo_results} we have derived analytical approximations for binary-state dynamics with algorithmic bias (for both homogeneous and modular networks) in terms of mean-field equations, which allow us to characterize the behavior of specific opinion formation models as a function of parameters $(Q,\alpha,b,p)$. In what follows we validate these theoretical results with numerical simulations in synthetic and real networks.

\subsection{Synthetic networks}\label{sec_synthetic}

Synthetic networks with heterogeneous, uncorrelated degrees and modular structure can be generated by using the configuration model \cite{Catanzaro:2005}, where we take as input degree sequences and the community each node belongs to
%the only relevant structural property is the degree distribution, but with the additional ingredient of modular structure, 
(see, e.g., Fig. \ref{fig_community_scheme}). We use the generated networks for numerical simulations and compute the corresponding pair approximation  
%As a matter of comparison we will take the theoretical results coming from the mean field Eqs.~(\ref{dyn_rho1}, \ref{dyn_rho2}) in the highly connected limit, and additionally the pair approximation results, i.e. 
(a more accurate analytical description developed in SM \cite{SM_ref}, Section~S2).
%{\color{blue} (see Supplemental Material at \cite{s2} for the development of the pair approximation, which corresponds to a more accurate analytical description than the mean field)}. 
In the case of homogeneous networks, we also solve the associated approximate master equations \cite{Gleeson:2013} numerically. We then compare these results with the mean-field Eqs.~(\ref{dyn_rho1})-(\ref{dyn_rho2}) to test the validity of the highly connected limit.

Comparing the pair approximation with the mean-field results (see Fig. \ref{fig_heterogeneous_bias}), we find the same qualitative behavior in networks with finite average degrees $z_{1}$, $z_{12}$, $z_{2}$, and $z_{21}$, with the same type of transitions. The transition lines shift, however, and the critical values depend on the connectivity parameters. In the pair approximation, the critical noise at the transition is smaller than in the mean-field case, and the network tends to destroy the discontinuous transition, reducing the parameter region where it can be found. For networks with $z \leq 5$, the discontinuous transition disappears completely. Something similar happens with the polarization transition; a network with finite connectivity reduces the parameter region where the polarized state can be found, as compared to the mean-field limit with the same $p$ value. For extremely sparse networks the polarized state disappears (see SM \cite{SM_ref}, Section~S4 and Fig. S9 in the case $z_{12}=1$ and $z_{1}=5$, $p=0.2$). In the SM we show how the network affects the transition values $Q_{c}$ and $Q_{p}$ for given model parameters and as a function of the connectivity $z_{1}$.
%{\color{blue}, see Supplemental Material at \cite{s4} and the corresponding Fig. S9 in the case $z_{12}=1$ and $z_{1}=5$, $p=0.2$. In the same section of the Supplemental Material we show how the network affects the transition values $Q_{c}$ and $Q_{p}$ for given model parameters and as a function of the connectivity $z_{1}$.}
Note that we vary the value of $z_{1}$ but keep $p=z_{12}/z_{1}$ ($N_{1}=N_{2}$) constant so that the mean-field results are independent of $z_{1}$. As shown in the SM \cite{SM_ref}, Section~S4, Fig. S9 and Fig. S11, the pair approximation is very accurate for the language model with $\alpha=2$ as compared to numerical simulations on $z-$regular networks with community structure, and we recover the mean-field limit for $z_{1} \rightarrow \infty$. In this case, the transition parameter values predicted by the pair approximation fall within error bars of computer simulations, thus validating our analytical results.
%{\color{blue} See Supplemental Material at \cite{s4}, Fig. S9 and Fig. S11, where we show that the pair approximation is very accurate for the language model with $\alpha=2$ as compared to numerical simulations on $z-$regular networks with community structure, and we recover the mean-field limit for $z_{1} \rightarrow \infty$. In this case, the transition parameter values predicted by the pair approximation fall within error bars of computer simulations, thus validating our analytical results.}

While the accuracy of the pair approximation is remarkable in most cases of interest, some discrepancies appear for high $\alpha \gtrsim z$, for extremely sparse networks ($z \approx 2$), and for the majority-vote model with large bias (see SM \cite{SM_ref}, Fig. S10 and Fig. S12).
%{\color{blue} (see Supplemental Material at \cite{s4}, Fig. S10 and Fig. S12)}.
These discrepancies are corrected by the approximate master equations, which predict the transition values within the error range of numerical simulations for all parameter values and models considered.

%{\color{red}JK: Can we make a general statement about the accuracy of the pair approximation?}

%\begin{figure}[ht!]
%\begin{center}

%\includegraphics[width=0.45\textwidth]{Phase_sim1.eps}
%\caption{Transition values $Q_{c}$ (black) and $Q_{p}$ (purple and green) as a function of the average degree $z_{1}$ in community $1$, for fixed parameter values $\alpha=2$, $b=0.5$, on degree regular symmetric communities $p_{1}=p_{2}=p$ with total average degree $z=(1+p)z_{1}$. The solid horizontal lines correspond to the mean field solution in the highly connected limit, independent of $z_{1}$. The black circle, purple triangle and green square dots correspond to the pair approximation results with fixed $p=0.1$ (purple) and $p=0.2$ (green and black). The red circle, triangle and square dots are the results obtained from numerical simulations of the model with the equivalent parameter values.}
%\label{fig_finite_network}
%\end{center}
%\end{figure}

\subsection{Real-world social networks}\label{sec_real}

Apart from heterogeneous degrees and a potentially stylized modular structure, real-world networks display higher-order (e.g., degree-degree) correlations and other mesoscopic properties not considered in the mean-field limit. We check the validity of our theoretical results by performing numerical simulations of the dynamics (see Section~\ref{sec_simulation}) on top of an empirical network structure, confirming that the effect of algorithmic bias on the transitions between coexistence, consensus and polarization is qualitatively the same even in the presence of more involved structural features.

We take the political blogs network \cite{Adamic:2005,asikainen2020cumulative}, where nodes represent liberal and conservative blogs around the time of the 2004 US presidential election. Links exist between blogs that refer to each other often, implying a strong interaction between blogs, i.e. the potential for information spreading. Based on existing metadata, the population can be divided in two groups of sizes $N_{1}=450$ (liberals) and $N_{2}=523$ (conservatives), with $N=N_{1}+N_{2}=973$ (after removing nodes with less than three links). Average degrees inside groups are $z_{1} = 31.74$ and $z_{2} = 29.37$, and between groups $z_{12} = 3.42$ and $z_{21}=2.94$. Thus, the conservative community is larger but less connected than the liberal group. Since average degrees are relatively large, we expect the mean-field Eqs.~(\ref{dyn_rho1})-(\ref{dyn_rho2}) (with connectivity parameters $p_{1}=0.1076$ and $p_{2}=0.1001$) to provide accurate results for the dynamics and stationary states of the social system. Note that the values of parameters $p_{1}$ and $p_{2}$ indicate that an individual in group $1$ (liberal) is (slightly) more likely to be convinced by an individual in group $2$ (conservative) than the other way around, suggesting that opinion dynamics might be driven by the conservative group.

We perform numerical simulations of the stochastic opinion dynamics using the noisy voter model with noise value $Q=0.01$ and a high bias intensity ($b=0.8$), which is close to the polarization maximum (see Fig. \ref{fig_heterogeneous_bias}). According to the mean-field theory there are two (symmetric and stable) polarized states with $\rho_{1}^{\text{st}} = 0.09$, $\rho_{2}^{\text{st}} = 0.92$, and global opinion $\rho_{\text{st}}=0.54$ (or $\rho_{1}^{\text{st}} = 0.91$, $\rho_{2}^{\text{st}} = 0.08$, and global opinion $\rho_{\text{st}}=0.46$). Note that opinion consensus in group $1$ is always weaker than in group $2$, a consequence of the asymmetry in group size and connectivity. Linear stability analysis determines that the polarized state is stable with eigenvalues and eigenvectors $\lambda_{1} = -0.16$, $\vec{v}_{1} = \begin{bmatrix} 0.86 \\ 0.51 \end{bmatrix}$, and $\lambda_{2} = -0.24$, $\vec{v}_{2} = \begin{bmatrix} -0.55 \\ 0.84 \end{bmatrix}$. There is a slow and a fast eigendirection, which means that the approach to the polarized state happens first in group $2$ ($\rho_{2}(t) \rightarrow \rho_{2}^{\text{st}}$), and afterwards in group $1$ ($\rho_{1}(t) \rightarrow \rho_{1}^{\text{st}}$). This also implies that group 1 is less resilient and more vulnerable to perturbations, dynamical fluctuations in the slow eigendirection will have a larger amplitude, and thus opinions will vary more in time.

\begin{figure*}[ht!]
\begin{center}

\includegraphics[width=0.55\textwidth]{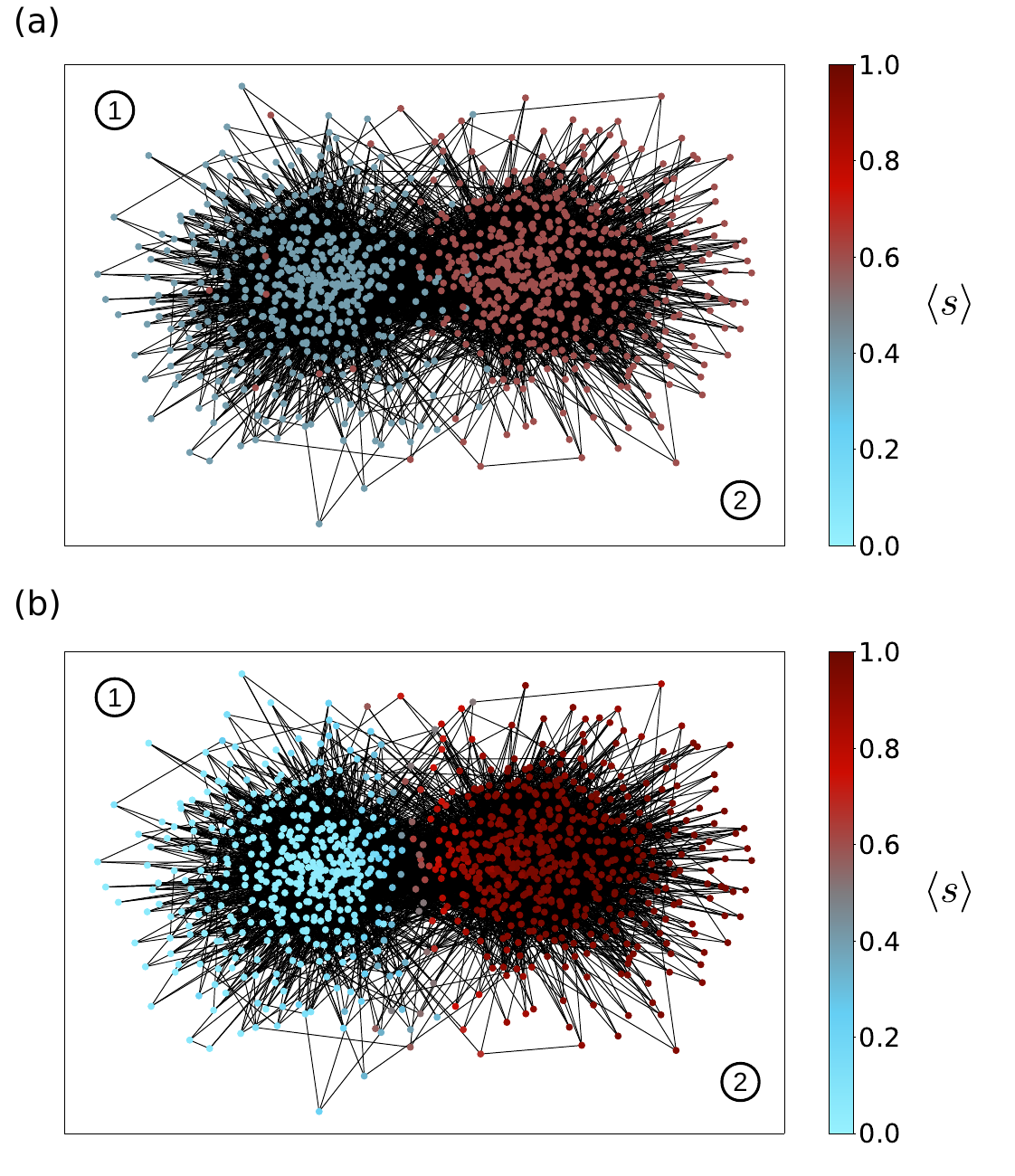}
\includegraphics[width=0.41\textwidth]{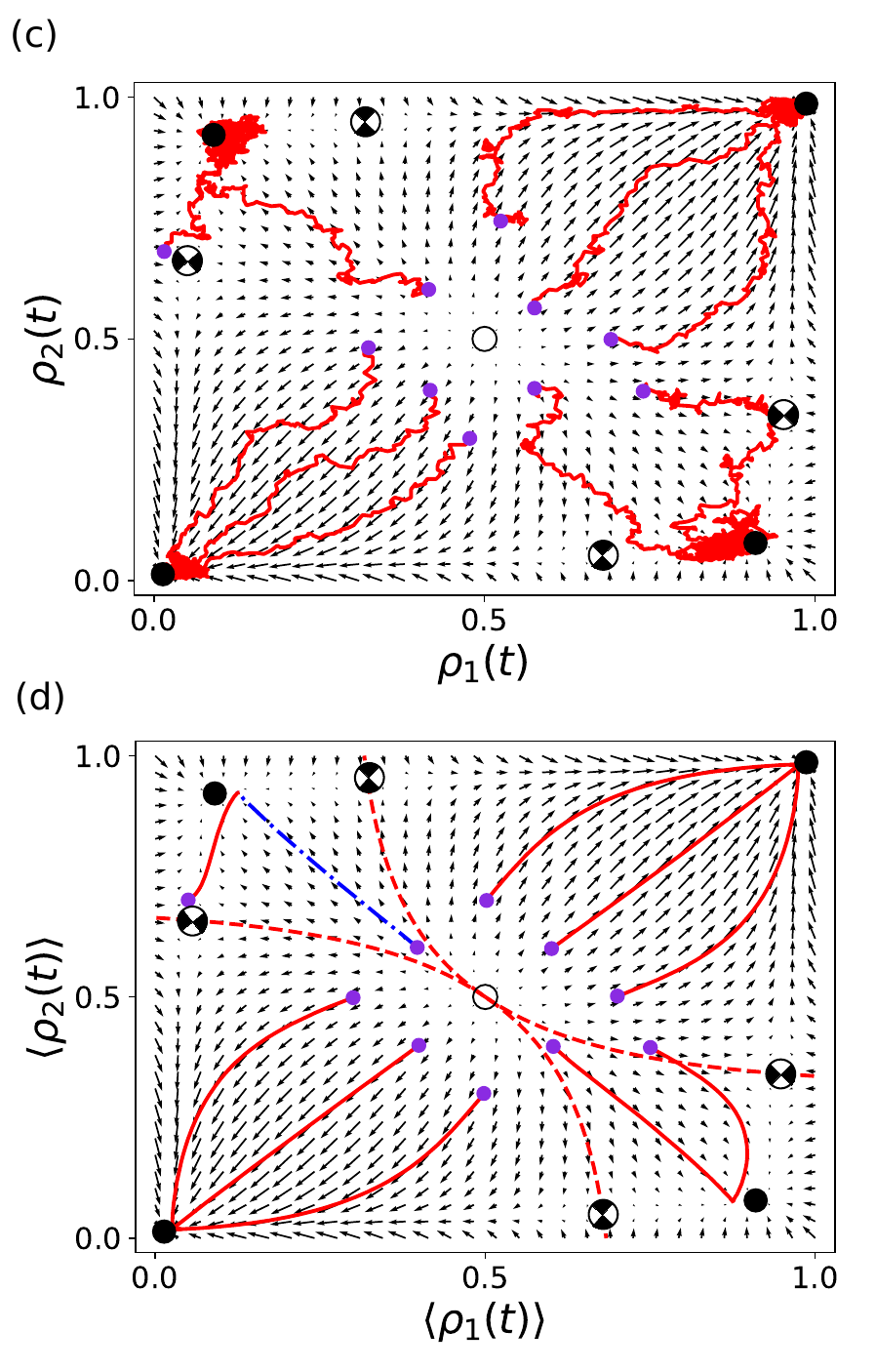}

\caption{Mean-field vector fields Eqs.~(\ref{dyn_rho1})-(\ref{dyn_rho2}) together with single trajectories (c) and average values over realizations (d) (solid lines) coming from numerical simulations of the noisy voter model in the political blogs network \cite{Adamic:2005,asikainen2020cumulative} with noise $Q=0.01$ and bias intensity $b=0.8$. Smaller dots are initial conditions, larger filled dots stable fixed points, empty dots unstable fixed points, and half-filled dots saddle points. Dashed lines in (d) delimit the basin of attraction of the polarization and consensus states. The dash-dotted line in (d) is a particular trajectory for which we show the average initial (a) and final (b) states of the simulated opinion dynamics over the empirical network. Node positions are calculated with the Louvain method of community detection.}
\label{fig_polblogs}
\end{center}
\end{figure*}

In Fig. \ref{fig_polblogs} we compare the vector field of the mean-field theory [coming from Eqs.~(\ref{dyn_rho1})-(\ref{dyn_rho2})] to single trajectories $\rho_{1}(t)$, $\rho_{2}(t)$ of numerical simulations with several initial conditions, as well as to restricted averages over realizations that end up in the same final state $\langle \rho_{1}(t) \rangle$, $\langle \rho_{2}(t) \rangle$. Numerical average values agree considerably well with the vector field of the theory, and also converge to the predicted final state, indicating that in the case of the political blogs network, a simple mean-field description is sufficient to describe the dynamical and stationary properties of binary-state information spreading under the effect of bias.

\section{Discussion}\label{sec_discussion}

\noindent
%- Importance of the problem\\
%- Usefulness of the approach (models and approximations)\\
%- Bias introduces a rich phenomenology\\
%- Importance of the character of the interaction (pair or group)\\
%- Here we focus on asymptotic stationarity but the approach is able to describe dynamics (only indicated here in the form of flows). E.g., it is possible to study the effect of a (small or major) perturbation\\
%- Future steps: What are the consequences of imbalances in the parameters? How does the phases and the transitions between them influence the relaxation dynamics? How to relate the parameters (e.g., $b$) to empirically observable quantities?

Algorithmic bias, an unexpected consequence of the content filtering tools behind most popular social media platforms used today, affects the dynamics of opinion formation and information spreading arising from digital interactions in non-trivial ways, ultimately leading to undesired collective phenomena like group polarization and opinion radicalization \cite{baumann2020modeling,cinelli2021echo}. While the role of algorithmic bias has been recently modeled in an array of concrete scenarios \cite{ciampaglia2018algorithmic,Sirbu:2019,Perra:2019}, a general theoretical framework systematically linking social dynamics, network structure and algorithmic filtering has been missing so far. Here we have put forward such a formalism by extending previous work on binary-state dynamics \cite{Gleeson:2011,Gleeson:2013,Ruan:2015,Porter:2016,Peralta:2018,Unicomb:2018,unicomb2021dynamics} with a notion of bias and applying it to synthetic and real-world social networks. While our formalism applies to any binary-state dynamics, we have showcased its flexibility by focusing on the noisy voter, language, and majority-vote models, which consider a (pairwise or group-based) copying or herding mechanism, alongside random or idiosyncratic changes of opinion.

We derived a set of deterministic, non-linear rate equations describing the macroscopic dynamics of the system with increasing levels of accuracy: mean-field, pair approximation, and approximate master equations. From these equations we determined the possible stationary states of the dynamics and their stability.
%as a function of the model parameters, particularly the bias intensity $b$. 
We observed a rich phenomenology with respect to various parameters (bias intensity $b$, noise $Q$, model tuning parameter $\alpha$, and inter-group connectivity $p$), including continuous and discontinuous phase transitions between stationary states of consensus, coexistence and polarization.

Algorithmic bias %(coded by the intensity $b$) 
plays a crucial role in the shape of the associated phase diagram. It promotes consensus (in homogeneous networks) and polarization of opinions (in modular networks) when pairwise interactions are predominant (the noisy-voter and language models with low $\alpha$). When the dynamics is driven instead by group-based interactions (majority-vote and language models with high $\alpha$), algorithmic bias promotes coexistence of opinions. The role of algorithmic bias on the coexistence-consensus transition has an intuitive interpretation: For pairwise interactions, bias makes it harder for individuals to copy opposing states, thus favoring consensus and polarization of opinions. For group interactions, however, opinion unanimity within a group is necessary for an individual to change state. Under algorithmic bias, the opinion of part of the group is hidden due to content filtering, making majorities harder to form. Fewer state changes due to copying mechanisms and an increasing role of noise leads to coexistence. The separation between pairwise and group based behaviors depends on the specific details of the rates of the model; for the language model the transition occurs smoothly for an intermediate value $\alpha \approx 2$, see Section S3.3 and Figure S2 of the SM \cite{SM_ref} for more details. For other binary-state models well-described by local transition rates we expect such a separation to exist (see \cite{peralta2021opinion} where these results are extended for asymmetric rates, and pair and group based behaviors are discussed in detail). We leave as further work the extension of our analysis to other opinion formation mechanisms beyond voter-like dynamics.

%These results can be interpreted intuitively as follows:
%For a group interaction: 
%Unanimity of the opposite opinion within the group is necessary in order for an individual to change state.
%When we introduce bias, a part of the group in the opposite opinion is not shown to the individual, thus it is very hard to reach unanimity in the group, thus the individual does not change state due to the copying mechanisms and the noise becomes more important, leading to coexistence.
%Heuristic, intuitive explanation of the effect of bias on pair and group interactions
%Non-trivial to fight against polarization
%In reality, high and low alpha, both emperically observable? Non-trivial issue

When algorithmic bias is strong, the coexistence-consensus transition becomes discontinuous. In addition to a `standard' polarized state (where each group is close to consensus in opposite states), for large $b$ and $\alpha$ we observed an additional, `half' polarized state where one group has opinion coexistence and the other one consensus. Note that a wide variety of other unstable or saddle points also exist depending on parameter values and initial conditions, regulating transient states in the dynamics. In most cases, the mean-field approximation presented here gives a good qualitative description of the phase diagram under the effect of algorithmic bias. Using higher-order approximations (pair and approximate master equations) gradually improves accuracy with respect to simulation results.
%The transition lines obtained from the theory, numerically and analytically, at the different approximation levels shows a good accuracy against numerical simulation on synthetic networks with modular structure, thus validating our approach.
We also explored the role of algorithmic bias in models of opinion dynamics over a real-world social network with strong modular structure, the political blogs network \citep{Adamic:2005}. In the parameter region where the polarized state is stable according to the mean field approximation, numerical simulations converge to a polarization of opinions that corresponds to the structural segregation of the network, provided the initial condition is in the basin of attraction of the associated fixed point. 
%The final state of the model at the node (individual) level coincides with the political opinion according to network data. 
We expect our mean-field results to work similarly well in other empirical networks that are link dense or modular.

%In a real system, both types of interaction (pair-wise and group based) may be possible, thus predicting and controlling the impact of algorithmic bias is a non-trivial issue and requires a detailed understanding of how people interact in the social platform. The methods developed in this work shed some light on this issue, as by observing how the system evolves macroscopically (at the global level) we may be able to infer the microscopic (individual level) behavior. Likewise, if we define an empirical protocol to discern the individual level behavior, we will be able to anticipate the evolution of the global system, and the impact of algorithmic bias. 

Our formalism provides a flexible way to parametrize a combination of algorithmic filters and mechanisms of social interactions in terms of transition rates. As such, it may help identify the features of algorithms that promote dynamical and structural polarization in online platforms arguably driven by a mixture of social processes, from homophily \cite{mcpherson2001birds,asikainen2020cumulative} to social contagion \cite{watts2002simple,unicomb2021dynamics}. This will require a validation of our theoretical framework by either fitting observational data with specific models, using automatic model selection \cite{gutmann2016bayesian,chen2019flexible} and statistical inference techniques \cite{peixoto2019network}, or uncovering causal relationships with controlled experiments of online social behavior \cite{salganik2006experimental,centola2010spread}.

The minimal implementation of algorithmic bias explored here can be extended to include more realistic traits of current online media platforms, such as an asymmetry in the state favored by bias, correlations between bias and individual traits (like network degree), a combination of potentially competing filtering algorithms, and the presence of algorithms that react adaptively to changes in human behavior. While we have focused on binary-state dynamics of opinion formation due to the simplicity of the related approximations and relevance to online platforms, we expect our framework to be straightforwardly generalized to models with more than two states or even continuous dynamical variables.

A key limitation of our framework is the consideration of static networks only. Real-world online social networks are instead temporal~\cite{holme2012temporal}, with both nodes and links changing in time due to a variety of mechanisms and external factors influencing how people use media platforms and choose their acquaintances. If extended to temporal networks \cite{unicomb2021dynamics}, our framework will potentially describe a feedback loop between algorithmic content filtering and network/state dynamics that segregates the social network into groups of similarly-minded people (as suggested by recent studies \cite{de2021modeling}), further promoting the polarization effect we already see in static networks. Even if our results focus on binary-state dynamics over simple networks, the rate-equation-based framework is flexible and can be extended to other dynamical descriptions, such as nonlinear dynamical systems with continuous variables~\cite{barzel2013universality,gao2016universal} and higher-order network models \cite{lambiotte2019networks}.

%In this work, we mainly focused on the stationary states and the different possible phases that can be obtained in the long time limit of the models.  The presented approach is able to describe the dynamics too, only indicated here in the form of flows. A thorough exploration of how the phases and the transitions between them influence the relaxation dynamics is a future work proposal. Also of interest is considering an asymmetry in the algorithmic bias intensities, in such a way that the online platform is able to unbalance the two possible opinions, favoring one more than the other, thus breaking the symmetry of the system. Additionally, an open issue for the future is to propose a procedure for the determination of the model parameters, with special attention to the bias intensity $b$, from empirically observable quantities.

The theoretical understanding of the dynamical feedback processes between filters, information transfer and network evolution provided by our framework can suggest heuristic techniques to correct bias. For example, promoting group-based information exchange in a platform dominated by pairwise contacts may decrease polarization and allow for coexistence. Such strategies can improve the chances of a population self-identifying the onset of processes that may reduce their robustness to undesired behavior, like the political swaying and polarization promoted by adversarial agents and social bots in online discussion forums~\cite{ferrara2016rise,bessi2016social}. We trust our results will be part of a larger trend providing scientific background for beneficial regulation and rules of practice in online social ecosystems.

\begin{acknowledgments}
We acknowledge support from AFOSR (Grant \#FA8655-20-1-7020), project EU H2020 Humane AI-net (Grant \#952026), and CHIST-ERA project SAI (FWF I 5205-N). J.K. is grateful for support by projects ERC DYNASNET Synergy (Grant \#810115) and EU H2020 SoBigData++ (Grant \#871042).
\end{acknowledgments}

\

\subsection*{Author contributions}

All authors conceived and designed the research. A.F.P. derived all analytical calculations, performed numerical simulations, and analyzed the data. M.N. and G.I. derived effective rates in presence of bias. A.F.P., J.K., and G.I. wrote the manuscript. All authors approved the manuscript. 

\bibliography{chap_intro_biblio}
\end{document}

\begin{center}
{\LARGE Supplemental Material for}\\[0.7cm]
{\Large \textbf{The effect of algorithmic bias and network structure on coexistence, consensus, and polarization of opinions}}\\[0.5cm]
{\large A. F. Peralta, M. Neri, J. Kertész, and G. Iñiguez}\\[0.7cm]
\end{center}

\addtocontents{toc}{\protect\setstretch{0.1}}
\tableofcontents

%\todo{Use sufix 'S' before all equations/figures, etc., i.e., Fig. S1, etc. [use latex macros 'eref' etc. at beginning of file]}

%\todo{Fix equation formatting. Many inline equations (i.e., between \$ and \$) look weird. Better to put them as equations in their own lines. Number all equations too!}

\section{Derivation of the mean field equations}

\subsection{Average rates and highly connected limit}\label{sec_highly_connected}

The derivation of the mean field equations of the models involves the definition of the average rates:
\begin{align}
\label{average_rate}
f[x] &= \sum_{k} \frac{P_{k} k}{z} \sum_{m=0}^{k} F_{k,m} B_{k,m}(x),\\
\label{average_rate_b}
f^{*}[x] &= \sum_{k} \frac{P_{k} k}{z} \sum_{m=0}^{k} \sum_{i=0}^{m} B_{m,i}(1-b) F_{k-m+i,i} B_{k,m}(x).
\end{align}
These definitions usually lead to complicated expressions for the average rates $f[x]$ and $f^{*}[x]$. Thus, we may use the highly connected limit, i.e., $k \rightarrow \infty$ and $m \rightarrow \infty$ with $m/k$ finite, to simplify the description. In this limit, the binomial functions $B_{m,i}(1-b)$ and $B_{k,m}(x)$ are highly peaked around the values $i = (1-b) m$ and $m = k x$ respectively, thus we have:
\begin{align}
\label{average_rate_hc}
f[x] &\approx \sum_{k} \frac{P_{k} k}{z} F_{k,k x},\\
\label{average_rate_b_hc}
f^{*}[x] & \approx \sum_{k} \frac{P_{k} k}{z} F_{k-k b x,(1-b) x k}.
\end{align}
If the rates $F_{k,m}$ only depend on the fraction $m/k$, the degree dependence of \eref{average_rate_hc} and \eref{average_rate_b_hc} disappears and the expressions simplify accordingly.

We can check the accuracy of the highly connected limit for finite $z$ with a simple example, i.e., $F_{k,m} = \left( \dfrac{m}{k} \right)^{\alpha}$ with $\alpha=1,2,3$:
\begin{align}
\alpha &= 1, \hspace{1cm} f[x] = x, \label{av_rate_1}\\
\alpha &= 2,  \hspace{1cm} f[x] = x^2 + \frac{x(1-x)}{z}, \label{av_rate_2}\\
 \alpha &= 3,  \hspace{1cm} f[x] = x^3 + \frac{3 x^2 (1-x)}{z} + \frac{x (1-3x+2x^2)}{z} \sum_{k} \frac{P_{k}}{k}.\label{av_rate_3}
\end{align}
According to \esref{av_rate_1}{av_rate_3} the average rate is equal to its highly connect limit plus a correcting factor of order $1/z$, in the limit $z \rightarrow \infty$ we recover \eref{average_rate_hc}, i.e., $f[x]=x^{\alpha}$ in this case.

Another approximation that we can use is to assume that the rates \eref{average_rate} and \eref{average_rate_b} are related as
\begin{equation}
\label{av_rate_relation}
f^{*}[x] \approx f \left[ \frac{(1-b) x}{1-b x} \right],
\end{equation}
which is valid in the highly connected limit \eref{average_rate_hc} and \eref{average_rate_b_hc}. For a finite $z$ it is not the case, but we may use the relation \eref{av_rate_relation} as a further simplification. Approximation \eref{av_rate_relation} is used to obtain the phase diagrams in Figs. 5 and 6 of the main text for the majority-vote model with finite $z=20$.

\subsection{Deterministic equations}
In the case without bias $b=0$, the individual rate of changing state from $s=0$ to $s=1$ is $F_{k,m}$ (infection rate), in the mean field assumption that is $f[x]$ with $x$ being the probability of finding a neighbor in state $1$, and from $s=1$ to $s=0$ it is $R_{k,m}=F_{k,k-m}$ (up-down symmetry condition), in the mean field assumption that is $f[1-x]$. When including bias $b>0$, we must replace $F_{k,m}$ by $F_{k,m}^{*}$ and $f[x]$ by $f^{*}[x]$.

\subsubsection{Homogeneous structure}

In the case of a homogeneous structure of $N$ nodes, the global rate of changing $s=0 \rightarrow 1$ is

\begin{equation}
\label{global_rate_up} 
W^{+}=(N-N\rho) f[x], \hspace{1cm} \rho \rightarrow \rho + \dfrac{1}{N},
\end{equation}
i.e., the number of individuals in state $0$ times the individual rate of changing to state $1$, and the rate of changing $s=1 \rightarrow 0$ is

\begin{equation}
\label{global_rate_down} 
W^{-}=N\rho f[1-x], \hspace{1cm} \rho \rightarrow \rho - \dfrac{1}{N},
\end{equation}
respectively. We can construct a deterministic evolution equation for the global fraction $\rho = N^{-1} \sum_{i=1}^{N} s_{i}$ as:

\begin{equation}
\label{mean_field_hom}
\frac{d \rho}{d t} = \frac{1}{N} \left[W^{+} - W^{-} \right] = (1-\rho) f[x] - \rho f[1-x],
\end{equation}
which is just the rate of the process times the change in the variable $\rho$ for that process, summed over all possible processes \cite{VanKampen:2007,Peralta_moments:2018}. It is important to understand that \eref{mean_field_hom} is for the average value over realizations/trajectories of the stochastic dynamics, neglecting fluctuations and finite-size effects in the limit $N \rightarrow \infty$.
When we include bias, we must replace $f[x]$ by $f^{*}[x]$, in the highly connected limit we may additionally use the relation \eref{av_rate_relation}. In the mean field description, we assume that the probability $x$ of finding a neighbor in state $1$ is equal to the fraction $\rho$ of nodes in state $1$. Replacing this in \eref{mean_field_hom} we obtain a closed differential equation for $\rho$, [see Eq. (5) in the main text].

\subsubsection{Modular structure}

In the case of modular structure we have two communities of sizes $N_{1}$ and $N_{2}$, and two description variables $\rho_{1}$ and $\rho_{2}$. There are four possibles processes and global rates: 

\begin{align}
\label{global_rate_up_1} 
W_{1}^{+}&=(N_{1}-N_{1} \rho_{1}) f[x_{1}], \hspace{1cm} \rho_{1} = \rho_{1} + \dfrac{1}{N_{1}},\\
\label{global_rate_down_1}
W_{1}^{-}&=N_{1} \rho_{1} f[1-x_{1}], \hspace{1.5cm} \rho_{1} = \rho_{1} - \dfrac{1}{N_{1}},\\
\label{global_rate_up_2} 
W_{2}^{+}&=(N_{2}-N_{2} \rho_{2}) f[x_{2}], \hspace{1cm} \rho_{2} = \rho_{2} + \dfrac{1}{N_{2}},\\
\label{global_rate_down_2}
W_{2}^{-}&=N_{2} \rho_{2} f[1-x_{2}], \hspace{1.5cm} \rho_{2} = \rho_{2} - \dfrac{1}{N_{2}}.
\end{align}
 We can construct the deterministic evolution equations for the variables $\rho_{1}$, $\rho_{2}$, as
\begin{align}
\label{mean_field_mod_1}
\frac{d \rho_{1}}{d t} &= \frac{1}{N_{1}} \left[W_{1}^{+} - W_{1}^{-} \right] = (1-\rho_{1}) f[x_{1}] - \rho_{1} f[1-x_{1}],\\
\label{mean_field_mod_2}
\frac{d \rho_{2}}{d t} &= \frac{1}{N_{2}} \left[W_{2}^{+} - W_{2}^{-} \right] = (1-\rho_{2}) f[x_{2}] - \rho_{2} f[1-x_{2}].
\end{align}
Again these are evolution equations for the average values over realizations, valid in the limit $N_{1} \rightarrow \infty$, $N_{2} \rightarrow \infty$, and when we include bias we must change $f[x]$ by $f^{*}[x]$. After replacing the probabilities $x_{1, 2}$ by their relations with the global fractions $\rho_{1}, \rho_{2}$, Eq. (6) in the main text, \eref{mean_field_mod_1} and \eref{mean_field_mod_2} become closed, leading to Eqs. (7)-(8) in the main text.

\section{Derivation of the pair approximation}\label{sec_pa}

\subsection{Homogeneous structure}

The pair approximation is a more detailed description as compared to the previous mean filed equations. The main difference is that we will consider a finite connectivity $z$ and avoid the highly connected limit $z \rightarrow \infty$. For the sake of simplicity, we will assume integer values of $z=1, 2, 3, \dots$ without degree heterogeneity, these are $z-$regular networks. The derivation presented next can be straightforwardly generalised to the case of an 
%general 
arbitrary degree distribution. For a homogeneous structure, we consider the description variables $\rho = \dfrac{1}{N} \sum_{i=1}^{N} s_{i}$, the fraction of nodes in state $1$ and $p=\dfrac{1}{z N} \sum_{i=1}^{N} \sum_{j=1}^{N} A_{ij} \left( s_{i} + s_{j} - 2 s_{i} s_{j} \right)$, the fraction of active links (links connecting nodes in different states). There are $2 (z+1)$ possible processes, either a node with $m = 0, 1, \dots, z$ neighbors in state $1$ changes state $s=0 \rightarrow 1$ or the other way around $s=1 \rightarrow 0$, with global rates: 
\begin{align}
\label{global_rate_up_PA} 
W_{z,m}^{+} &= N (1-\rho) P^{(-)}_{z,m} F_{z,m}, \hspace{0.8cm} \rho \rightarrow \rho + \dfrac{1}{N}, \hspace{1.0cm} p \rightarrow p + \frac{2(z-2 m)}{N z},\\
\label{global_rate_down_PA} 
W_{z,m}^{-} &= N \rho P^{(+)}_{z,m} R_{z,m}, \hspace{1.63cm} \rho \rightarrow \rho - \dfrac{1}{N}, \hspace{1.0cm} p \rightarrow p - \frac{2(z-2 m)}{N z},
\end{align}
where $P^{(-/+)}_{z,m}$ is the fraction of nodes with $m$ neighbors in state $1$ that connect to a node in state $0/1$. We can construct the deterministic evolution equations for the variables $\rho$, $p$, as
\begin{align}
\label{pa_hom_rho}
\frac{d \rho}{d t} &= \frac{1}{N} \sum_{m=0}^{z} \left[W_{z,m}^{+} - W_{z,m}^{-} \right],\\
\label{pa_hom_p}
\frac{d p}{d t} &= \frac{2}{N z} \sum_{m=0}^{z}(z-2 m) \left[W_{z,m}^{+} - W_{z,m}^{-} \right].
\end{align}
The pair approximation \cite{Vazquez:2008a, Peralta_pair:2018} assumes that the probabilities $P_{z,m}^{(-)}=B_{z,m}(p_{-})$ and $P_{z,m}^{(+)}=B_{z,z-m}(p_{+})$ are binomial distributions with single event probabilities 

\begin{equation}
\label{prob_single}
p_{-}=\dfrac{p}{2(1-\rho)}, \hspace{1.0cm} p_{+}=\dfrac{p}{2 \rho},
\end{equation}
which are 
%nothing but 
the ratios between the number of active links and the number of links connected to nodes in state $0$ or $1$. Introducing this in \eref{pa_hom_rho} and \eref{pa_hom_p} and assuming up-down symmetric rates $R_{z,m}=F_{z,z-m}$, we can rearrange the evolution equations as
\begin{align}
\label{pa_hom_rho_t}
\frac{d \rho}{d t} &= (1-\rho) \sum_{m=0}^{z} B_{z,m}(p_{-}) F_{z,m} - \rho \sum_{m=0}^{z} B_{z,m}(p_{+}) F_{z,m},\\
\label{pa_hom_p_t}
\frac{d p}{d t} &= (1-\rho) \sum_{m=0}^{z} \frac{2(z-2 m)}{z} B_{z,m}(p_{-}) F_{z,m} \notag\\
&+ \rho \sum_{m=0}^{z} \frac{2(z-2 m)}{z} B_{z,m}(p_{+}) F_{z,m}.
\end{align}

\subsection{Modular structure}

For a modular structure, we consider four different (finite) connectivity parameters defined in the main text, i.e., $z_{1}$, $z_{12}$, $z_{2}$, $z_{21}$, and we will assume that they take integer values without degree heterogeneity, in the $z-$regular fashion. We define six description variables $\rho_{1}$, $\rho_{2}$, $p_{1}$, $p_{2}$, $p_{12}^{01}$, $p_{12}^{10}$ as: 
\begin{equation}
\label{def_rho_1}   
\rho_{1} = \dfrac{1}{N_{1}} \sum_{i=1}^{N_{1}} s_{i},
\end{equation}
the fraction of nodes with state $1$ in community $1$; 
\begin{equation}
\label{def_rho_2}   
\rho_{2} = \dfrac{1}{N_{2}} \sum_{i=N_{1}+1}^{N} s_{i},
\end{equation}
the fraction of nodes with state $1$ in community $2$; 
\begin{equation}
\label{def_p_1}
p_{1}=\dfrac{1}{N_{1} z_{1}} \sum_{i=1}^{N_{1}} \sum_{j=1}^{N_{1}} A_{ij} \left( s_{i} + s_{j} - 2 s_{i} s_{j} \right),
\end{equation}
the fraction of active links in community $1$; 
\begin{equation}
\label{def_p_2}
p_{2}=\dfrac{1}{N_{2} z_{2}} \sum_{i=N_{1}+1}^{N_{2}} \sum_{j=N_{1}+1}^{N_{2}} A_{ij} \left( s_{i} + s_{j} - 2 s_{i} s_{j} \right),
\end{equation}
the fraction of active links in community $2$; 
\begin{equation}
\label{def_p_12_01}
p^{01}_{12}=\dfrac{1}{N_{1} z_{12}} \sum_{i=1}^{N_{1}} \sum_{j=N_{1}+1}^{N_{2}} A_{ij} \left( s_{j} - s_{i} s_{j} \right),
\end{equation}
the fraction of links connecting states $0 - 1$ between communities $1 - 2$; and 
\begin{equation}
\label{def_p_12_10}
p^{10}_{12}=\dfrac{1}{N_{1} z_{12}} \sum_{i=1}^{N_{1}} \sum_{j=N_{1}+1}^{N_{2}} A_{ij} \left( s_{i} - s_{i} s_{j} \right),
\end{equation}
the fraction of links connecting states $1 - 0$ between communities $1 - 2$. Due to the symmetry of the adjacency matrix $A_{ij} = A_{ji}$, other quantities can be expressed as a function of the already defined variables, these are $N_{1} z_{12} = N_{2} z_{21}$, $p_{21}^{01}=p_{12}^{10}$ and $p_{21}^{10}=p_{12}^{01}$.

There are $2(z_{1}+1) \times (z_{12}+1)+2(z_{2}+1) \times (z_{21}+1)$ possibles processes and global rates: 

\begin{equation}
\label{global_rate_up_PA_1}
W_{z_{1},m_{1},z_{12},m_{12}}^{+}=N_{1} (1-\rho_{1}) P^{(1-)}_{z_{1},m_{1}} P^{(12-)}_{z_{12},m_{12}} F_{z_{1}+z_{12},m_{1}+m_{12}},
\end{equation}
with changes in the variables
\begin{align}
\label{changes_up_PA_1}
\rho_{1} &= \rho_{1} + \dfrac{1}{N_{1}}, \hspace{1.7cm} p_{1}=p_{1}+ \dfrac{2(z_{1}-2 m_{1})}{N_{1} z_{1}},\notag\\
p_{12}^{01} &= p_{12}^{01}-\dfrac{m_{12}}{N_{1} z_{12}}, \hspace{1.0cm} p_{12}^{10} = p_{12}^{10}-\dfrac{z_{12}-m_{12}}{N_{1} z_{12}};
\end{align}
\begin{equation}
\label{global_rate_down_PA_1}
W_{z_{1},m_{1},z_{12},m_{12}}^{-}=N_{1} \rho_{1} P^{(1+)}_{z_{1},m_{1}} P^{(12+)}_{z_{12},m_{12}} R_{z_{1}+z_{12},m_{1}+m_{12}},
\end{equation}
with changes in the variables
\begin{align}
\label{changes_down_PA_1}
\rho_{1} &= \rho_{1} - \dfrac{1}{N_{1}}, \hspace{1.7cm} p_{1}=p_{1}- \dfrac{2(z_{1}-2 m_{1})}{N_{1} z_{1}},\notag\\
p_{12}^{01} &= p_{12}^{01}+\dfrac{m_{12}}{N_{1} z_{12}}, \hspace{1.0cm} p_{12}^{10} = p_{12}^{10}+\dfrac{z_{12}-m_{12}}{N_{1} z_{12}};
\end{align}
\begin{equation}
\label{global_rate_up_PA_2}
W_{z_{2},m_{2},z_{21},m_{21}}^{+}=N_{2} (1-\rho_{2}) P^{(2-)}_{z_{2},m_{2}} P^{(21-)}_{z_{21},m_{21}} F_{z_{2}+z_{21},m_{2}+m_{21}},
\end{equation}
with changes in the variables
\begin{align}
\label{changes_up_PA_2}
\rho_{2} &= \rho_{2} + \dfrac{1}{N_{2}}, \hspace{1.7cm} p_{2}=p_{2}+ \dfrac{2(z_{2}-2 m_{2})}{N_{2} z_{2}},\notag\\
p_{12}^{01} &= p_{12}^{01}+\dfrac{z_{21}-m_{21}}{N_{2} z_{21}}, \hspace{1.0cm} p_{12}^{10}=p_{12}^{10}-\dfrac{m_{21}}{N_{2} z_{21}};
\end{align}
and
\begin{equation}
\label{global_rate_down_PA_2}
W_{z_{2},m_{2},z_{21},m_{21}}^{-}=N_{2} \rho_{2} P^{(2+)}_{z_{2},m_{2}} P^{(21+)}_{z_{21},m_{21}} R_{z_{2}+z_{21},m_{2}+m_{21}},
\end{equation}
with changes in the variables
\begin{align}
\label{changes_down_PA_2}
\rho_{2} &= \rho_{2} - \dfrac{1}{N_{2}}, \hspace{1.7cm} p_{2}=p_{2}- \dfrac{2(z_{2}-2 m_{2})}{N_{2} z_{2}},\notag\\
p_{12}^{01} &= p_{12}^{01}-\dfrac{z_{21}-m_{21}}{N_{2} z_{21}}, \hspace{1.0cm} p_{12}^{10}=p_{12}^{10}+\dfrac{m_{21}}{N_{2} z_{21}}.
\end{align}

We can construct the deterministic evolution equations of the variables $\rho_{1}$, $\rho_{2}$, $p_{1}$, $p_{2}$, $p_{12}^{01}$, $p_{12}^{10}$ as
\begin{align}
\label{pa_com_rho1}
\frac{d \rho_{1}}{d t} &= \frac{1}{N_{1}} \sum_{m_{1}=0}^{z_{1}} \sum_{m_{12}=0}^{z_{12}} \left[ W_{z_{1},m_{1},z_{12},m_{12}}^{+}-W_{z_{1},m_{1},z_{12},m_{12}}^{-} \right],\\
\label{pa_com_rho2}
\frac{d \rho_{2}}{d t} &= \frac{1}{N_{2}} \sum_{m_{2}=0}^{z_{2}} \sum_{m_{21}=0}^{z_{21}} \left[ W_{z_{2},m_{2},z_{21},m_{21}}^{+}-W_{z_{2},m_{2},z_{21},m_{21}}^{-} \right],\\
\label{pa_com_p1}
\frac{d p_{1}}{d t} &= \frac{2}{N_{1} z_{1}} \sum_{m_{1}=0}^{z_{1}} \sum_{m_{12}=0}^{z_{12}} (z_{1}-2 m_{1})\left[ W_{z_{1},m_{1},z_{12},m_{12}}^{+}-W_{z_{1},m_{1},z_{12},m_{12}}^{-} \right],\\
\label{pa_com_p2}
\frac{d p_{2}}{d t} &= \frac{2}{N_{2} z_{2}} \sum_{m_{2}=0}^{z_{2}} \sum_{m_{21}=0}^{z_{21}} (z_{2}-2 m_{2}) \left[ W_{z_{2},m_{2},z_{21},m_{21}}^{+}-W_{z_{2},m_{2},z_{21},m_{21}}^{-} \right],\\
\label{pa_com_p12_01}
\frac{d p^{01}_{12}}{d t} &= \frac{1}{N_{1} z_{12}} \sum_{m_{1}=0}^{z_{1}} \sum_{m_{12}=0}^{z_{12}}\left[ -m_{12} W_{z_{1},m_{1},z_{12},m_{12}}^{+}+ m_{12} W_{z_{1},m_{1},z_{12},m_{12}}^{-} \right.\notag\\
& \left. +(z_{21}-m_{21}) W_{z_{1},m_{1},z_{12},m_{12}}^{+} - (z_{21}-m_{21}) W_{z_{1},m_{1},z_{12},m_{12}}^{-} \right],\\
\label{pa_com_p12_10}
\frac{d p^{10}_{12}}{d t} &= \frac{1}{N_{1} z_{12}} \sum_{m_{1}=0}^{z_{1}} \sum_{m_{12}=0}^{z_{12}}\left[ -m_{21} W_{z_{2},m_{2},z_{21},m_{21}}^{+} + m_{21} W_{z_{2},m_{2},z_{21},m_{21}}^{-}  \right.\notag\\
& \left.-(z_{12}-m_{12}) W_{z_{1},m_{1},z_{12},m_{12}}^{+}+ (z_{12}-m_{12}) W_{z_{1},m_{1},z_{12},m_{12}}^{-} \right].
\end{align}
The pair approximation assumes that the probabilities
\begin{align}
\label{probabilities_comm}
P_{z_{1},m_{1}}^{(1-)}&=B_{z_{1},m_{1}}(p_{1-}), \hspace{1.5cm} P_{z_{12},m_{12}}^{(12-)}=B_{z_{12},m_{12}}(p_{12-}),\notag\\
P_{z_{1},m_{1}}^{(1+)}&=B_{z_{1},z_{1}-m_{1}}(p_{1+}), \hspace{1cm} P_{z_{12},m_{12}}^{(12+)}=B_{z_{12},z_{12}-m_{12}}(p_{12+}),\notag\\
P_{z_{2},m_{2}}^{(2-)}&=B_{z_{2},m_{2}}(p_{2-}), \hspace{1.5cm} P_{z_{21},m_{21}}^{(21-)}=B_{z_{21},m_{21}}(p_{21-}),\notag\\
P_{z_{2},m_{2}}^{(2+)} &=B_{z_{2},z_{2}-m_{2}}(p_{2+}), \hspace{1cm} P_{z_{21},m_{21}}^{(21+)}=B_{z_{21},z_{21}-m_{21}}(p_{21+}),
\end{align}
are binomial distributions with single event probabilities 

\begin{align}
\label{single_prob_comm}
p_{1-} &=\dfrac{p_{1}}{2(1-\rho_{1})}, \hspace{1cm} p_{1+}=\dfrac{p_{1}}{2 \rho_{1}},\notag\\
p_{2-} &=\dfrac{p_{2}}{2(1-\rho_{2})}, \hspace{1cm} p_{2+}=\dfrac{p_{2}}{2 \rho_{2}},\notag\\
p_{12-} &=\dfrac{p^{01}_{12}}{1- \rho_{1}}, \hspace{1.3cm} p_{12+}=\dfrac{p^{10}_{12}}{\rho_{1}},\notag\\
p_{21-} &=\dfrac{p^{01}_{21}}{1- \rho_{2}}, \hspace{1.3cm} p_{21+}=\dfrac{p^{10}_{21}}{\rho_{2}},
\end{align}
with $p_{21}^{01}=p_{12}^{10}$, $p_{21}^{10}=p_{12}^{01}$. Introducing the properties \eref{probabilities_comm} and \eref{single_prob_comm} in \esref{pa_com_rho1}{pa_com_p12_10} and assuming up-down symmetric rates $R_{z,m}=F_{z,z-m}$, we rearrange the evolution equations of  the system as

\begin{align}
\label{pair_eq_rho1_c}
& \frac{d \rho_{1}}{d t} = (1-\rho_{1})  \sum_{m_{1}=0}^{z_{1}} B_{z_{1},m_{1}}(p_{1-}) \sum_{m_{12}=0}^{z_{12}} B_{z_{12},m_{12}}(p_{12-}) F_{z_{1}+z_{12} , m_{1} + m_{12}} \notag\\
 &- \rho_{1} \sum_{m_{1}=0}^{z_{1}} B_{z_{1},m_{1}}(p_{1+}) \sum_{m_{12}=0}^{z_{12}} B_{z_{12},m_{12}}(p_{12+})  F_{z_{1}+z_{12},m_{1}+m_{12}},\\
\label{pair_eq_rho2_c}
& \frac{d \rho_{2}}{d t} = (1-\rho_{2})  \sum_{m_{2}=0}^{z_{2}} B_{z_{2},m_{2}}(p_{2-}) \sum_{m_{21}=0}^{z_{21}} B_{z_{21},m_{21}}(p_{21-}) F_{z_{2}+z_{21} , m_{2} + m_{21}} \notag\\
 &- \rho_{2} \sum_{m_{2}=0}^{z_{2}} B_{z_{2},m_{2}}(p_{2+}) \sum_{m_{21}=0}^{z_{21}} B_{z_{21},m_{21}}(p_{21+})  F_{z_{2}+z_{21},m_{2}+m_{21}},
\end{align}
\begin{align}
\label{pair_eq_p1_c}
& \frac{d p_{1}}{d t} =  \frac{2(1-\rho_{1})}{z_{1}} \sum_{m_{1}=0}^{z_{1}} B_{z_{1},m_{1}}(p_{1-}) \sum_{m_{12}=0}^{z_{12}} B_{z_{12},m_{12}}(p_{12-}) (z_{1} - 2 m_{1}) F_{z_{1}+z_{12} , m_{1} + m_{12}} \notag\\
 &+ \frac{2 \rho_{1}}{z_{1}} \sum_{m_{1}=0}^{z_{1}} B_{z_{1},m_{1}}(p_{1+}) \sum_{m_{12}=0}^{z_{12}} B_{z_{12},m_{12}}(p_{12+}) (z_{1}- 2 m_{1})  F_{z_{1}+z_{12},m_{1}+m_{12}},\\
\label{pair_eq_p2_c}
& \frac{d p_{2}}{d t} = \frac{2 (1-\rho_{2})}{z_{2}}  \sum_{m_{2}=0}^{z_{2}} B_{z_{2},m_{2}}(p_{2-}) \sum_{m_{21}=0}^{z_{21}} B_{z_{21},m_{21}}(p_{21-}) (z_{2} - 2 m_{2}) F_{z_{2}+z_{21} , m_{2} + m_{21}} \notag\\
&+ \frac{2  \rho_{2}}{z_{2}} \sum_{m_{2}=0}^{z_{2}} B_{z_{2},m_{2}}(p_{2+}) \sum_{m_{21}=0}^{z_{21}} B_{z_{21},m_{21}}(p_{21+}) (z_{2} - 2 m_{2})  F_{z_{2}+z_{21},m_{2}+m_{21}},
\end{align}
\begin{align}
\label{pair_eq_p12_01}
&\frac{d p^{01}_{12}}{d t} = -\frac{1-\rho_{1}}{z_{12}}  \sum_{m_{1}=0}^{z_{1}} B_{z_{1},m_{1}}(p_{1-}) \sum_{m_{12}=0}^{z_{12}} B_{z_{12},m_{12}}(p_{12-}) m_{12} F_{z_{1}+z_{12} , m_{1} + m_{12}} \notag\\
 &+ \frac{\rho_{1}}{z_{12}} \sum_{m_{1}=0}^{z_{1}} B_{z_{1},m_{1}}(p_{1+}) \sum_{m_{12}=0}^{z_{12}} B_{z_{12},m_{12}}(p_{12+}) (z_{12}-m_{12})  F_{z_{1}+z_{12},m_{1}+m_{12}}\notag\\
&+\frac{1-\rho_{2}}{z_{21}}  \sum_{m_{2}=0}^{z_{2}} B_{z_{2},m_{2}}(p_{2-}) \sum_{m_{21}=0}^{z_{21}} B_{z_{21},m_{21}}(p_{21-}) (z_{21}-m_{21}) F_{z_{2}+z_{21} , m_{2} + m_{21}} \notag\\
 &- \frac{\rho_{2}}{z_{21}} \sum_{m_{2}=0}^{z_{2}} B_{z_{2},m_{2}}(p_{2+}) \sum_{m_{21}=0}^{z_{21}} B_{z_{21},m_{21}}(p_{21+}) m_{21} F_{z_{2}+z_{21},m_{2}+m_{21}},\\
\label{pair_eq_p12_10}
&\frac{d p^{10}_{12}}{d t} = \frac{1-\rho_{1}}{z_{12}}  \sum_{m_{1}=0}^{z_{1}} B_{z_{1},m_{1}}(p_{1-}) \sum_{m_{12}=0}^{z_{12}} B_{z_{12},m_{12}}(p_{12-})  (z_{12}-m_{12}) F_{z_{1}+z_{12} , m_{1} + m_{12}} \notag\\
 &- \frac{\rho_{1}}{z_{12}} \sum_{m_{1}=0}^{z_{1}} B_{z_{1},m_{1}}(p_{1+}) \sum_{m_{12}=0}^{z_{12}} B_{z_{12},m_{12}}(p_{12+}) m_{12}  F_{z_{1}+z_{12},m_{1}+m_{12}}\notag\\
&-\frac{1-\rho_{2}}{z_{21}}  \sum_{m_{2}=0}^{z_{2}} B_{z_{2},m_{2}}(p_{2-}) \sum_{m_{21}=0}^{z_{21}} B_{z_{21},m_{21}}(p_{21-}) m_{21} F_{z_{2}+z_{21} , m_{2} + m_{21}} \notag\\
 &+ \frac{\rho_{2}}{z_{21}} \sum_{m_{2}=0}^{z_{2}} B_{z_{2},m_{2}}(p_{2+}) \sum_{m_{21}=0}^{z_{21}} B_{z_{21},m_{21}}(p_{21+}) (z_{21}-m_{21}) F_{z_{2}+z_{21},m_{2}+m_{21}}.
\end{align}
When we include bias we must replace $F_{z,m}$ by $F^{*}_{z,m}$. For an initial condition with $\rho_{1}(0)$ and $\rho_{2}(0)$, the other description variable are determined as follows:

\begin{align}
\label{ic_comm}    
p_{1}(0)&=2 \rho_{1}(0) (1-\rho_{1}(0)), \hspace{1cm} p_{2}(0)=2 \rho_{2}(0) (1-\rho_{2}(0)), \notag\\
p_{12}^{01}(0) &=(1-\rho_{1}(0))\rho_{2}(0), \hspace{1cm} p_{12}^{10}(0)=\rho_{1}(0) (1-\rho_{2}(0)).
\end{align}

\section{Characterization of the phase transitions}

It is possible to obtain analytical closed expressions of the transition parameter values and the phase diagrams for some cases of interest. This can be performed trough studying the fixed points and stability of the evolution equations derived in \sref{sec_pa}, at different levels of approximation, and considering homogeneous and modular structures.

\subsection{Homogeneous structure}\label{sec_critical_hom}
First consider a system with homogeneous structure and without bias $b=0$, governed by the evolution equation:
\begin{equation}
\label{hom_no_bias}
\frac{d \rho}{d t}=(1-\rho) f[\rho] - \rho f[1-\rho] \equiv \mu[\rho].
\end{equation}
The fixed points come determined by the condition $\mu[\rho_{\text{st}}]=0$, and the stability by the sign of the first derivative $\mu'[\rho_{\text{st}}]$, i.e., $\mu'[\rho_{\text{st}}]>0$ unstable and $\mu'[\rho_{\text{st}}]<0$ stable.  The transition parameter values, where the stability of the fixed point changes and new fixed points appear, are determined by the condition $\mu[\rho_{\text{st}}]=0$ ($\rho_{\text{st}}$ is a fixed point) and $\mu'[\rho_{\text{st}}]=0$ (marginal stability condition).

The coexistence state $\rho_{\text{st}}=1/2$ is always a possible stationary solution of \eref{hom_no_bias}. This solution changes its stability when 
\begin{equation}
\label{condition_critical}
-2 f \left[1/2 \right]+f'[1/2]=0,
\end{equation}
and from this condition we determine the critical parameter value $Q_{c}$ at the coexistence-consensus transition.  Other possible fixed points and transitions are also possible and can be determined numerically (and analytically in some cases of interest) with the condition $\mu[\rho_{\text{st}}]=\mu'[\rho_{\text{st}}]=0$.

For the noisy voter model with rate $f[\rho]=Q+(1-2 Q) \rho$, we obtain the critical value $Q_{c}=0$. Thus, for $Q>Q_{c}$ we have $\rho_{\text{st}}=1/2$ (stable) as the only fixed point of the dynamics, i.e., the noisy voter model does not have a well defined coexistence-consensus transition.

For the majority-vote model with rate $f[\rho]=Q+(1-2 Q) \theta(\rho-1/2)$, where $\theta(x)$ is the Heaviside step function using the half-maximum convention $\theta(1/2)=1/2$, we obtain the critical value $Q_{c}=1/2$. For $Q \geq Q_{c}$ we have $\rho_{\text{st}}=1/2$ (stable) as the only fixed point, while for $Q<Q_{c}$ the coexistence state $\rho_{\text{st}}=1/2$ becomes unstable and two new symmetrical fixed points, the consensus states, appear $\rho_{\text{st}}=Q$ and $\rho_{\text{st}}=1-Q$, both stable. It is important to understand that in this case the condition \eref{condition_critical} cannot be used due to the discontinuity in the rates of the majority-vote model (the derivative $f'[1/2]$ is not well defined), we must then resort to a graphical analysis of \eref{hom_no_bias} to obtain the fixed points and their stability.

For the language model with rates $f[\rho]=Q+(1-2 Q) \rho^{\alpha}$ we obtain, using \eref{condition_critical}, the critical value

\begin{equation}
\label{critical_language_nobias}
a_{c} \equiv \dfrac{Q_{c}}{1-2 Q_{c}}=2^{-\alpha}(\alpha-1).
\end{equation}
If we expand the function $\mu[\rho]$ around the coexistence solution $\rho_{\text{st}}=1/2$, it only has odd powers $\left(\rho-1/2 \right)^{n}$, $n=1,3,5, \dots$ Depending on the sign of the third derivative $\mu'''[1/2]$, the coexistence-consensus transition at $Q=Q_{c}$ is continuous (supercritical ptichfork) or discontinuous (subcritical pitchfork). The tricritical point, can be determined using the condition $\mu'[1/2]=\mu'''[1/2]=0$ or equivalently
\begin{equation}
\label{condition_tricritical}
-6 f'' \left[1/2 \right]+f'''[1/2]=0,
\end{equation}
which separates the continuous from the discontinuous transition lines, this leads to $\alpha=\alpha_{t}=5$. In the continuous transition case $\alpha \leq \alpha_{t}$, at the transition $Q_{c}$, the coexistence state changes from stable to unstable and two consensus stable states appear for $Q<Q_{c}$. In the discontinuous transition case $\alpha \geq \alpha_{t}$, there are two transition lines $Q_{c}$ and $Q_{t}$, such that in the region $Q_{c} < Q < Q_{t}$ the coexistence and consensus states are all stable, for $Q < Q_{c}$ the consensus states are stable (coexistence is unstable) and for $Q > Q_{t}$ the coexistence state is the only stable state. In the region $Q_{c} < Q < Q_{t}$, there are two additional consensus states that are unstable. The transition value $Q_{t}$ can be determined numerically from the condition $\mu'[\rho_{\text{st}}]=0$ around the consensus states $\mu[\rho_{\text{st}}]=0$.

It is possible to repeat the previous analysis including bias $b>0$, replacing $f[\rho]$ by $f^{*}[\rho]$ using \eref{av_rate_relation} in \eref{hom_no_bias}. For the noisy voter model we obtain the fixed points $\rho_{\text{st}}=\dfrac{1}{2}$ (stable) for $Q>Q_{c}$, and $\rho_{\text{st}}=\dfrac{1}{2}$ (unstable),
\begin{equation}
\rho_{\text{st}} = \dfrac{1}{2} \pm \sqrt{\dfrac{ a (2-b)^2 - b (1-b)}{b (a b - 1 + b)}}  \hspace{0.2cm} (\text{stable}),
\end{equation}
for $Q<Q_{c}$, with $a \equiv \dfrac{Q}{1-2 Q}$ and 
\begin{equation}
\label{critical_noisy_bias}  
a_{c} = \dfrac{Q_{c}}{1- 2 Q_{c}}= \dfrac{b(1-b)}{(2-b)^2},
\end{equation}
with a maximum of $Q_{c}(b)$ at $b=2/3$. 

For the majority-vote model we have a discontinuous transition for $b>0$, that is $\rho_{\text{st}}=\dfrac{1}{2}$ (stable), $\rho_{\text{st}}=Q$ (stable), $\rho_{\text{st}}=1-Q$ (stable) for $Q<Q_{t}$, and $\rho_{st}=\dfrac{1}{2}$ (stable) for $Q>Q_{t}$, with
\begin{equation}
\label{critical_majority_hc}
Q_{t}=\dfrac{1-b}{2-b}.
\end{equation} 
For $Q < Q_{t}$ there are two additional consensus (unstable) fixed points $\rho_{\text{st}}=Q_{t}$ and $\rho_{\text{st}}=1-Q_{t}$.

For the language model, the transition parameter values change as a function of the bias intensity $b$ as:
\begin{equation}
\label{critical_language_bias}
a_{c} = \dfrac{Q_{c}}{1- 2 Q_{c}}= \left( \dfrac{1-b}{2-b} \right)^{\alpha} \dfrac{2(\alpha-1)+b}{2-b},
\end{equation}
\begin{equation}
\label{tricritical_language_bias_b}
b_{t} = \frac{-3(3 \alpha-5) \pm \sqrt{3(1+\alpha)(11 \alpha-5)}}{12},
\end{equation}
\begin{equation}
\label{tricritical_language_bias_alpha}
\alpha_{t} = \frac{-3(3 b-4) \pm \sqrt{64-96b+33b^2}}{4}.
\end{equation}
The transition is continuous for $b < b_{t}$ (or $\alpha < \alpha_{t}$) and discontinuous for $b > b_{t}$ (or $\alpha > \alpha_{t}$). $Q_{c}(b)$ has a maximum as a function of $b$ at $b=\dfrac{2}{3}(2-\alpha)$. If $\alpha > 2$ there is no maximum and $Q_{c}(b)$ is a decreasing function.

The analysis in a finite connectivity $z$ network can be performed using the pair approximation \esref{pa_hom_rho_t}{pa_hom_p_t}. The results show an equivalent phenomenology to the previous mean field analysis, but the transition lines shift as a function of $z$. It is possible, see \cite{Gleeson:2013, Vieira:2020}, to obtain a condition for the critical parameter value $Q_{c}$ in the pair approximation:

\begin{equation}
\label{pair_critical}
\sum_{m=0}^{z} \left( 1 - \frac{2 m}{z} \right) F_{z,m} B_{z,m} \left( \frac{z-2}{2 z- 2} \right) = 0,
\end{equation}
in the limit $z \rightarrow \infty$ the condition \eref{pair_critical} is equivalent to \eref{condition_critical}.
The condition for the tricritical point can be also calculated approximately as:
\begin{eqnarray}
\label{pair_tricritical}
\frac{z-2}{z- 1}  \sum_{m=0}^{z}  F_{z,m} B'''_{z,m} \left( \frac{z-2}{2 z- 2} \right)- 6 \sum_{m=0}^{z}  F_{z,m} B''_{z,m} \left( \frac{z-2}{2 z- 2} \right) = 0,
\end{eqnarray}
in the limit $z \rightarrow \infty$ the condition \eref{pair_tricritical} is equivalent to \eref{condition_tricritical}.

\begin{figure}[h!]
\centering
\includegraphics[width=0.80\textwidth]{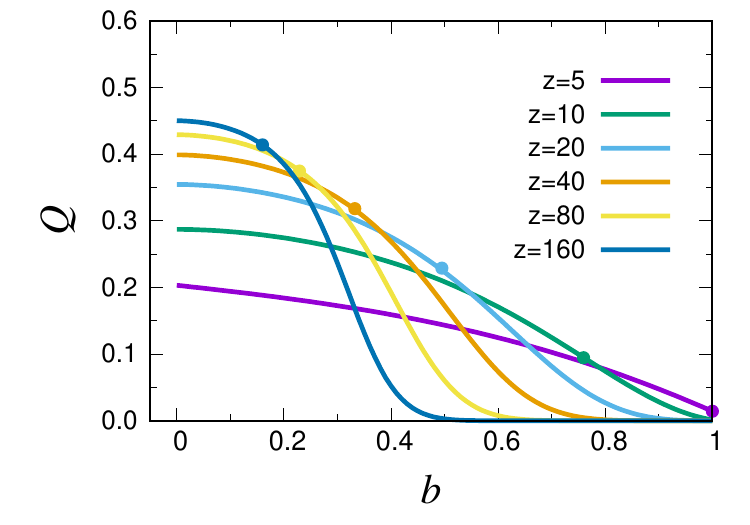}
\caption{Transition values $Q_{c}$ of the majority-vote model as a function of the bias $b$ (colored solid lines) and tricritical points (colored points) for networks with different connectives $z$, as predicted by the pair approximation \eref{majority_z_regular} and \eref{majority_z_regular_tricritical}.}
\label{fig:Majority_network}
\end{figure}

The determination of the critical and tricritical points from \eref{pair_critical} and \eref{pair_tricritical} corrects the lacks of the highly connected approximation in finite $z$ networks. For the noisy voter and language models we expect a correcting factor of order $1/z$, as shown in \sref{sec_highly_connected}, which is small for networks with a relatively large average degree. For the majority-vote model, however, due to the discontinuous nature of the rates the correcting factor is more important and we will show later to be order $1/\sqrt{z}$ (see \eref{critical_majority}). For this reason, we will focus only on this last model to apply the pair approximation, rewriting the transition rates with bias as
\begin{equation}
\label{majority_efective_rate}
F^{*}_{k,m} = \begin{cases}
Q \hspace{0.2cm} &\text{if} \hspace{0.5cm}  m < k/2, \\
Q + (1-2 Q) G_{k,m} \hspace{0.2cm} &\text{if} \hspace{0.5cm} m \geq k/2,
\end{cases}
\end{equation}
with
\begin{equation}
\label{Gkm}
G_{k,m} =  \frac{1}{2} B_{m,k-m}(1-b) + \sum_{i=k-m+1}^{m} B_{m,i}(1-b).
\end{equation}
From \eref{pair_critical} we obtain the critical point
\begin{equation}
\label{majority_z_regular}
Q_{c} =  -\frac{ \sum_{m = \left[ \frac{z+1}{2} \right]}^{z} \left(  1 - \frac{2 m}{ z}\right) B_{k,m} \left( p_{c} \right) G_{k,m}}{\sum_{m = 0}^{\left[ \frac{z+1}{2} \right]} \left(  1 - \frac{2 m}{ z}\right) B_{k,m} \left( p_{c} \right) + \sum_{m = \left[ \frac{z+1}{2} \right]}^{z} \left(  1 - \frac{2 m}{ z}\right) B_{k,m} \left( p_{c} \right)(1-2 G_{k,m})},
\end{equation}
and the tricritical point from \eref{pair_tricritical}
\begin{eqnarray}
\label{majority_z_regular_tricritical}
 p_{c}  \sum_{m = \left[ \frac{z+1}{2} \right]}^{z}  G_{k,m} B'''_{k,m} \left( p_{c} \right)- 3 \sum_{m = \left[ \frac{z+1}{2} \right]}^{z}  G_{k,m} B''_{k,m} \left( p_{c} \right) = 0,
\end{eqnarray}
with $ p_{c} = \dfrac{z-2}{2z-2}$. In \fref{fig:Majority_network} we plot the critical $Q_{c}$ and tricritical points for different values of $z$, and a strong dependence of $Q_{c}$ on $z$ is observed.

In order to understand the dependence of $Q_{c}$ on $z$, we will take the simpler mean field description without resorting to highly connected limit and without bias $b=0$, with average rate:
\begin{equation}
\label{rate_def}
f[\rho]=\sum_{m=0}^{z} F_{z,m} B_{z,m}(\rho).
\end{equation}
For the majority-vote model with $z$ even this is:
\begin{align}
\label{rate_majority}
f[\rho] &= Q+(1-2 Q) g[\rho], \notag\\
g[\rho] &= \frac{1}{2} B_{z,z/2}(\rho) + \sum_{m=z/2+1}^{z} B_{z,m}(\rho).
\end{align}
It is possible to find an analytical expression for $Q_{c}$ in this case:
\begin{equation}
\label{critical_majority}
Q_{c} = \frac{1}{2} \left(1 - \frac{\sqrt{\pi}}{2} \frac{\Gamma \left( \frac{z}{2} \right)}{\Gamma \left( \frac{z+1}{2} \right)} \right) \approx \frac{1}{2} \left(1 - \sqrt{\frac{\pi}{2 z} } \right),
\end{equation} 
in the limit $z \rightarrow \infty$ we recover $Q_{c}=1/2$.

\subsection{Modular structure}

We now consider a modular structure and focus on solutions of \eref{mean_field_mod_1} and  \eref{mean_field_mod_2} for communities of equal sizes $N_{1}=N_{2}=N/2$, that is:

\begin{align}
\label{dyn_mod_1}
\frac{d \rho_{1}}{d t} &= (1-\rho_{1}) f\left[ \frac{\rho_{1}+p \rho_{2}}{1+p} \right] - \rho_{1} f\left[ \frac{1-\rho_{1}+p (1-\rho_{2})}{1+p} \right] \equiv \mu_{1}[\rho_{1},\rho_{2}],\\
\label{dyn_mod_2}
\frac{d \rho_{1}}{d t} &= (1-\rho_{2}) f\left[ \frac{\rho_{2}+p \rho_{1}}{1+p} \right] - \rho_{2} f\left[ \frac{1-\rho_{2}+p (1-\rho_{1})}{1+p} \right] \equiv \mu_{2}[\rho_{1},\rho_{2}].
\end{align}
The fixed points of this system fulfill $\mu_{1}[\rho^{\text{st}}_{1},\rho^{\text{st}}_{2}]=\mu_{2}[\rho^{\text{st}}_{1},\rho^{\text{st}}_{2}]=0$. The Jacobian matrix can be calculated as:

\begin{align}
J_{11} &= \frac{\partial \dot{\rho_{1}}}{\partial \rho_{1}} = - f\left[ \frac{\rho_{1}+p \rho_{2}}{1+p} \right] - f\left[ \frac{1-\rho_{1}+p (1-\rho_{2})}{1+p} \right] \notag\\
&+ \frac{1-\rho_{1}}{1+p} f'\left[ \frac{\rho_{1}+p \rho_{2}}{1+p} \right] + \frac{\rho_{1}}{1+p} f'\left[ \frac{1-\rho_{1}+p (1-\rho_{2})}{1+p} \right],\\
J_{12} &= \frac{\partial \dot{\rho_{1}}}{\partial \rho_{2}} = \frac{p(1-\rho_{1})}{1+p} f'\left[ \frac{\rho_{1}+p \rho_{2}}{1+p} \right] + \frac{ p \rho_{1}}{1+p} f'\left[ \frac{1-\rho_{1}+p (1-\rho_{2})}{1+p} \right],\\
J_{21} &= \frac{\partial \dot{\rho_{2}}}{\partial \rho_{1}} = \frac{p(1-\rho_{2})}{1+p} f'\left[ \frac{\rho_{2}+p \rho_{1}}{1+p} \right] + \frac{ p \rho_{2}}{1+p} f'\left[ \frac{1-\rho_{2}+p (1-\rho_{1})}{1+p} \right],\\
J_{22} &= \frac{\partial \dot{\rho_{2}}}{\partial \rho_{2}} = - f\left[ \frac{\rho_{2}+p \rho_{1}}{1+p} \right] - f\left[ \frac{1-\rho_{2}+p (1-\rho_{1})}{1+p} \right] \notag\\
&+ \frac{1-\rho_{2}}{1+p} f'\left[ \frac{\rho_{2}+p \rho_{1}}{1+p} \right] + \frac{\rho_{2}}{1+p} f'\left[ \frac{1-\rho_{2}+p (1-\rho_{1})}{1+p} \right].
\end{align}
Equivalently to the homogeneous case \eref{hom_no_bias}, the coexistence solution $\rho^{\text{st}}_{1}=\rho^{\text{st}}_{2}=1/2$ is always possible. The stability of this solution is determined by the Jacobian matrix:
\begin{align}
J_{11}&=-2 f[1/2] + \frac{1}{1+p} f'[1/2], \hspace{0.8cm}
J_{12}=\frac{p}{1+p} f'[1/2],\notag\\
J_{21}&=J_{12}, \hspace{4.0cm}
J_{22}=J_{11}.
\end{align}
The eigenvalues and eigenvectors of the Jacobian matrix are:
\begin{align}
\lambda_{1} &= J_{11} + J_{12} = -2 f[1/2] + f'[1/2], \hspace{1.15cm} \vec{v}_{1}= \frac{1}{\sqrt{2}}\begin{bmatrix}
           1 \\
           1
         \end{bmatrix},\\
\lambda_{2} &= J_{11} - J_{12} = -2 f[1/2] + \frac{1-p}{1+p} f'[1/2], \hspace{0.3cm} \vec{v}_{2}= \frac{1}{\sqrt{2}}\begin{bmatrix}
           1 \\
           -1
         \end{bmatrix}.
\end{align}
If $J_{11} + J_{12} = 0$ the stability of the coexistence solution changes along the homogeneous direction $\rho_{1}(t)=\rho_{2}(t)$, this relation determines the critical value $Q_{c}$, the same that we obtained in \sref{sec_critical_hom}. If $J_{11} - J_{12} = 0$ the stability of the coexistence solution changes along the polarized direction $\rho_{1}(t)+\rho_{2}(t)=1$, this relation determines the critical value $Q_{p}^{*}$ where two new polarized fixed points appear. 

We analyze now the stability of the fixed points in the polarized line $\rho_{1}+\rho_{2} = 1$. The Jacobian matrix elements in this case read:

\begin{align}
J_{11} &= - f\left[\frac{\rho_{1}+ p \rho_{2}}{1+p}\right] - f\left[\frac{\rho_{2}+ p \rho_{1}}{1+p}\right] \notag\\
&+ \frac{1}{1+p} \left( \rho_{1}f'\left[\frac{\rho_{2}+ p \rho_{1}}{1+p}\right] +\rho_{2} f'\left[\frac{\rho_{1}+ p \rho_{2}}{1+p}\right] \right),\\
J_{12} &=\frac{p}{1+p} \left( \rho_{1} f'\left[\frac{\rho_{2}+ p \rho_{1}}{1+p}\right] +\rho_{2} f'\left[\frac{\rho_{1}+ p \rho_{2}}{1+p}\right] \right),
\end{align}
with $J_{21}=J_{12}$ and $J_{22}=J_{11}$.

The eigenvalues and eigenvectors are:
\begin{align}
\lambda_{1} &= J_{11} + J_{12} = - f\left[\frac{\rho_{1}+ p \rho_{2}}{1+p}\right] -f\left[\frac{\rho_{2}+ p \rho_{1}}{1+p}\right] \notag\\
&+  \rho_{1} f'\left[\frac{\rho_{2}+ p \rho_{1}}{1+p}\right] +\rho_{2} f'\left[\frac{\rho_{1}+ p \rho_{2}}{1+p}\right], \hspace{1.7cm} \vec{v}_{1}= \frac{1}{\sqrt{2}}\begin{bmatrix}
           1 \\
           1
         \end{bmatrix},\\
\lambda_{2} &= J_{11} - J_{12} = - f \left[\frac{\rho_{1}+ p \rho_{2}}{1+p}\right] - f \left[\frac{\rho_{2}+ p \rho_{1}}{1+p}\right] \notag\\
&+  \frac{1-p}{1+p} \left(\rho_{1} f'\left[\frac{\rho_{2}+ p \rho_{1}}{1+p}\right] +\rho_{2} f'\left[\frac{\rho_{1}+ p \rho_{2}}{1+p}\right] \right), \hspace{0.3cm} \vec{v}_{2}= \frac{1}{\sqrt{2}}\begin{bmatrix}
           1 \\
           -1
         \end{bmatrix}.
\end{align}
If $J_{11} + J_{12} = 0$ the polarized solution becomes stable along both direction, thus it is a stable fixed point. This relation determines the critical value $Q_{p}$, where the polarized solution becomes stable. These three transition lines, $Q_{c}$, $Q^{*}_{p}$, $Q_{p}$ (as a function of $b$) are the main focus of this section, but other fixed points outside the polarization line are also possible and can be computed numerically and its stability analyzed using the Jacobian matrix.

We start with the noisy voter model, for which we will be able to compute analytically all transition lines. The solutions in the polarization line fulfill $\mu_{1}[\rho_{\text{st}},1-\rho_{\text{st}}]=0$ and closed expressions can be obtained as a function of the parameters of the model. These are: for $Q_{c}>Q>Q^{*}_{p}$ we have $\rho_{st}=\dfrac{1}{2}$ (unstable in $\vec{v}_{1}$, stable in $\vec{v}_{2}$), for $Q<Q^{*}_{p}$ we have $\rho_{st}=\dfrac{1}{2}$ (unstable in $\vec{v}_{1}$, unstable in $\vec{v}_{2}$) and two new polarized fixed points appear:

\begin{equation}
\label{pol_fixed_noisy}
\rho_{st} = \frac{1}{2} \left( 1 \pm \dfrac{\sqrt{1+p}}{1-p}  \sqrt{\dfrac{a (2-b)^2 (1+p) - (1-b)(b+(b-4)p)}{ b (a b - 1 + b)}} \right),
\end{equation}
with $a = \dfrac{Q}{1-2 Q}$ and

\begin{equation}
\label{noisy_critical_ps}
a^{*}_{p} = \dfrac{Q^{*}_{p}}{1- 2 Q^{*}_{p}}= \dfrac{(1-b)(b+(b-4)p)}{(2-b)^2(1+p)}.
\end{equation}
The transition $Q^{*}_{p}$ exists if $b>\dfrac{4 p}{1+p}$, it has a maximum at $b=\dfrac{2(1+p)}{3-p}$, and for $p>\dfrac{1}{3}$ there is no transition. The polarized solution \eref{pol_fixed_noisy} is unstable in $\vec{v}_{1}$ and stable in $\vec{v}_{2}$ if $Q_{p}<Q<Q^{*}_{p}$, and stable in both directions if $Q<Q_{p}$ with

\begin{equation}
\label{noisy_critical_p}
a_{p} \equiv \dfrac{Q_{p}}{1-2 Q_{p}} = \dfrac{b(1-b)+(-6+8b-2b^2)p+(-2+3b-b^2)p^2}{(2-b)^2 (1+p)^2}.
\end{equation}
The polarization transition $Q_{p}$ exists if $b>\dfrac{2p (3+p)}{(1+p)^2}$, it has a maximum at $b=\dfrac{2(1+p)^2}{3+p^2}$, and for $p>\sqrt{5}-2 \approx 0.236$ there is no polarization transition. In \fref{fig:vector_field1} and \fref{fig:vector_field2} we show the vector fields \eref{dyn_mod_1} and \eref{dyn_mod_2} and fixed points as we cross the different transition values $Q_{c}$, $Q^{*}_{p}$ and $Q_{p}$ that can be determined straightforwardly using the expressions that we calculated previously. In \fref{fig:vector_field1} with fixed $b=2/3$ and $p=0.1$, that is $Q_{c}=0.1$, $Q^{*}_{p}=0.05102$ and $Q_{p}=0.02734$. In \fref{fig:vector_field2} with fixed $Q=0.01$ and $p=0.1$, that is $b_{c}=0.04083$, $b^{*}_{p}=0.4073$, $b_{p}=0.5605$.

For the majority-voter model, the highly connected assumption is not a good approximation to obtain the transition lines of the polarized states. We thus use a numerical analysis with the average rate \eref{rate_majority} with a finite $z$ ($z=20$ in Fig. 6 of the main text) to obtain the phase diagram of the model. The reason why the highly connected assumption is not a good approximation is that it predicts $Q_{c}=Q^{*}_{p}=Q_{p}=1/2$ in the case without bias, while when using the average rate \eref{rate_majority} with finite $z$ it is
\begin{equation}
\label{critical_p_majority}
Q^{*}_{p} = \frac{1}{2} \left(\frac{1-p}{1+p} - \frac{\sqrt{\pi}}{2} \frac{\Gamma \left( \frac{z}{2} \right)}{\Gamma \left( \frac{z+1}{2} \right)} \right).
\end{equation}
Taking the limit $z \rightarrow \infty$ of \eref{critical_p_majority} we obtain $Q^{*}_{p} = \dfrac{1-p}{2(1+p)}$ which does not coincide with the previous highly connected result. We thus deduce that, the highly connected limit must be done carefully in this case and that taking $z \rightarrow \infty$ in the average rates before doing the calculation may lead to incorrect results (in order words, the limits do not commute).

For the language model with a general value of $\alpha$, the polarized fixed points fulfilling $\mu_{1}[\rho_{\text{st}},1-\rho_{\text{st}}]=0$ can be obtained numerically. It is possible to calculate the transition parameter values:
\begin{equation}
\label{critical_language_bias_com}
a^{*}_{p} = \dfrac{Q^{*}_{p}}{1- 2 Q^{*}_{p}}= \left( \dfrac{1-b}{2-b} \right)^{\alpha} \dfrac{2(1-p)\alpha-(2-b)(1+p)}{(2-b)(1+p)},
\end{equation}
\begin{equation}
\label{tricritical_language_bias_b_com}
b^{*}_{t} = \frac{-3(3 \alpha-5) + 3 p (1+\alpha) \pm \sqrt{3(1+\alpha)(11 \alpha-5+p(14+3p+(3p-2)\alpha))}}{12},
\end{equation}
\begin{equation}
\label{tricritical_language_bias_alpha_com}
\alpha^{*}_{t} = \frac{12-3b(3-p) \pm \sqrt{16(2+p)^2-24b(4+p+p^2)+3b^2(11-2p+3p^2)}}{4(1-p)}.
\end{equation}
This transition $Q^{*}_{p}$ exists if $b>2\dfrac{1-\alpha+p(1+\alpha)}{1+p}$. There is a maximum of $Q^{*}_{p}$ at $b=\dfrac{2(2-(1-p)\alpha)}{3-p}$. If $p>\dfrac{2 \alpha-1}{2\alpha+1}$ there is no transition. If $p>\dfrac{\alpha-2}{\alpha}$, $Q^{*}_{p}$ has a maximum as a function of $b$, and for $p>\dfrac{\alpha-1}{\alpha+1}$ the transition appears with increasing $b$. The transition $Q^{*}_{p}$ can be characterized as follows, for $Q_{c}>Q>Q^{*}_{p}$ we have $\rho_{st}=\dfrac{1}{2}$ (unstable in $\vec{v}_{1}$, stable in $\vec{v}_{2}$), for $Q<Q^{*}_{p}$ we have $\rho_{st}=\dfrac{1}{2}$ (unstable in $\vec{v}_{1}$, unstable in $\vec{v}_{2}$) and two new polarized fixed points appear. The values $b^{*}_{t}$ and $\alpha^{*}_{t}$ separate the continuous from the discontinuous transitions. Unfortunately, it is not possible to obtain an analytical expression for the polarization transition value $Q_{p}$, where the polarized states become stable fixed points in both eigendirections. We thus have to resort to the numerical evaluation of the fixed points and their stability. For certain specific values of the parameters it is still possible to obtain analytical expressions, for example in the case without noise $Q=0$ and $\alpha=2$ we can determine the polarized fixed points $\rho_{1}^{\text{st}}=\rho_{\text{st}}$ and $\rho_{2}^{\text{st}}=1-\rho_{\text{st}}$:
\begin{equation}
\rho_{\text{st}} = \frac{1}{2} \left( 1 \pm \frac{\sqrt{(1-3p)(1+p)}}{1-p} \right),
\end{equation}
with transition values $p^{*} = \frac{1}{3} \approx 0.333$ and $p=\sqrt{5}-2 \approx 0.236$. In \fref{fig:vector_field3} we analyze the vector fields in this case, and in \fref{fig:vector_field4} another example of the language model with fixed $\alpha=2.5$, $Q=0.1$ and $p=0.1$, with critical parameter values $b_{c}=0.5668$, $b^{*}_{p}=0.4411$, $b_{p}=0.3240$. 

Other fixed points outside the polarization line are also possible and they may become stable for certain parameter values, thus defining additional transition lines. The fixed points $\rho^{\text{st}}_{1}$, $\rho^{\text{st}}_{2}$ in this case must be obtained numerically from the equations
\begin{equation}
\label{fixed_points_pol}
\mu_{1}[\rho^{\text{st}}_{1},\rho^{\text{st}}_{2}]=0, \hspace{1cm} \mu_{2}[\rho^{\text{st}}_{1},\rho^{\text{st}}_{2}]=0,
\end{equation}
as a function of the parameters $(Q, b, p)$. The stability of the fixed points comes determined by the eigenvalues:
\begin{equation}
\lambda_{1,2}=\dfrac{1}{2} \left( \tau \pm \sqrt{\tau^2-4 \Delta} \right), \hspace{0.4cm} \Delta=\lambda_{1} \lambda_{2}, \hspace{0.4cm} \tau=\lambda_{1} + \lambda_{2},
\end{equation}
where $\tau = J_{11} + J_{22}$ is the trace and $\Delta = J_{11} J_{22} - J_{12} J_{21}$ is the determinant of the Jacobian matrix evaluated at the fixed point $\rho^{\text{st}}_{1}$, $\rho^{\text{st}}_{2}$ for which we want to evaluate the stability. If $\tau^2 > 4 \Delta$ the eigenvalues only have real part, which will be the case for the fixed points found in the models under study. The condition for the transition is then that one of the eigenvalues becomes equal to zero or equivalently $\Delta[\rho^{\text{st}}_{1},\rho^{\text{st}}_{2}]=0$. This condition together with \eref{fixed_points_pol} will determine all transition lines.

In \fref{fig:vector_field5} we show that other stable fixed points are possible, for the majority-vote model with fixed $b=0.8$ and $p=0.1$, using the average rate \eref{rate_def} and \eref{rate_majority} with $z=20$. These (four) new states are close to $\rho_{1}^{\text{st}} \approx 0$, $\rho_{2}^{\text{st}} \approx 0.5$; $\rho_{1}^{\text{st}} \approx 0.5$, $\rho_{2}^{\text{st}} \approx 0$; $\rho_{1}^{\text{st}} \approx 0.5$, $\rho_{2}^{\text{st}} \approx 1$ and $\rho_{1}^{\text{st}} \approx 1$, $\rho_{2}^{\text{st}} \approx 0.5$, that is one community in the coexistence state and the other in one of the two consensus states. The region in the parameter space where these new states are stable is shown in the phase diagrams of the main text Fig. 6, and also in \fref{fig:phase_diagrams} for the language model with high $\alpha$.

\subsection{Special cases: intermediate $\alpha$ and very high $\alpha>5$}

In the main text, we divide the phenomenology in three types of behavior, these are: low $\alpha$ (voter-like), high alpha $\alpha \leq 5$ (majority-vote like) and very high alpha $\alpha > 5$, and we mainly focused on the first two behaviors. We will discuss now the case of an intermediate value of $\alpha$, in between the low and high $\alpha$ behaviors, and also the very high $\alpha > 5$ case.

\begin{figure}[h!]
\centering
\includegraphics[width=0.49\textwidth]{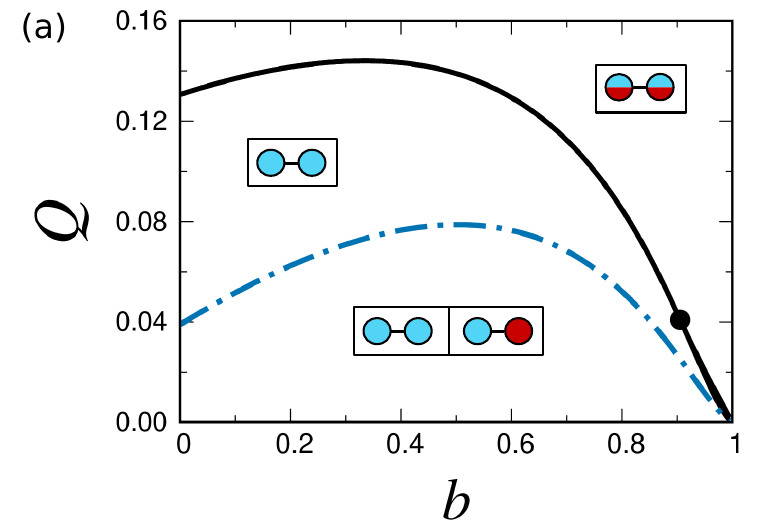}
\includegraphics[width=0.49\textwidth]{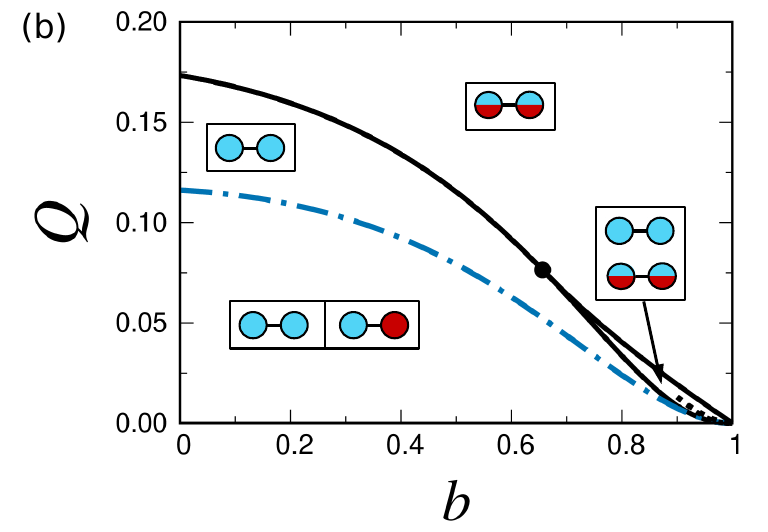}
\includegraphics[width=0.49\textwidth]{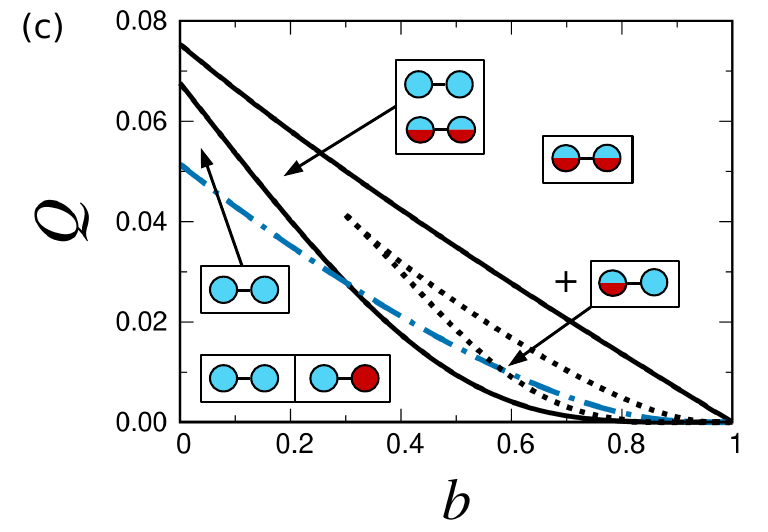}
\caption{Phase diagrams, in the noise-bias parameter space $(Q, b)$, of the homogeneous and polarized solutions for the language model with $\alpha=1.5$ (panel a), $\alpha=2.5$ (panel b), and $\alpha=6$ (panel c) with the inclusion of algorithmic bias and community structure with fixed $p=0.1$. There are two different polarized states: (i) standard polarization (different color circles), delimited by the dash-dotted (blue), and (ii) partial polarization (mixed with full color circles), delimited by the black dotted line. Symmetric states, obtained by exchanging blue and red colors, are always possible in the same parameter region. The final state $t \rightarrow \infty$ in a parameter region with several possible stable states is determined by the initial condition. The average rates used to obtain the transition lines are calculated using \eref{average_rate_b_hc} in the highly connected limit.}
\label{fig:phase_diagrams}
\end{figure}

According to our discussion, in the low $\alpha$ case $Q_{c}(\alpha, b)$ and $Q_{p}(\alpha, b, p)$ increase as a function $b$ up to a maximum, while in the high $\alpha$ case they decrease as a function of $b$. If we focus on $Q_{c}(\alpha, b)$ \eref{critical_language_bias}, the separation of the two behaviors occurs at $\alpha=2$, such that a maximum exists for $\alpha <2$ while for $\alpha>2$ it disappears. However, if we focus on $Q_{p}(\alpha, b, p)$, the transition value of $\alpha$ depends on $p$. As we do not have an expressions for $Q_{p}(\alpha, b, p)$, we will use $Q^{*}_{p}(\alpha, b, p)$ \eref{critical_language_bias_com} alternatively, obtaining the value $\alpha=2/(1-p)$ when the maximum appears/disappears, for $p=0.1$ it is $\alpha = 20/9\approx 2.22$. In the very high $\alpha>5$ case, the transition is always discontinuous. In \fref{fig:phase_diagrams}, we show the phase diagrams in some cases of interest with intermediate and very high $\alpha$ values ($\alpha=1.5$, $\alpha=2.5$ and $\alpha=6$).

\subsection{Examples of fixed points and vector fields}

\begin{figure}[h!]
\centering
\includegraphics[width=0.32\textwidth]{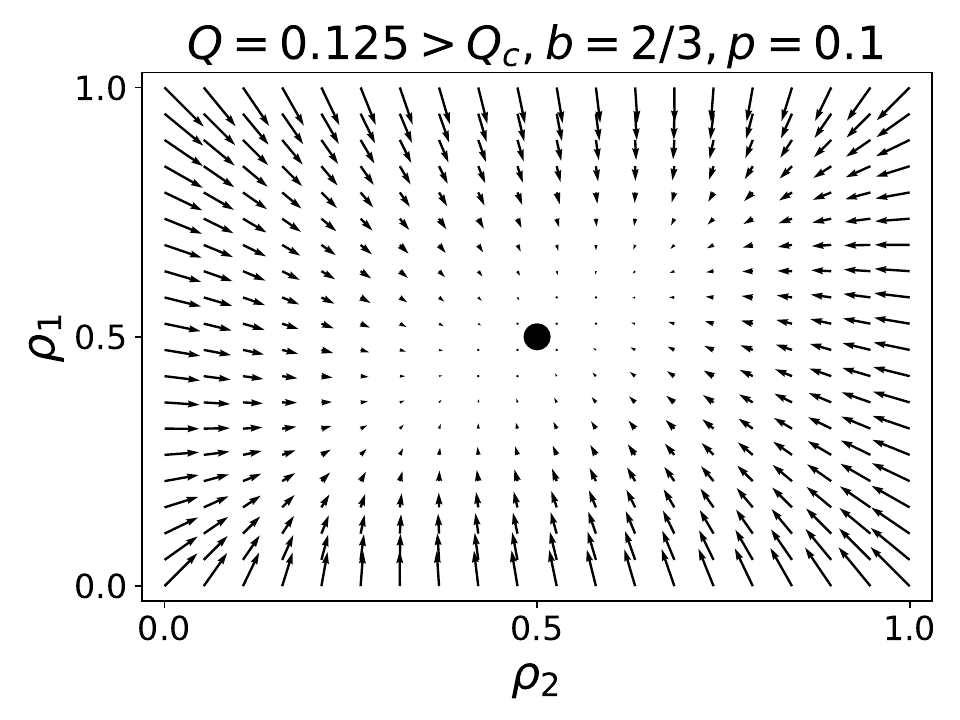}
\includegraphics[width=0.32\textwidth]{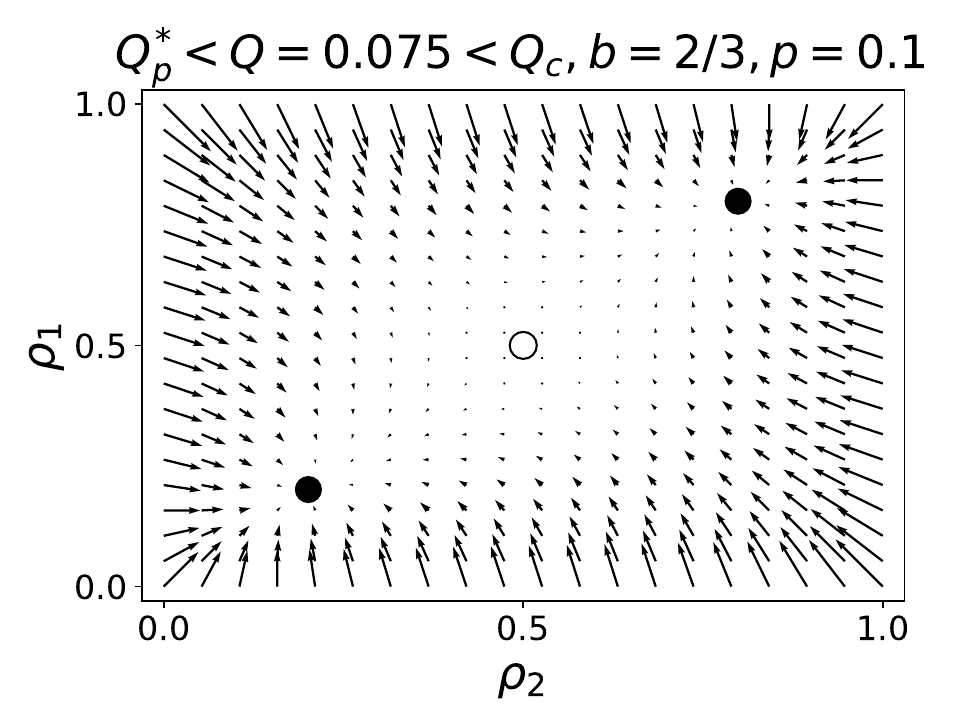}
\includegraphics[width=0.32\textwidth]{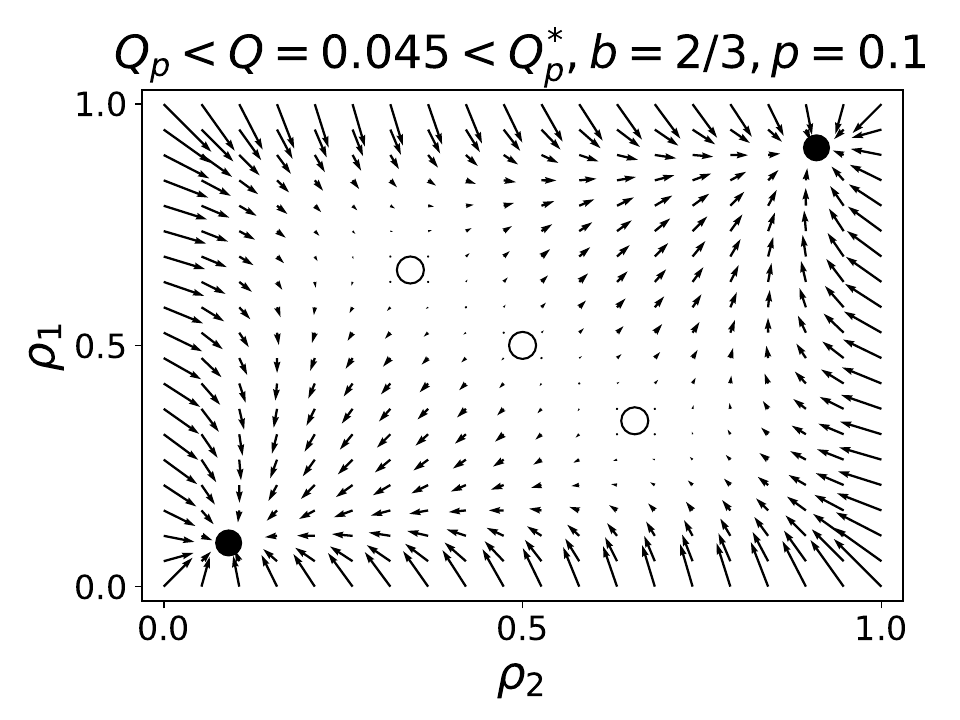}

\includegraphics[width=0.32\textwidth]{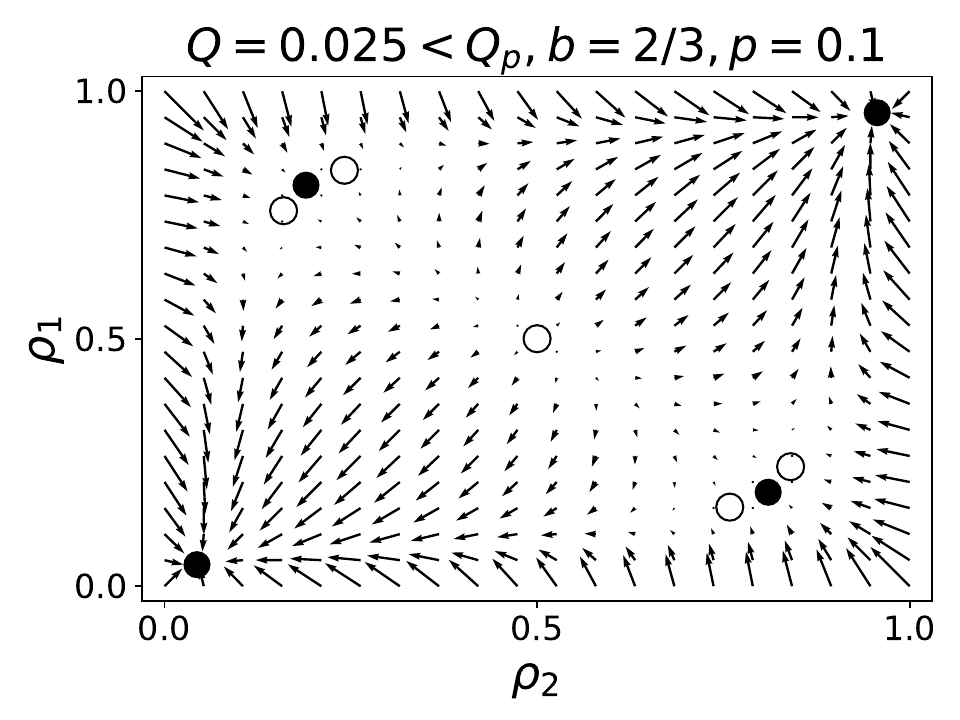}
\includegraphics[width=0.32\textwidth]{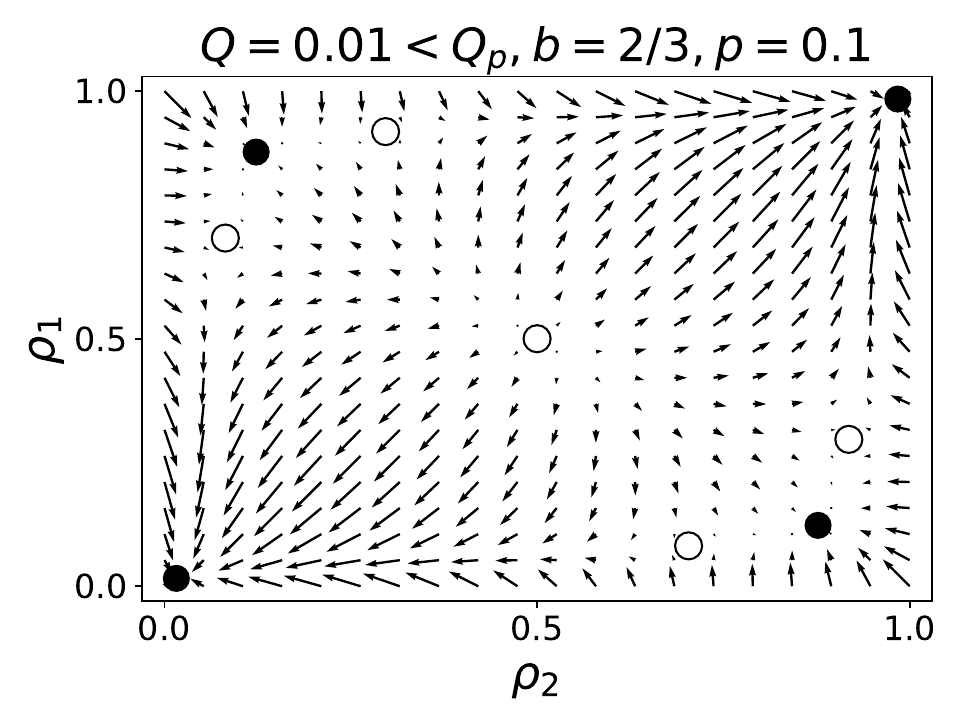}
\includegraphics[width=0.32\textwidth]{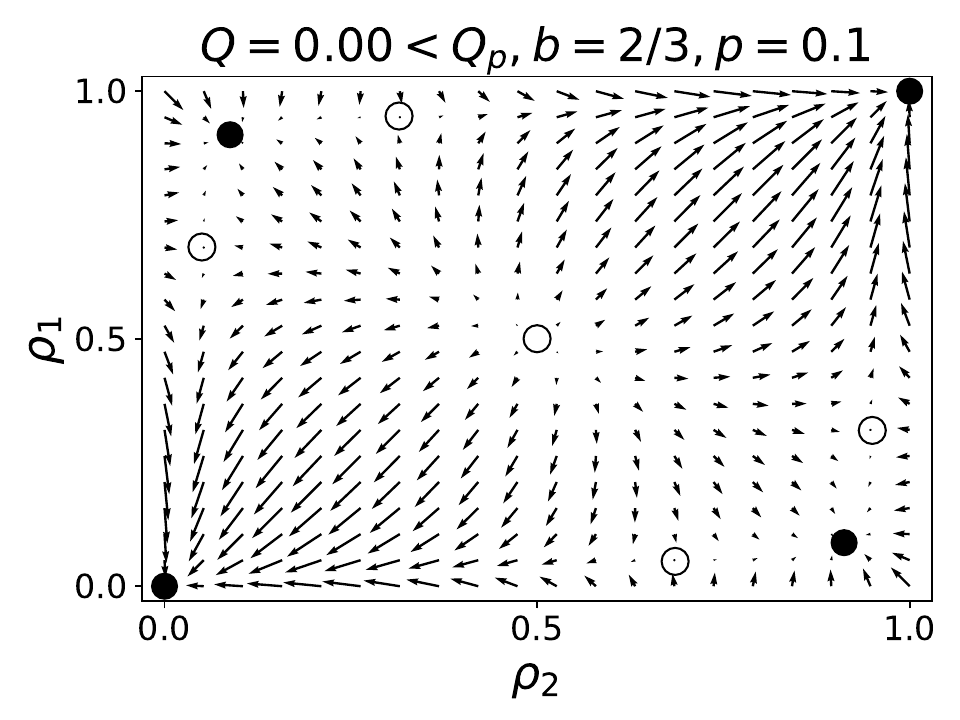}
\caption{Vector fields and fixed points for the noisy voter model with different values of $Q$ specified in the titles, fixed $b=2/3$ and $p=0.1$.}
\label{fig:vector_field1}
\end{figure}

\begin{figure}[h!]
\centering
\includegraphics[width=0.32\textwidth]{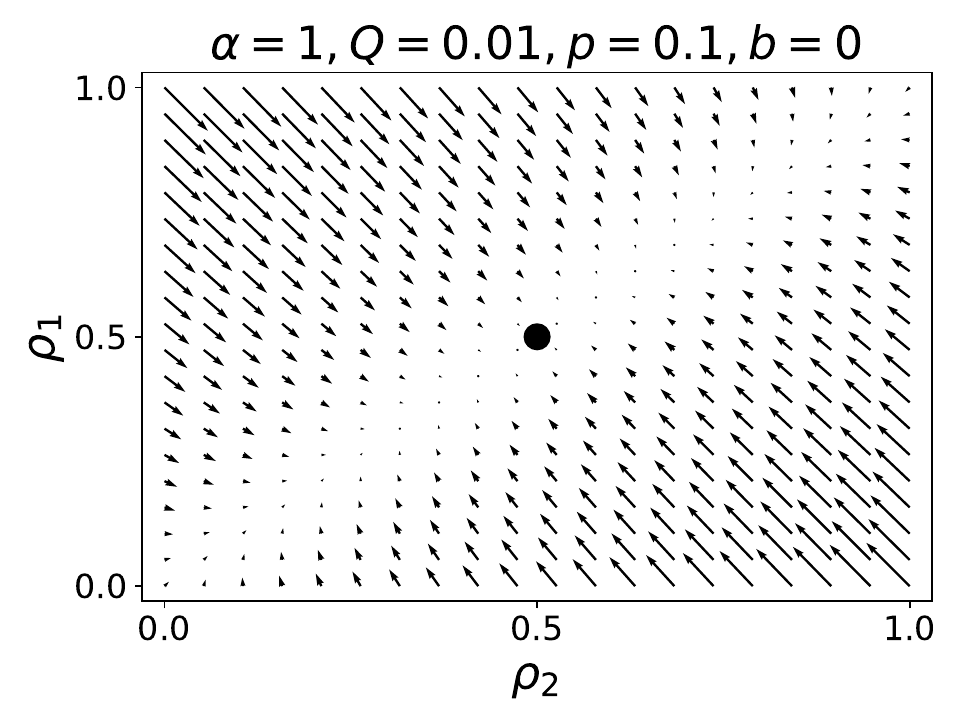}
\includegraphics[width=0.32\textwidth]{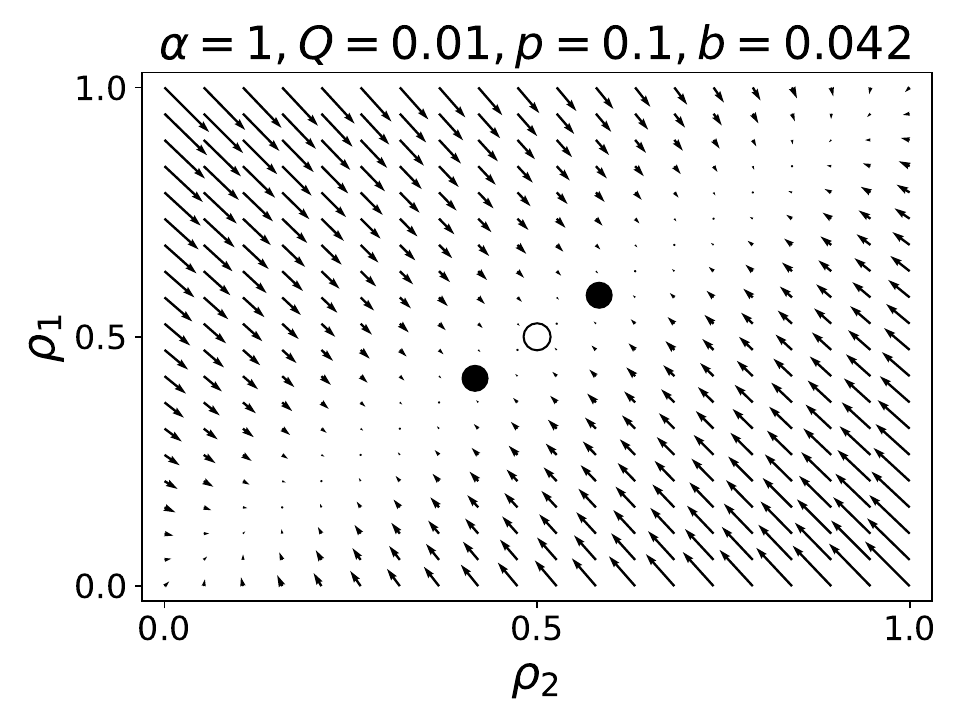}
\includegraphics[width=0.32\textwidth]{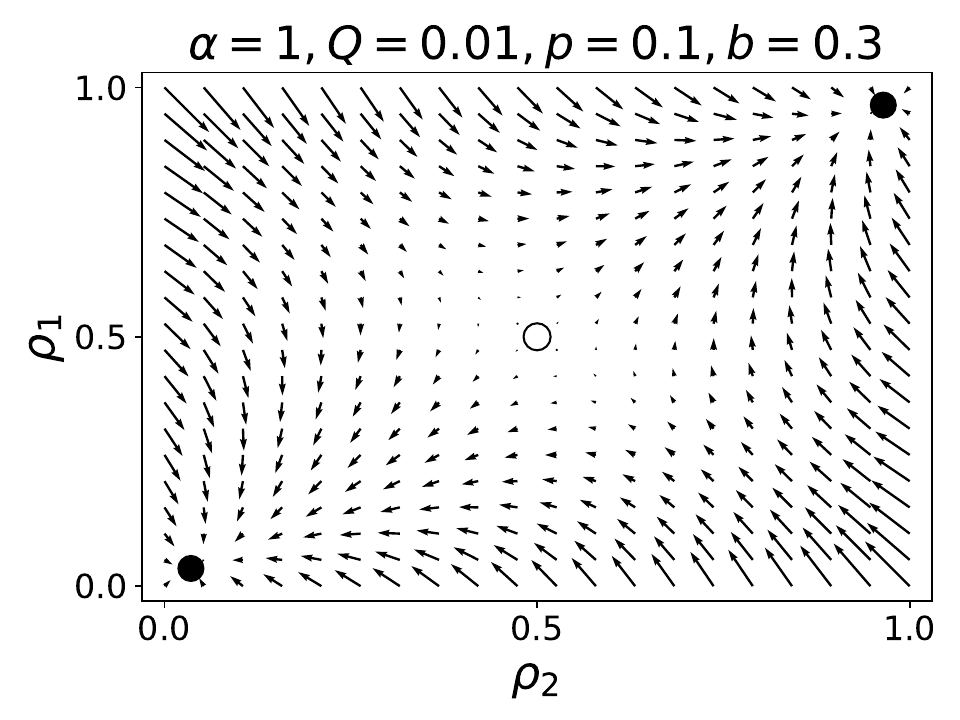}

\includegraphics[width=0.32\textwidth]{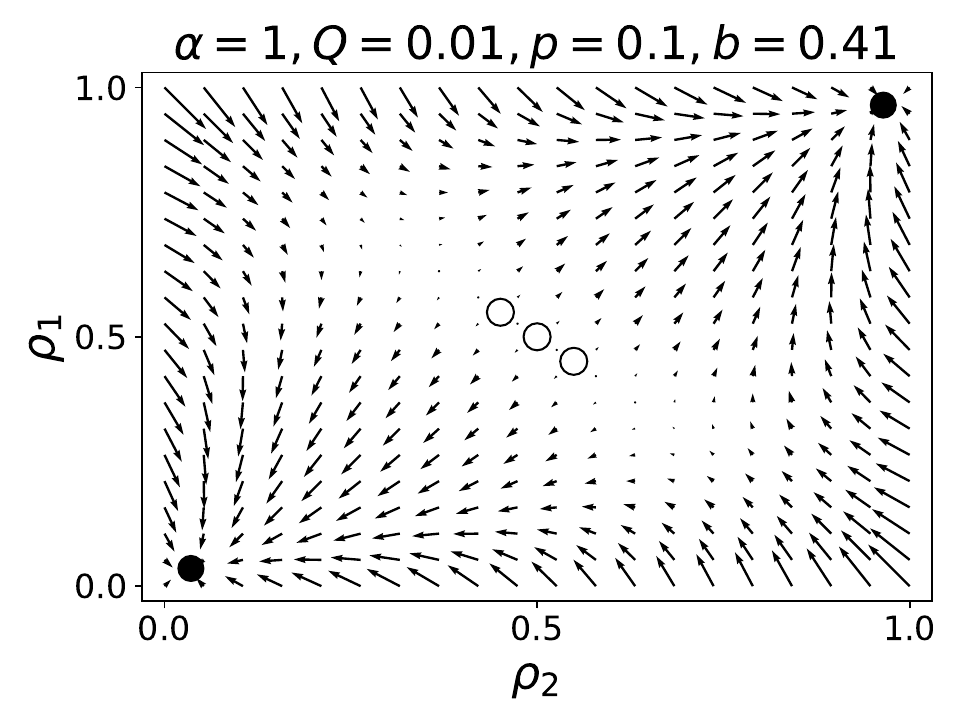}
\includegraphics[width=0.32\textwidth]{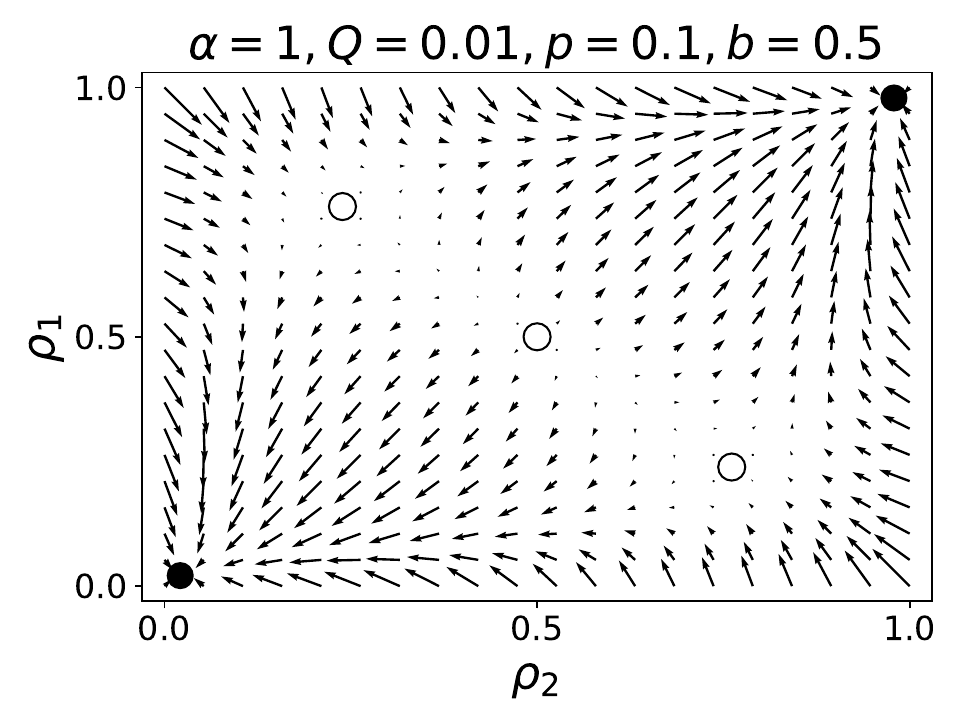}
\includegraphics[width=0.32\textwidth]{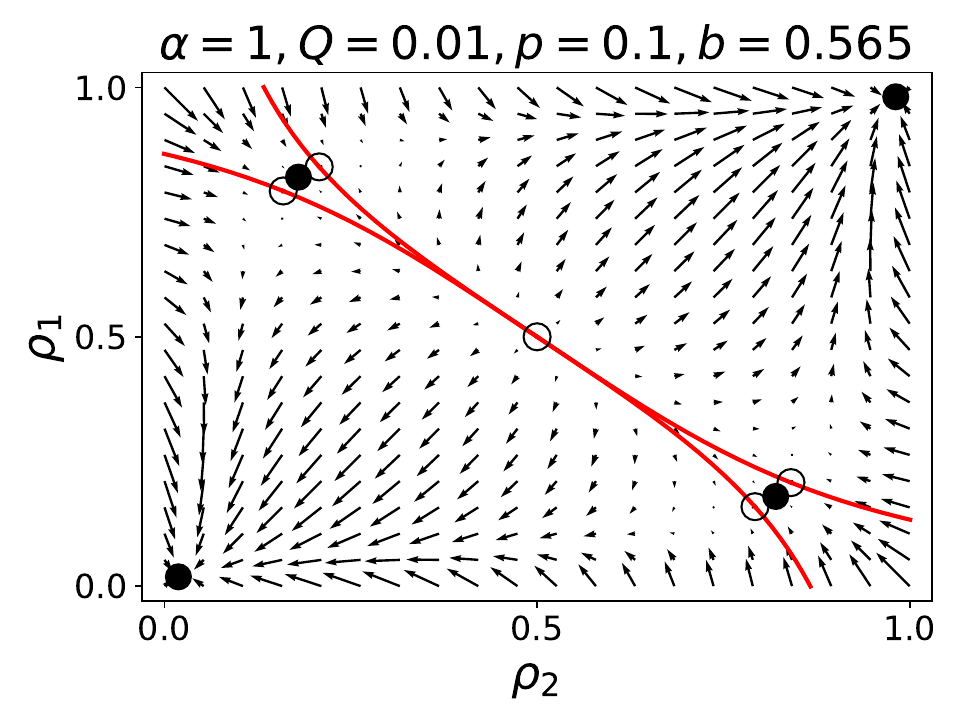}

\includegraphics[width=0.32\textwidth]{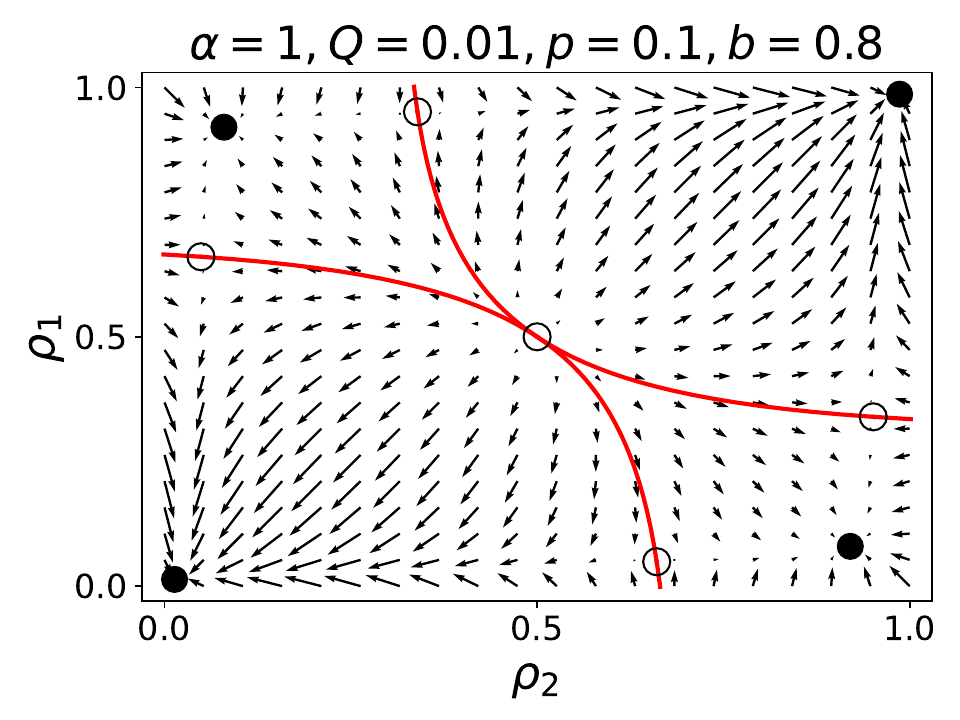}
\caption{Vector fields and fixed points for the noisy voter model with different values of $b$ specified in the titles, fixed $Q=0.01$ and $p=0.1$. The red lines indicated the basin of attraction of the polarized state.}
\label{fig:vector_field2}
\end{figure}

\begin{figure}[h!]
\centering
\includegraphics[width=0.32\textwidth]{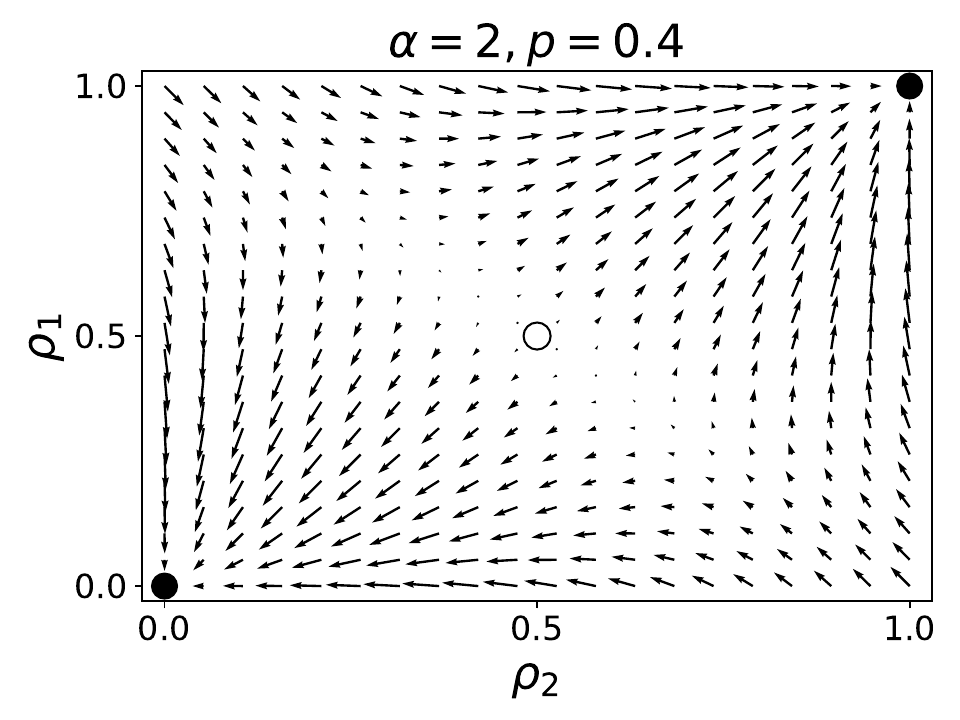}
\includegraphics[width=0.32\textwidth]{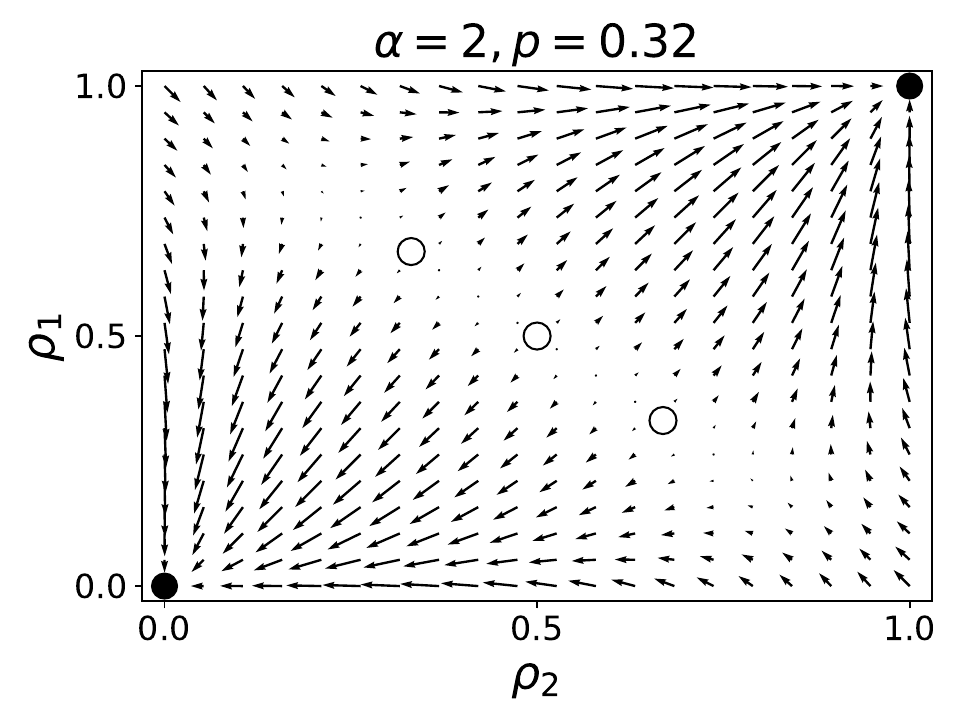}
\includegraphics[width=0.32\textwidth]{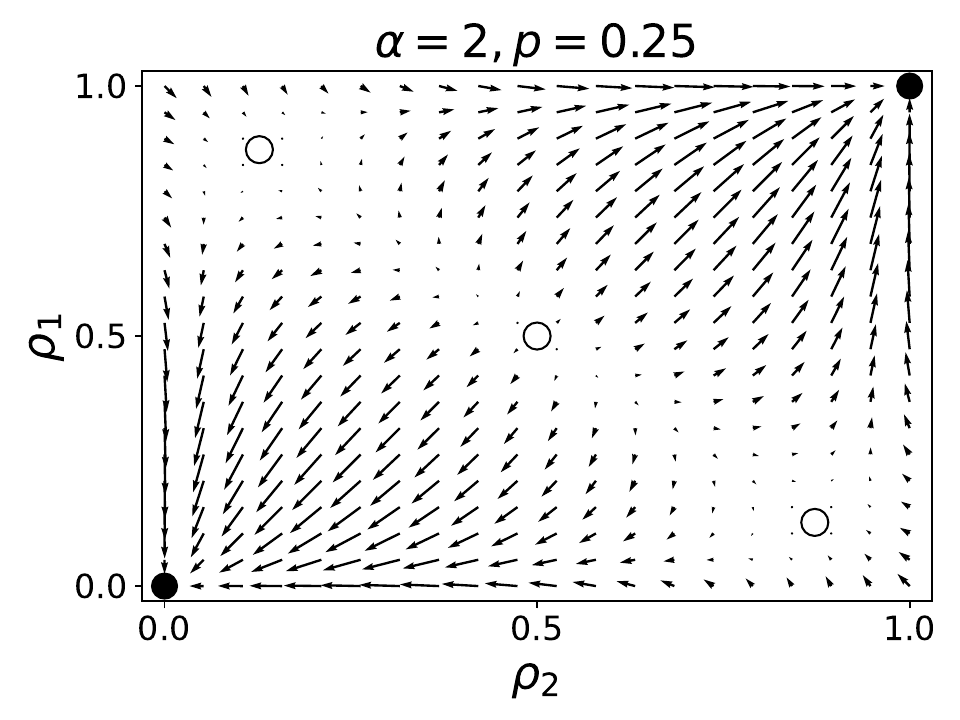}

\includegraphics[width=0.32\textwidth]{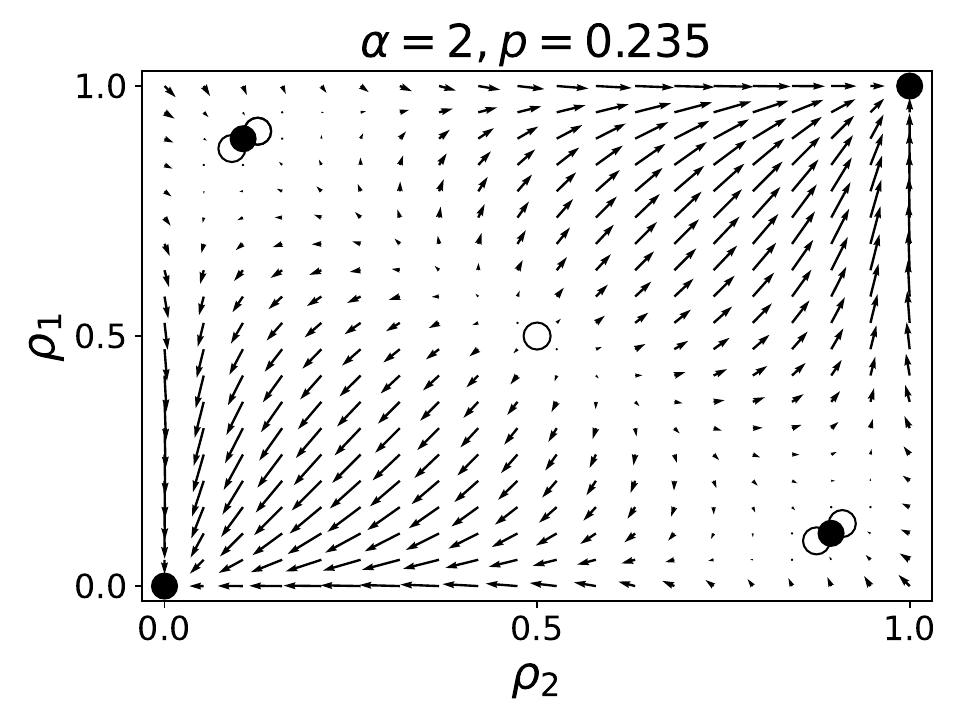}
\includegraphics[width=0.32\textwidth]{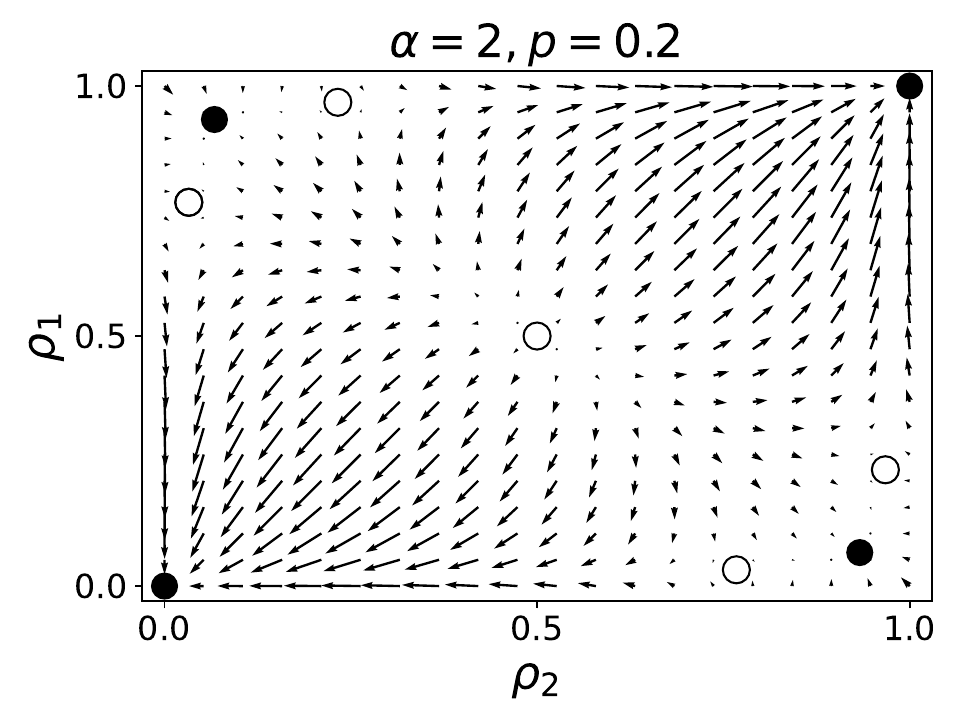}
\includegraphics[width=0.32\textwidth]{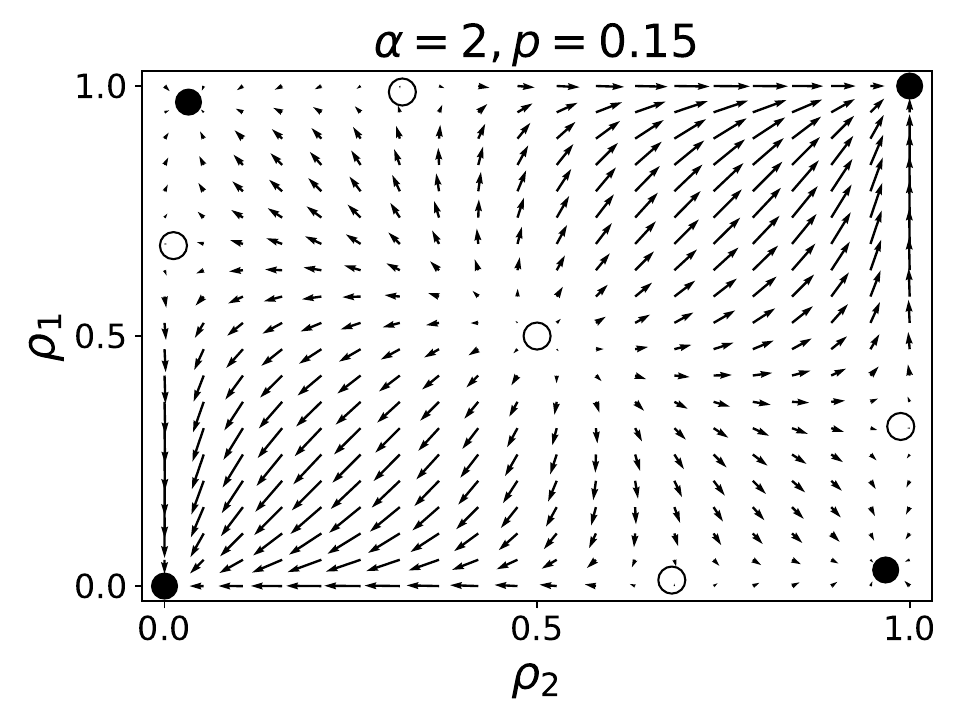}
\caption{Vector fields and fixed points for the language model with different values of $p$ specified in the titles and fixed $\alpha=2$, $Q=0$ and $b=0$.}
\label{fig:vector_field3}
\end{figure}

\begin{figure}[h!]
\centering
\includegraphics[width=0.32\textwidth]{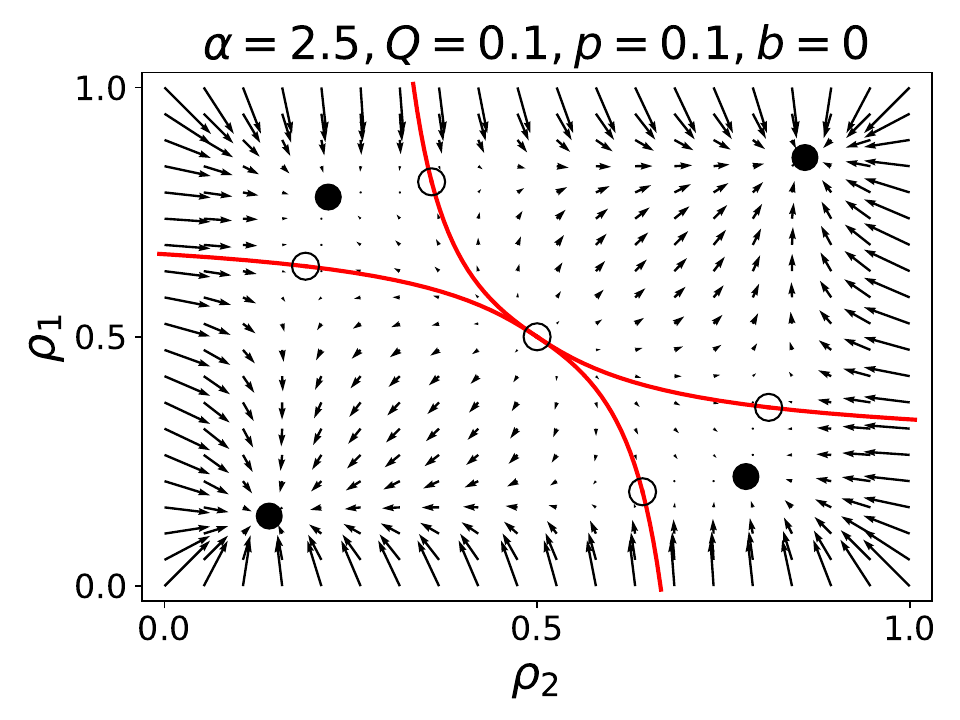}
\includegraphics[width=0.32\textwidth]{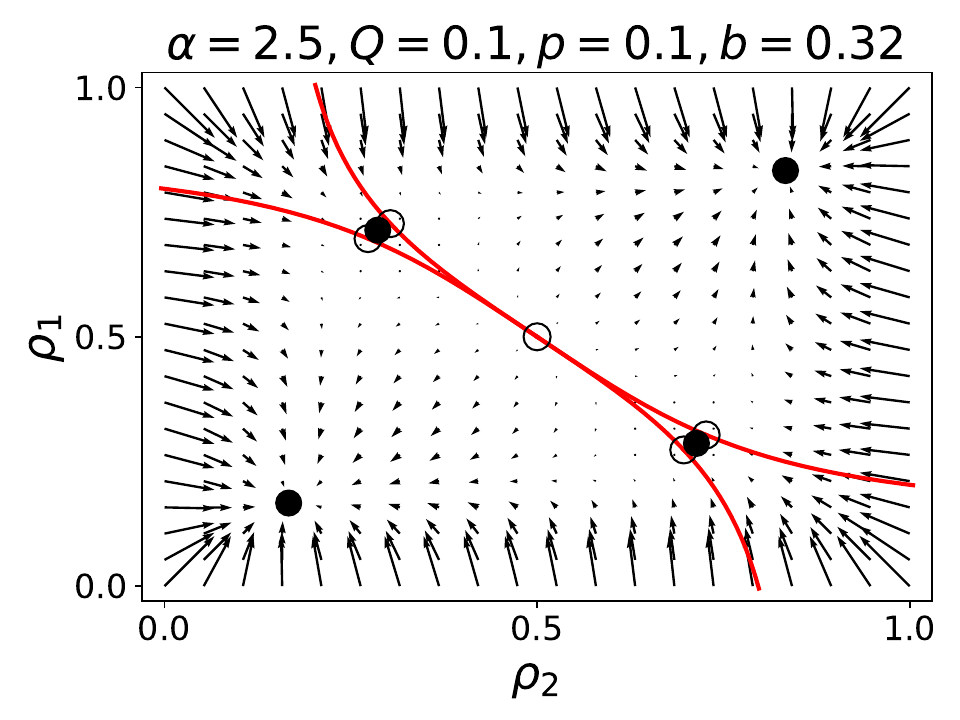}
\includegraphics[width=0.32\textwidth]{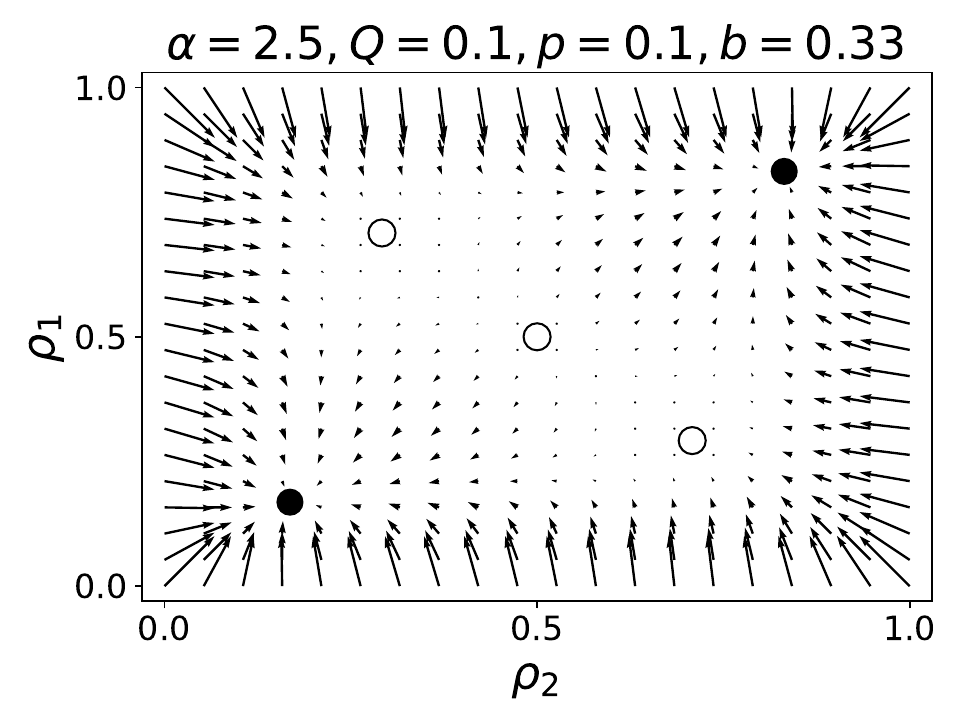}

\includegraphics[width=0.32\textwidth]{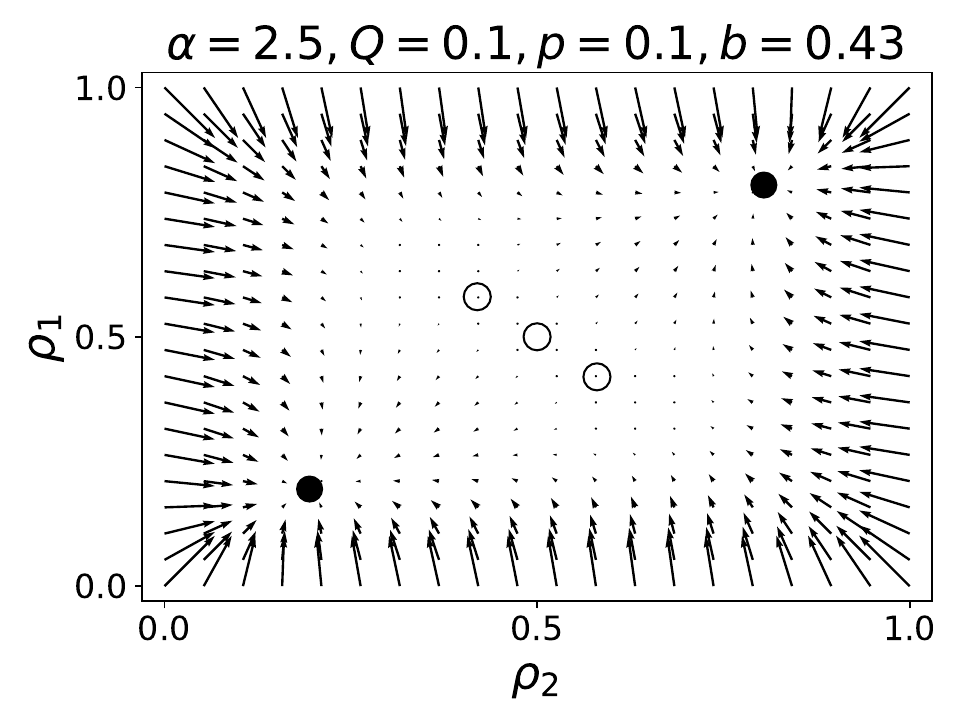}
\includegraphics[width=0.32\textwidth]{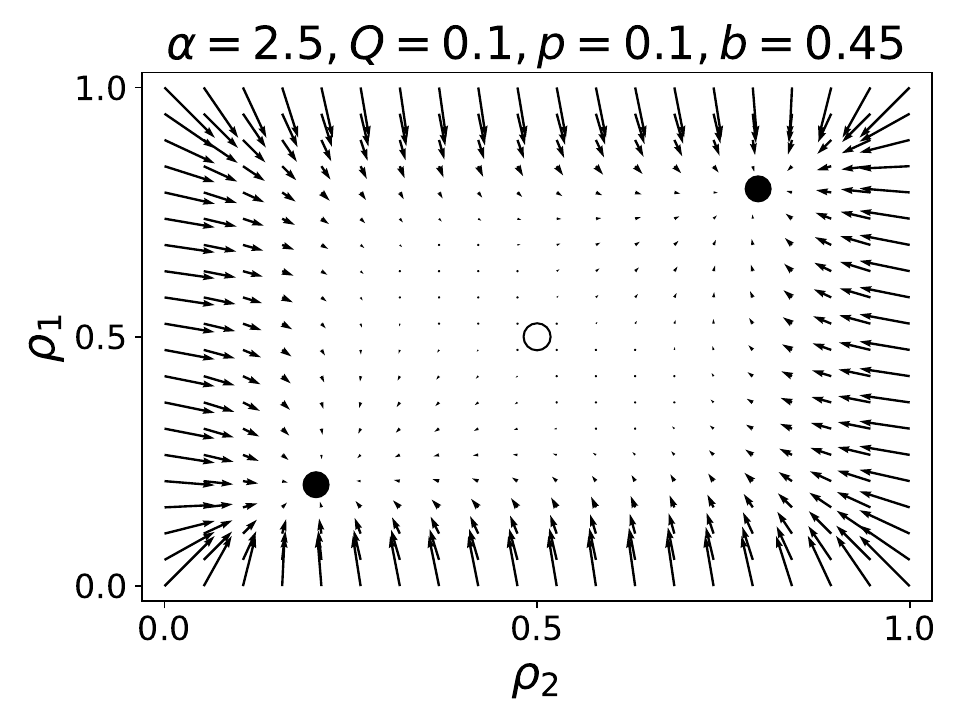}
\includegraphics[width=0.32\textwidth]{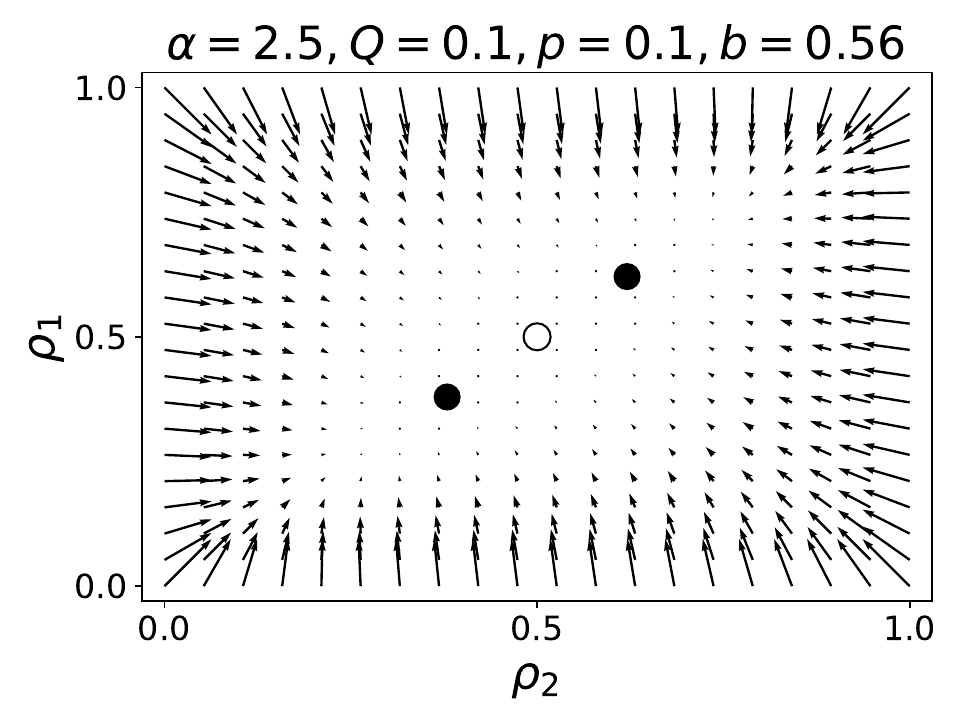}

\includegraphics[width=0.32\textwidth]{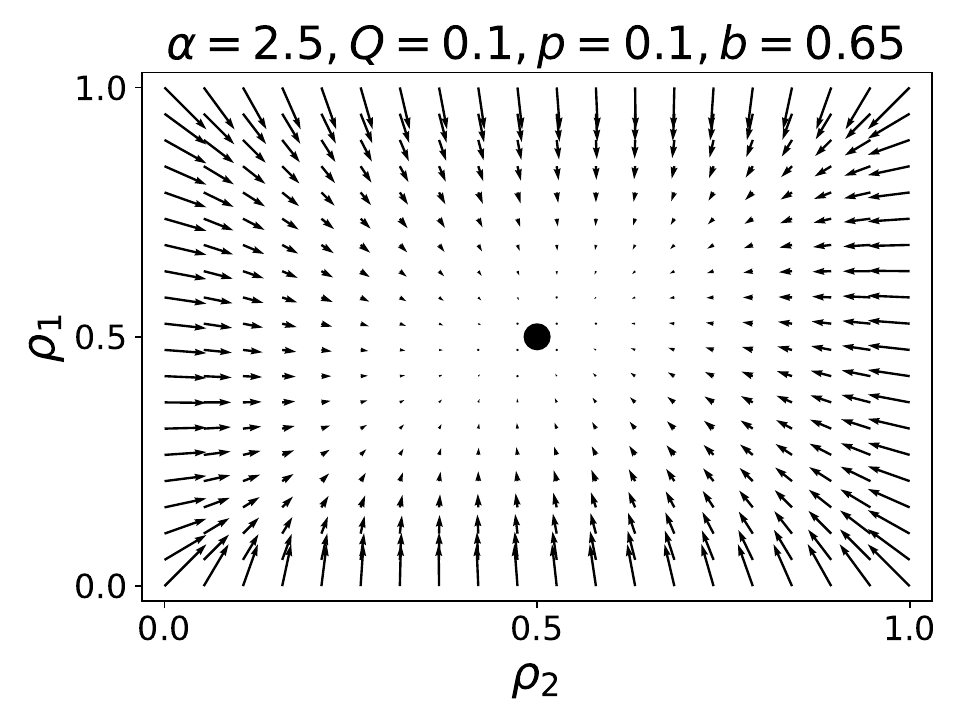}
\caption{Vector fields and fixed points for the language model with different values of $b$ specified in the titles, fixed $\alpha=2.5$, $Q=0.1$ and $p=0.1$. The red lines indicated the basin of attraction of the polarized state.}
\label{fig:vector_field4}
\end{figure}

\begin{figure}[h!]
\centering
6\includegraphics[width=0.45\textwidth]{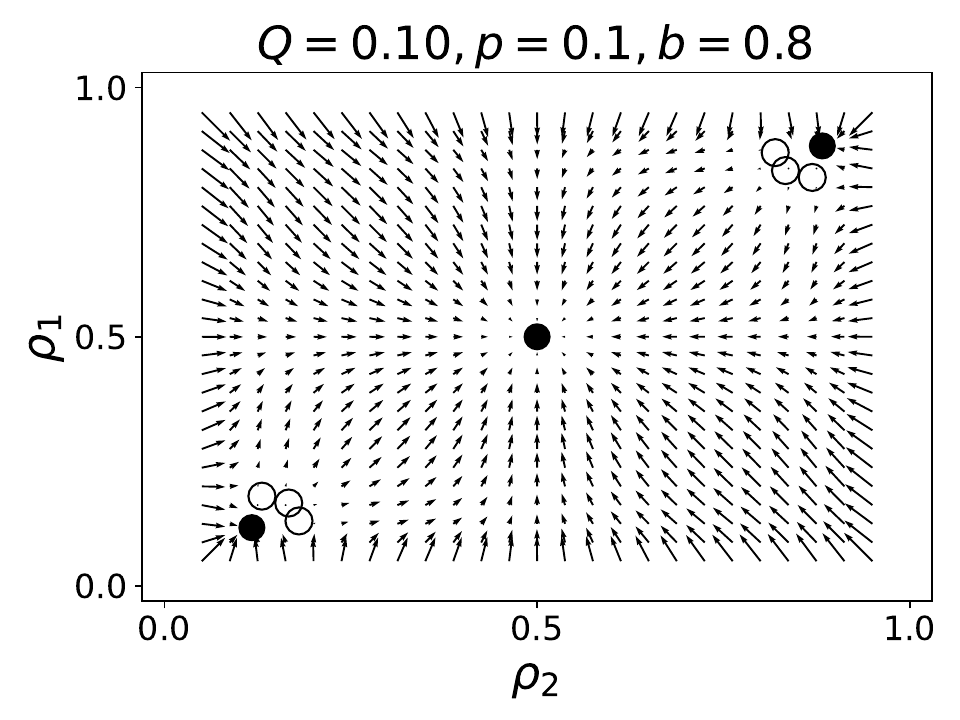}
\includegraphics[width=0.45\textwidth]{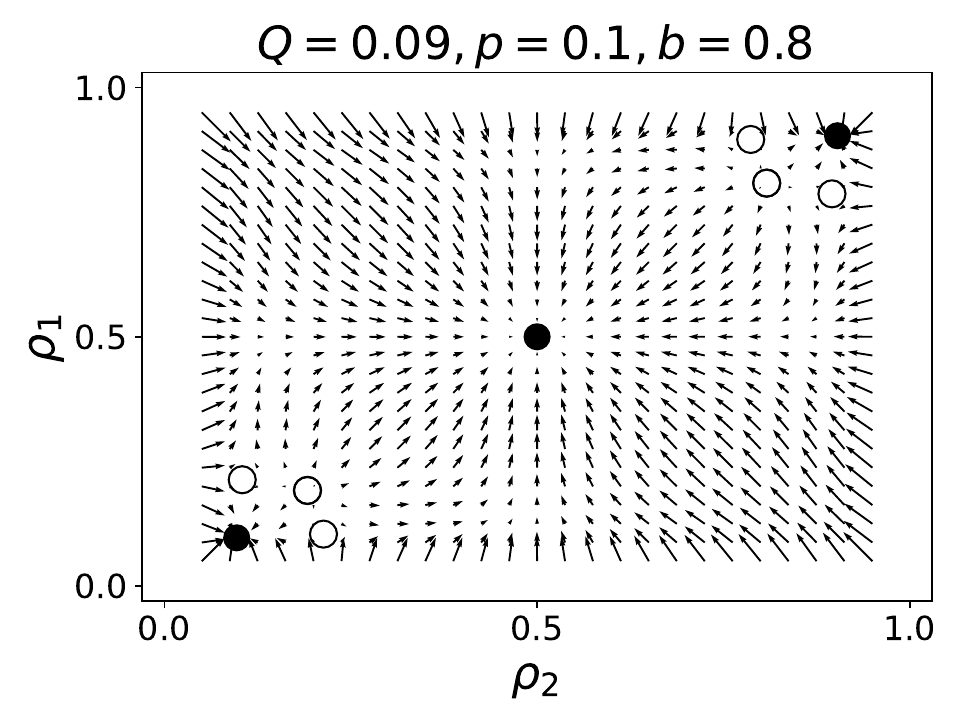}

\includegraphics[width=0.45\textwidth]{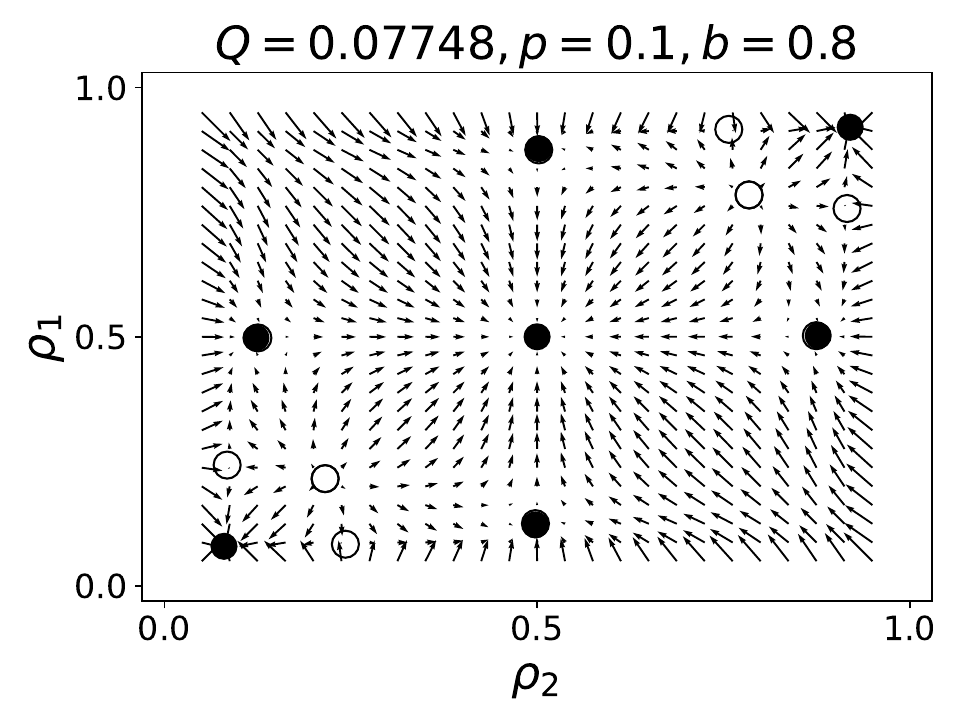}
\includegraphics[width=0.45\textwidth]{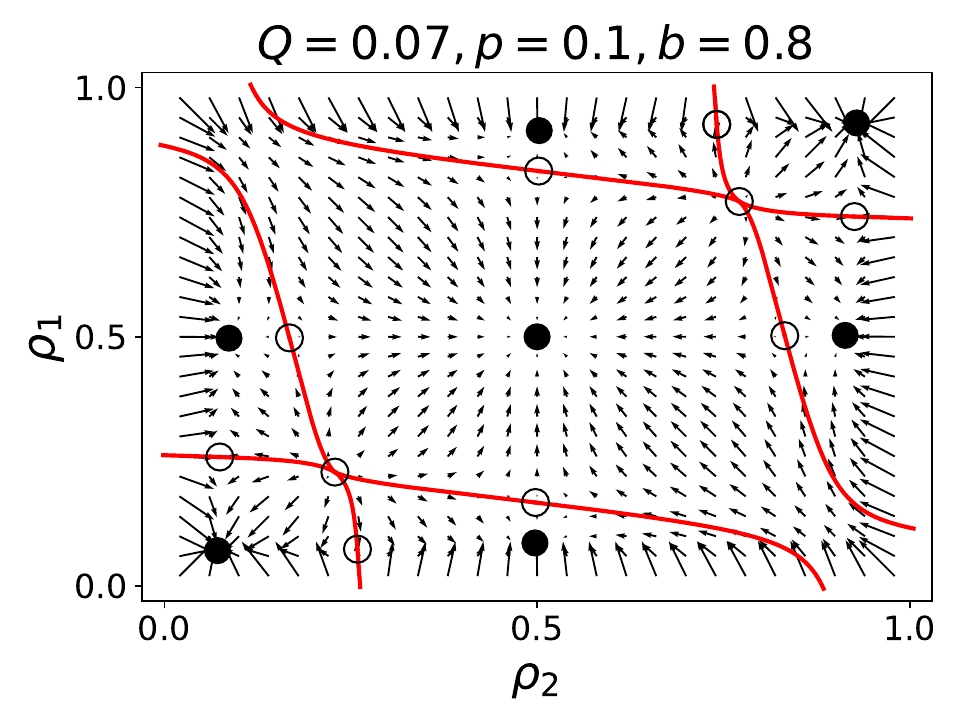}

\includegraphics[width=0.45\textwidth]{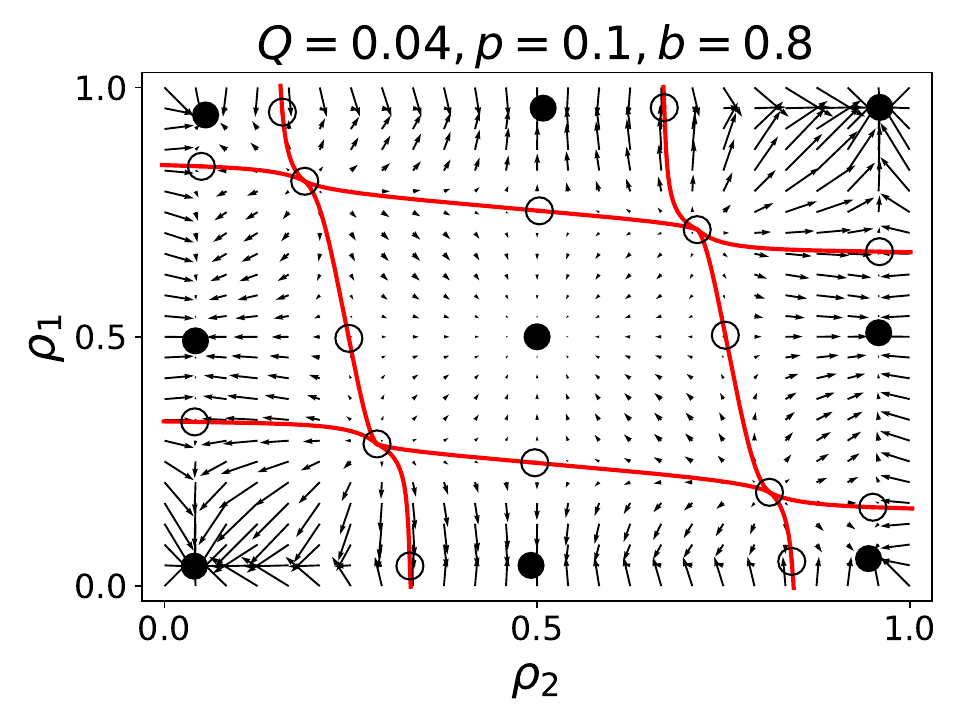}
\includegraphics[width=0.45\textwidth]{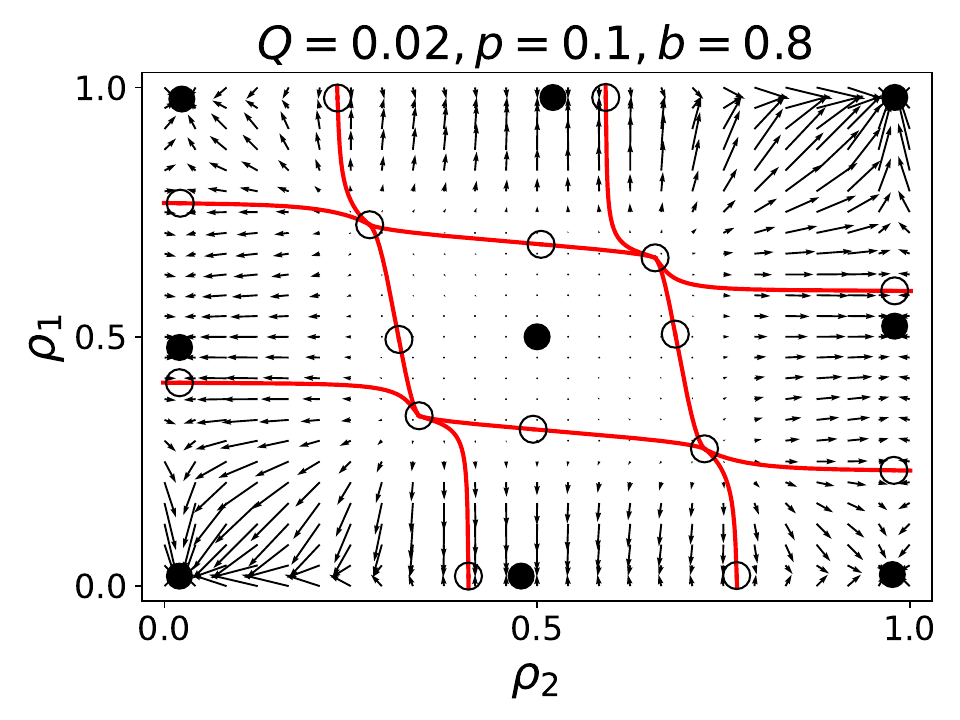}

\caption{Vector fields and fixed points for the majority-vote model with different values of $Q$ specified in the titles, fixed $b=0.8$ and $p=0.1$. The average rates \eref{rate_def} and \eref{rate_majority} are used with $z=20$. The red lines indicated the basin of attraction of the different stable states.}
\label{fig:vector_field5}
\end{figure}

In \fsref{fig:vector_field1}{fig:vector_field6}, we show different examples of the fixed points and vector fields \eref{dyn_mod_1} and \eref{dyn_mod_2}. In the main text, only an schematic representation was shown, while in this case the exact calculation for different parameter values is provided. The fixed points are classified as stable and plotted as full dots, and unstable or saddles as empty dots. Note in the figures that when the different transition parameter values are crossed, a fixed point changes stability and new fixed points appear at the same position of that fixed point. The basin of attraction is also indicated, only for some cases of interest, as red lines.

\begin{figure}[h!]
\centering

\includegraphics[width=0.45\textwidth]{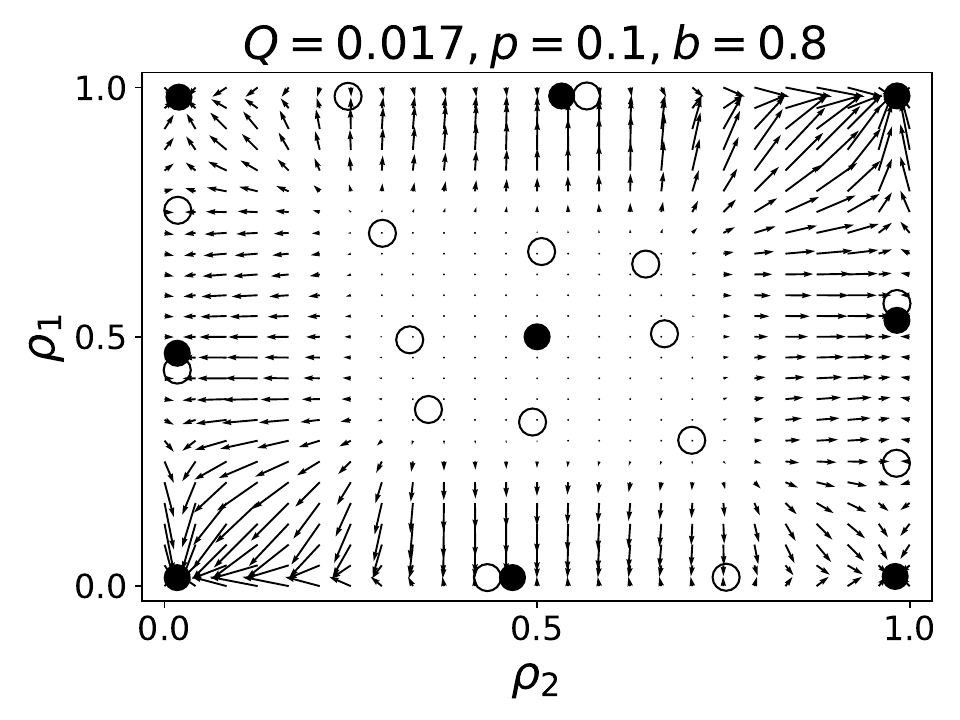}
\includegraphics[width=0.45\textwidth]{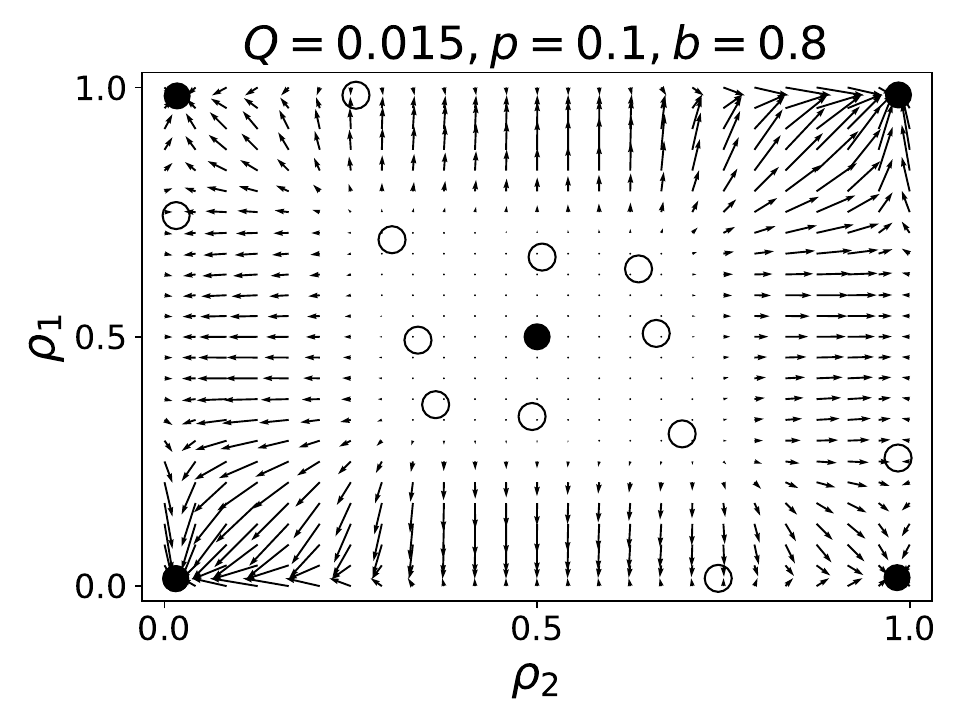}

\includegraphics[width=0.45\textwidth]{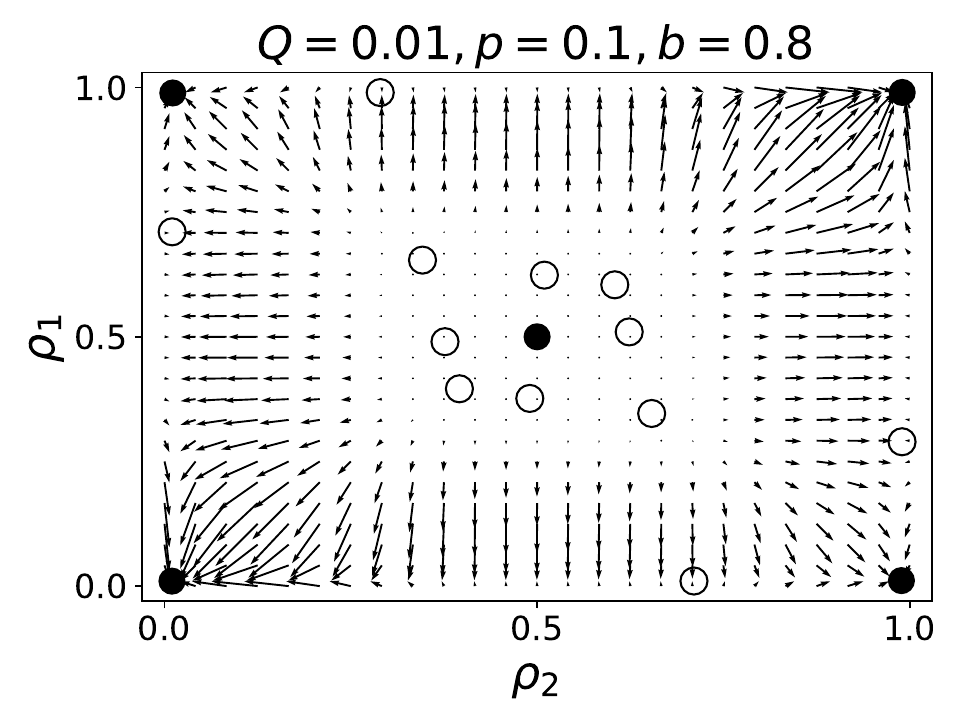}
\includegraphics[width=0.45\textwidth]{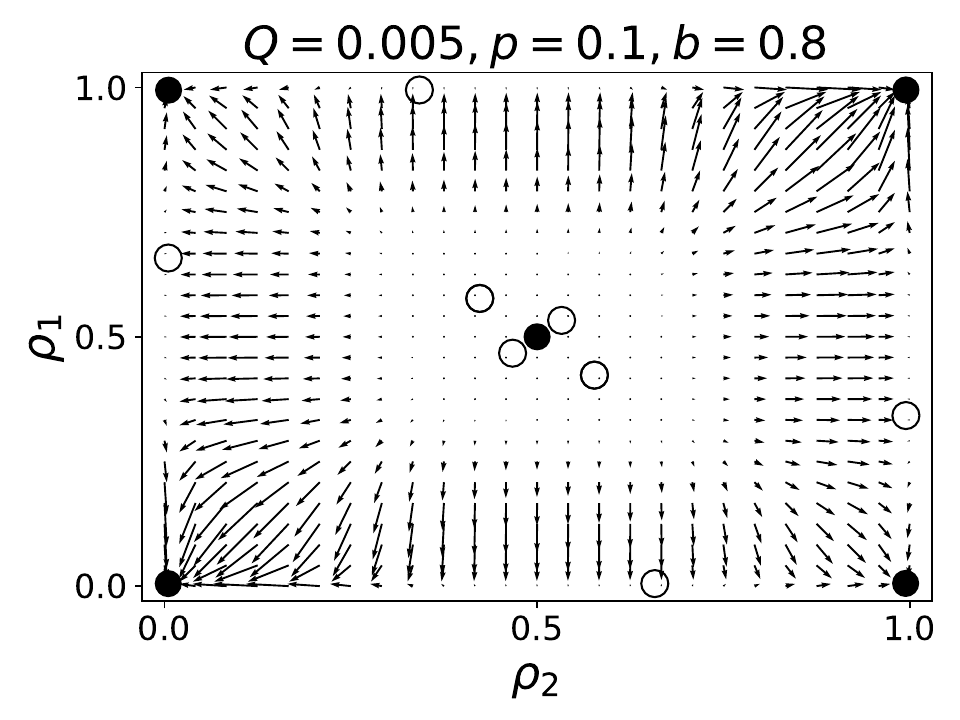}

\caption{Vector fields and fixed points for the majority-vote model with different values of $Q$ specified in the titles, fixed $b=0.8$ and $p=0.1$. The average rates Eqs. (\ref{rate_def}, \ref{rate_majority}) are used with $z=20$.}
\label{fig:vector_field6}
\end{figure}

\section{Comparison with numerical simulations}

\begin{figure}[ht!]
\begin{center}

\includegraphics[width=0.7\textwidth]{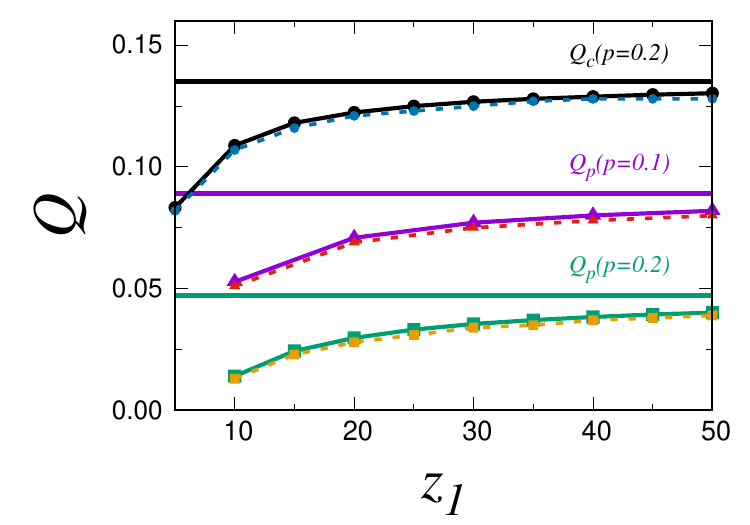}
\caption{Transition values $Q_{c}$ (black) and $Q_{p}$ (purple and green) as a function of the average degree $z_{1}$ in community $1$, for the language model with fixed parameter values $\alpha=2$, $b=0.5$, on degree regular symmetric communities $p_{1}=p_{2}=p$ with total average degree $z=(1+p)z_{1}$. The solid horizontal lines correspond to the mean field solution in the highly connected limit, independent of $z_{1}$. The black circle, purple triangle and green square dots correspond to the pair approximation results with fixed $p=0.1$ (purple) and $p=0.2$ (green and black). The circle (blue), triangle (red) and square (yellow) dots are the results obtained from numerical simulations of the model with the equivalent parameter values.}
\label{fig_finite_network}
\end{center}
\end{figure}

In order to check the validity and accuracy of the proposed theoretical approaches, we will compute the stationary states and transition parameter values using also a Monte Carlo simulation method, see Section III. A of the main text. We generate a long trajectory starting with a given initial condition, close to the final state that we want to analyze, and take averages sampling each Monte Carlo step (after thermalizing). The averages are restricted to a confined zone of the phase space where the trajectory fluctuates, in order to avoid stochastic jumps from one final state to another, for example jumps between the symmetric consensus states. In \fref{fig_finite_network} we computed the transition parameter values $Q_{c}$ and $Q_{p}$ for networks with different degree values (specified in the figure) and fixed $p=0.1$ and $p=0.2$, for the language model with an intermediate value of $\alpha=2$ and bias $b=0.5$. We show the predictions of the pair approximation, developed in \sref{sec_pa}, the highly connected limit, developed in \sref{sec_highly_connected}, and the Monte Carlo simulations. The pair approximation shows a good accuracy as compared to numerical simulations, with small discrepancies (e.g., it predicts a higher value of $Q_{p}$) that falls in most cases in the error range of the simulation, i.e., $Q_{p}\pm 0.001$. In \fref{fig_finite_network_bif} we show the bifurcation diagrams in a case of interest with $z_{1}=20$, under the same other conditions of \fref{fig_finite_network}, for the pair approximation and comparing to numerical simulations. The pair approximation captures almost perfectly the dependence of the order parameter $\vert \rho - 1/2 \vert$ with respect to the noise intensity $Q$, reproducing the nature of the different coexistence-consensus and polarized transitions. In all cases, for $z_{1} \rightarrow \infty$ the highly connected limit is recovered by the pair approximation and the Monte Carlo simulations.

\begin{figure}[ht!]
\begin{center}

\includegraphics[width=0.7\textwidth]{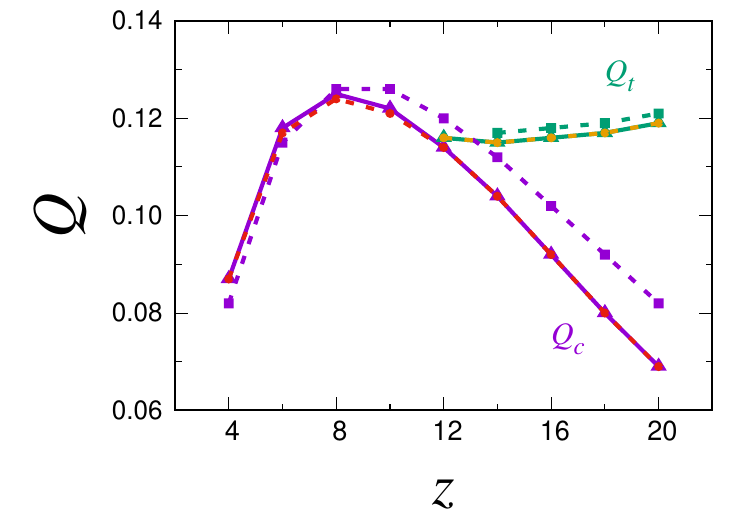}
\caption{Transition values $Q_{c}$ (purple) and $Q_{t}$ (green) as a function of the average degree $z$, for the majority-vote model with fixed bias $b=0.7$ in a $z-$regular network. The square dots correspond to the pair approximation, the triangle dots are the approximate master equation, and the circle dots (red and yellow) the Monte Carlo simulation results.}
\label{fig_finite_network2}
\end{center}
\end{figure}

\begin{figure}[ht!]
\begin{center}
\includegraphics[width=0.7\textwidth]{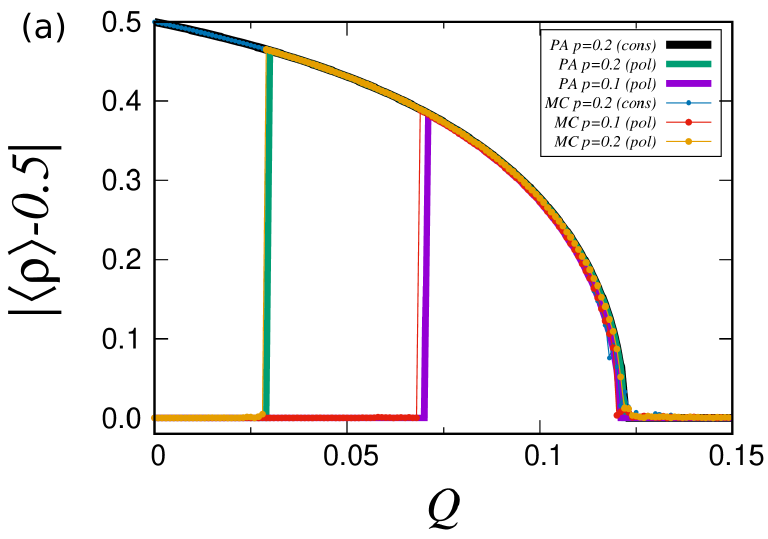}

\includegraphics[width=0.7\textwidth]{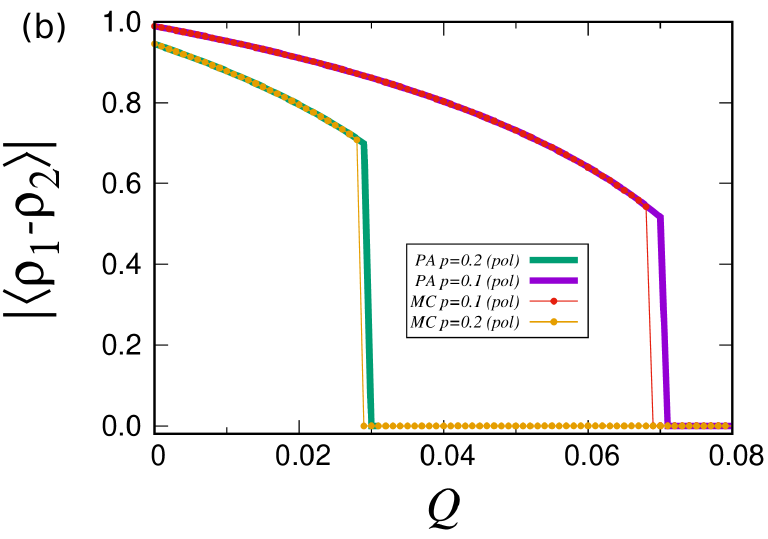}
\caption{Order parameters $\vert \rho - 1/2 \vert$, panel (a), and $\vert \rho_{1}-\rho_{2} \vert$, panel (b), as a function of the noise intensity $Q$ for the language model with $\alpha=2$, $b=0.5$ on a $z-$regular network with modular structure and $z_{1} = z_{2} = 20$, $z = (1+p) z_{1}$, the same case of Fig. S\ref{fig_finite_network} (a point in that figure). The color reference is also equivalent to Fig. S\ref{fig_finite_network}, that is: purple ($p=0.1$) and green ($p=0.2$) lines for the polarized initial condition and black ($p=0.2$) (covered by red and yellow points) for the homogeneous initial condition, all in the pair approximation description. Red points correspond to $p=0.1$ with polarized initial condition, yellow points to $p=0.2$ with polarized i.c., and for the blue points it is $p=0.2$ with homogeneous i.c., coming from numerical (Monte Carlo) simulations averaged over a long trajectory with final time $5 \times 10^{3}$ (after a thermalization of $5 \times 10^{3}$ steps) and system size $N=4 \times 10^{4}$.}
\label{fig_finite_network_bif}
\end{center}
\end{figure}

\begin{figure}[ht!]
\begin{center}
\includegraphics[width=0.7\textwidth]{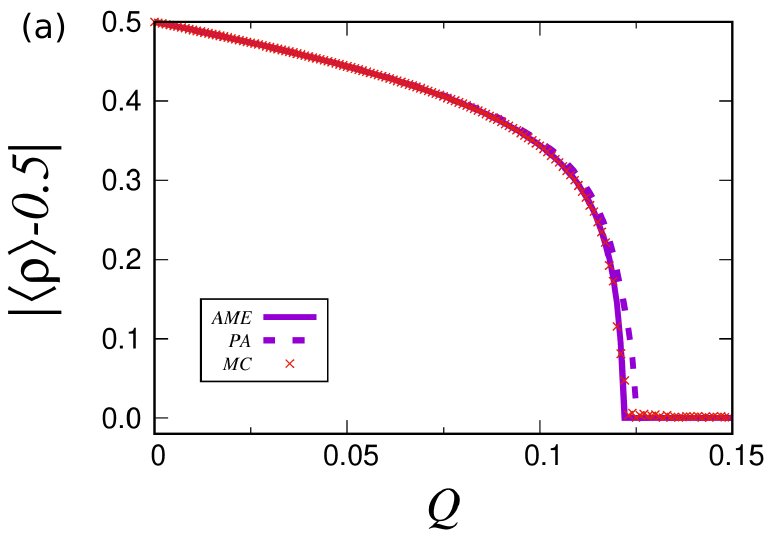}
\includegraphics[width=0.7\textwidth]{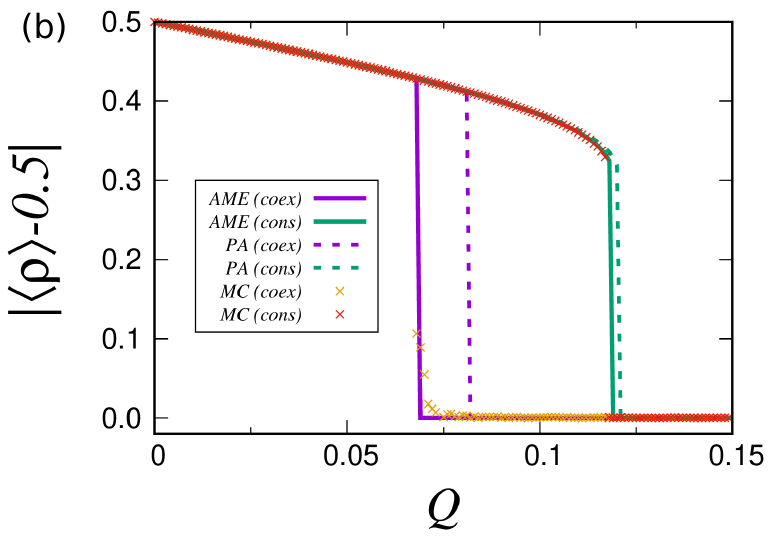}
\caption{Order parameter $\vert \rho - 1/2 \vert$ as a function of the noise intensity $Q$ for the majority-vote model with $b=0.7$ on a $z-$regular (homogeneous) network with $z =10$ (panel a) and $z=20$ (panel b), the same case of Fig. S\ref{fig_finite_network2} (two points in that figure, one case of continuous and another of discontinuous transitions). The color reference is also equivalent to Fig. S\ref{fig_finite_network2}, that is: solid (dashed) purple lines for the approximate master equation (pair approximation) prediction with coexistence initial condition, and solid (dashed) green lines for the approximate master equation (pair approximation) prediction with consensus initial condition. In panel (a), independently on the initial condition we obtain the same result. Red and yellow points are the equivalent obtained from numerical (Monte Carlo) simulations, averaging over a long trajectory with final time $5 \times 10^{3}$ (after a thermalization of $10^4$ steps) and system size $N=10^{4}$.}
\label{fig_finite_network_bif2}
\end{center}
\end{figure}

In \fref{fig_finite_network2} we show the transition parameter values $Q_{c}$ and $Q_{t}$ as a function of the mean degree $z$ for the majority-vote model with $b=0.7$ on a homogeneous $z-$regular network, determined using the pair approximation, the approximate master equation developed in \cite{Gleeson:2013}, and Monte Carlo simulations. In \fref{fig_finite_network_bif2} we show the bifurcation diagram for two cases of interest with $z=10$, continuous coexistence-consensus transition, and $z=20$, discontinuous transition. In these cases, some discrepancies of the pair approximation are observed, specially for the determination of $Q_{c}$ when the transition is discontinuous. These small discrepancies are well corrected by the approximate master equation approach, which can be directly applied from the methods in \cite{Gleeson:2013}, showing a perfect agreement compared to numerical simulations.

%explain colors better
%legends colors
%check references
%all section should be mentioned in the main text too
%maybe reference figures too

\bibliographystyle{unsrt}
\bibliography{chap_intro_biblio}